\documentclass[11pt]{article}
\usepackage{hyperref}
\usepackage{graphicx}
\usepackage{graphics}
\usepackage{dsfont}
\usepackage{epsfig}
\usepackage{amsmath,amssymb,amsthm,amscd}
\usepackage{float}
\usepackage{verbatim} 
\allowdisplaybreaks

\usepackage{tikz}
\usetikzlibrary{decorations.pathmorphing}
\usetikzlibrary{arrows,decorations.markings}

\newcommand{\be}{\begin{equation}}
\newcommand{\ee}{\end{equation}}

\numberwithin{equation}{section}

\begin{document}
\begin{titlepage}
{}~ \hfill\vbox{ \hbox{} }\break

\rightline{USTC-ICTS/PCFT-20-32}
\rightline{BONN-TH-2020-07}

\vskip 1.5 cm

\centerline{
\Large \bf Towards Refining the Topological Strings 
 } 
\vskip 0.2 cm
\centerline{\Large 
\bf 
on Compact Calabi-Yau 3-folds }   
\vskip 0.4 cm

\renewcommand{\thefootnote}{\fnsymbol{footnote}}
\vskip 25pt \centerline{ {\large \rm Min-xin Huang
\footnote{minxin@ustc.edu.cn}, Sheldon Katz\footnote{katz@math.uiuc.edu} and Albrecht Klemm
\footnote{aklemm@th.physik.uni-bonn.de}  } } \vskip .5cm \vskip 25pt

\begin{center}
{\em $^*$ Interdisciplinary Center for Theoretical Study,  \\ \vskip 0.2cm  University of Science and Technology of China,  Hefei, Anhui 230026, China
 \\ \vskip 0.2 cm
 $^*$ Peng Huanwu Center for Fundamental Theory,   Hefei, Anhui 230026, China} 
\\ [3 mm]
{\em $^\dagger$   Department of Mathematics,  University of Illinois at Urbana-Champaign, \\ \vskip 0.2 cm 
 1409 W.\ Green St., Urbana, IL 
61801, USA  } \\ [3 mm]
{\em $^\ddagger$ Bethe Center for Theoretical Physics (BCTP),   \\ \vskip 0.2 cm 
Physikalisches Institut, Universit\"at Bonn,  53115 Bonn, Germany}
\end{center}

\setcounter{footnote}{0}
\renewcommand{\thefootnote}{\arabic{footnote}}
\vskip 40pt
\begin{abstract}
We make a proposal for calculating refined Gopakumar-Vafa numbers (GVN)  on 
elliptically fibered Calabi-Yau 3-folds based on refined holomorphic anomaly 
equations. The key examples are smooth elliptic fibrations over (almost) 
Fano surfaces. We include a detailed review of existing mathematical 
methods towards
defining and calculating the (unrefined) Gopakumar-Vafa invariants (GVI) and the GVNs on compact Calabi-Yau 3-folds using moduli of stable sheaves, in a language that should be accessible to physicists.   In 
particular, we discuss the dependence of the GVNs on the complex structure 
moduli  and on the choice of an orientation.  We calculate  the GVNs  in many 
instances  and compare the B-model predictions with the geometric calculations. We also derive the modular anomaly equations from the holomorphic anomaly equations by analyzing the quasi-modular properties of the propagators. We speculate about the 
physical relevance of the mathematical choices that can be made for the orientation.                      
\end{abstract}

\end{titlepage}
\vfill \eject


\newpage

\baselineskip=16pt

\tableofcontents

\section{Introduction}

Our aim in this paper is twofold.  First, we review the refinement~\cite{GV2}  of the  Gopakumar-Vafa 
invariants   (GVI)~\cite{GV1} from the respective perspectives of M theory, type II string theory and geometry, 
and compare these perspectives.   The definition of the GVI~\cite{GV2} quintessentially combines  $N=2$ 
heterotic/type II duality to 4d~\cite{Kachru:1995wm,Klemm:1995tj}, with arguments from instanton counting in 5d 
gauge theory~\cite{Lawrence:1997jr}, concrete calculations of heterotic BPS amplitudes~\cite{Antoniadis:1995zn} 
and applications of heterotic/type II duality to higher genus world-sheet counting~\cite{Marino:1998} on 
$K3$ fibered Calabi-Yau spaces. It adds a geometrical interpretation taken up in~\cite{KKV,MR1849482} that 
has inspired major developments  in mathematics  as cited in detail below.  As a second point, we also extend the results in our previous paper~\cite{Huang:2015sta} to the refined case in concrete calculations, since the geometric description is still incomplete in particular as far as the refinement goes.  The main example is still a smooth elliptic fibration over $\mathbb{P}^2$ with a single section, but we 
also extend the analysis to similar fibrations  over toric del Pezzo surfaces of high degree.  

So far on compact Calabi-Yau spaces only the reduced GVI's in the $N=4$ geometry  $K3\times T2$ studied in~\cite{KKV} 
have been refined~\cite{KKP}  for curve classes in the K3. Previous works on refined topological string theory mostly focus on 
non-compact or local Calabi-Yau spaces, where the problem of defining the refined partition function 
and  calculating it have been successfully solved as described below.  
In the two  cases above the problem of defining the refinement of  the GV invariants~\cite{GV2} 
gets considerably easier due to a global  $U(1)_R$ isometry, which allows the extraction of the refined invariants from a five dimensional supersymmetric index, 
twisted  by the  global  $U(1)_R$ symmetry; see~\cite{Lawrence:1997jr,MR1957555} and~\cite{Nekrasov:2014yra,Choi:2012jz} for reviews. 
In the general  global case one does not expect the refined  Gopakumar-Vafa invariants  to be invariant 
under either complex  structure deformations or under K\"ahler deformations. Moreover, even for 
fixed deformation parameters  the  mathematical definition depends on additional orientation data.  For these reasons we call the 
refined Gopakumar-Vafa invariants in the global case  Gopakumar-Vafa  numbers (GVN).  
One  can view the  Gopakumar-Vafa numbers  $N_{\beta}^{j_L,j_R}$ as dimensions of 
cohomology groups  of the moduli spaces $\widehat {\cal M}_\beta$ of $M2$ brane 
ground states, parametrized by all its zero modes,  that are split according to representations $[j_L]$ and $[j_R]$ w.r.t. the 5D little group 
${\rm Spin}(4)\simeq \mathrm{SU}(2)_L\times \mathrm{SU}(2)_R$  Lefschetz--type actions on the total cohomology~\cite{GV2,KKV}.
Note that by this definition the $N_{\beta}^{j_L,j_R}\in \mathbb{N}$ are natural numbers.  As usual one can identify the total 
cohomology  of $\widehat {\cal M}_\beta$ with the Hilbert space of a twisted supersymmetric $\sigma$-model, 
which is identified up to to a pre factor $\left(\left[\frac{1}{2},0\right]\oplus 2 [0,0]\right)$ with the Hilbert space  
$H_\beta=\bigoplus_{j_L,j_R}  N_{\beta}^{j_L,j_R} [j_L][j_R]$ of states associated to  $M2$ brane that wraps a 
holomorphic curve ${\cal C}_\beta$ in  the  class $\beta\in H_2(X,\mathbb{Z})$. As it will be explained in 
Section \ref{sec:Physics}  it is important to distinguish between zero modes that parametrize the 
deformation space  ${\cal M}_\beta$ of  ${\cal C}_g$ and the the ones that parametrize the 
gauge field configurations on the brane and that there is a natural projection from 
${\widehat {{\cal M}_\beta}}$  to ${\cal M}_\beta$.  In the  type IIA language the $M2$ brane corresponds to the 
lowest energy bound states of $D2$-- and $D0$--branes. 
The charge of the $D2$- $D0$--brane  system is identified with $(\chi, \beta)$, where $\chi$ is the 
number of $D0$ branes.  The mass of the bound stated proportional to the volume of that curve number and $\chi$. 
One obtains the GVI's $n^g_\beta $  from  the GVN's  as  partial indices defined for fixed combination of 
$\mathrm{SU}(2)_L$ representations $[L_g]=\left(2 [0_L]+\left[\frac{1}{2}_L\right]\right)^{\otimes g}$ by 
taking alternating sums over the degeneracies of the representations of the right $\mathrm{SU}(2)_R$
\begin{equation} 
{\rm Tr}_{[j_R]\subset H_\beta} (-1)^{F_R} =  \sum_{j_R,j_L} (-1)^{2 j_R}(2 j_R+1) N_{\beta}^{j_L,j_R}[ j_L]=\sum_{g=0}^\infty n^g_\beta \left[L_g \right]\ . 
\label{eq:rightsum}
\end{equation}
This particular combination of left representations recovers the genus $g$ of the curve ${\cal C}$. 
Based on physical arguments the GVI's $n^g_\beta\in \mathbb{Z}$ determine  the genus expansion 
of the A-model topological string and determine and are determined by the genus $g$ Gromov-Witten invariants $r^\beta_g\in \mathbb{Q}$. 
Since the latter are complex  structure  deformation invariant the 
$n^g_\beta\in \mathbb{Z}$ are expected to be invariant.  Moreover since the $n^\beta_g$ capture all 
multi-covering contributions to the $r^\beta_g$ there are only finitely many in a given curve class $\beta$.     
One of the main problems that we address in this paper is to review the state of affairs of the attempt to provide 
mathematical definitions of these concepts when the moduli spaces are not smooth. 

Of course  in the decompactification limit, the  GVN's should approach the  locally well defined refined GVI's, so let us 
review the results on the latter. For local toric  Calabi-Yau manifolds the calculations can be performed in the 
A-model using an array of techniques: localization techniques and the virtual Bia\l ynicki-Birula 
decomposition~\cite{Choi:2012jz};  large N-techniques in Chern-Simons theories leading to topological 
vertex~\cite{Aganagic:2003db} can be refined if there is a preferred direction in the torus action~\cite{Iqbal:2007ii}. 
Geometries with this property admit gauge theory interpretations and therefore in 
particular  for the calculation of  the 5d Nekrasov partition function general $\Omega$ backgrounds  
using  localization and the blowup equations~\cite{MR2183121}\cite{MR2472477}.  It has been 
observed in~\cite{Huang:2017mis} that the blow up equations apply to all toric local  Calabi-Yau 
spaces\footnote{That is also the ones without a preferred direction. Note that with blow downs 
the latter geometries can be related to those with a preferred direction.}.  There is a wide class  
of local Calabi-Yau geometries of interest to construct 6d supersymmetric theories and in particular 
super conformal  theories from F-theory.  In these geometries the compact part is an elliptically fibered 
surface $S$~\cite{Klemm:1996hh}, not necessarily with a smooth fibration, or a contractable, intersecting 
configurations of  such surfaces~\cite{Heckman:2015bfa}.  In these cases 
there are three techniques that provide a complete  solution for the refined invariants. 
The modular bootstrap developed in~\cite{Huang:2015sta} for the unrefined compact case
reconstructs the all genus instanton counting functions in the fibre as certain meromorphic  
Jacobi forms recursively in the base degrees. A remarkable feature is that the 
topological string coupling $\lambda$ becomes the elliptic  parameter 
of the  Jacobi forms and yields the complete all genus answer for fixed base degree.       
 This approach  becomes completely  solvable for the local F-theory 
geometries~\cite{Huang:2013,Huang:2015sta}.  Moreover one can define a refinement by 
extending the elliptic parameter $\lambda$ to two elliptic parameters $\epsilon_\pm$. The numerator of the 
Jacobi forms has finite weight and a finite quadratic index associated to $\epsilon_\pm$  and the 
theory stays solvable by virtue of its boundary conditions~\cite{Gu:2017ccq,DelZotto:2017mee}. 
In some cases one knows a dual 2d gauged linear sigma quiver models in which the elliptic genera, calculated by Jeffrey-Kirwan 
residues~\cite{Benini:2013xpa},  provides the  above Jacobi forms~\cite{Haghighat:2014vxa,Kim:2016foj}.    
As it turns out, the so called  elliptic blow up equations~\cite{Gu:2018gmy,Gu:2019dan,Gu:2019pqj,Gu:2020fem} 
are the most widely applicable technique to solve the refined string on this class of local elliptic  
geometries.  In~\cite{KKP} also a suggestions has been made how to refine the GVI in the fibre 
of $K3$ fibered Calabi-Yau 3-folds.   
 
Also the B-model approach using a  refinement~\cite{Huang:2010kf,Krefl:2010,Huang:2011qx}  of BCOV (Bershadsky-Cecotti-Ooguri-Vafa) holomorphic anomaly equations~\cite{BCOV}  
supplemented by boundary conditions at the points 
of parabolic monodromy combined with the modular ansatz  can be extended to calculate the refined GVI's 
using the refined holomorphic anomaly equations and refined boundary conditions~\cite{Huang:2010kf,Huang:2011qx} 
very efficiently.  It applies to local (toric) Calabi-Yau geometries~\cite{Huang:2010kf,Huang:2011qx},  
which have B-model mirror whose compact part is a Riemann surface with a meromorphic differential; see 
also~\cite{Huang:2013,Klemm:2015iya}.  To make predictions for the refined invariants  in the 
global case, the idea is to extend the ansatz for the refined holomorphic anomaly equation \cite{Huang:2010kf} and the 
boundary conditions.  Since our examples are elliptically fibered  this must also be consistent with the 
refinement of the meromorphic Jacobi form ansatz.    The generalization of refined topological 
string theory to the case of  compact elliptic fibrations is not unmotivated, because on the 
torus away from the singular fibers as well as the base in the local limit one can define the  
$U(1)_R$ actions that might be used to define the twisted  5d index, at least in principle. 

There are some modular anomaly equations well known for elliptic fibered Calabi-Yau geometries, e.g. described in \cite{Klemm:2012, Alim:2012ss}. It is generally believed that they are related to the BCOV holomorphic anomaly equations. However, there are some notable differences. The modular anomaly equations apply for the A-model topological free energy expanded on the degrees of base Kahler class and appear already at genus zero, while the  BCOV holomorphic anomaly equations apply for B-model on a general (not necessarily elliptic fibered) Calabi-Yau geometry and only appear at higher genus. The modular anomalies are the basis of weak Jacobi form ansatz for the topological partition function in \cite{Huang:2015sta}. In this paper, building on previous work \cite{Huang:2015sta}, we consider some examples of  compact elliptic Calabi-Yau geometries and provide a derivation of the modular anomaly equations from BCOV holomorphic anomaly equations. The keys of the derivation are some modular anomaly equations for the BCOV propagators. The derivations work also straightforwardly for the refined theory. However, as we see in comparing with geometric calculations, the refined holomorphic anomaly equations may be only partially valid in the compact Calabi-Yau examples, so the derived refined modular anomaly equations are also only valid to a limited extent.

While this paper focuses  on the mathematical challenges posed by the definition and 
calculations of GVN on compact Calabi-Yau threefolds one should note that  the relation  
between the GVN and the GVI affect some key issues in the understanding of bound states
quantum gravity questions in the simplifying context of $N=2$ theories. For example 
if one tries to explain  the microscopic entropy  of black holes in terms of the GVN one 
usually has only access to the GVI~\cite{KKV} to test any predictions~\cite{Denef:2007vg}\cite{Huang:2007sb}.
Clearly if one talks about state counts the $N^{j_L,j_R}_{\beta}$  are the natural quantity. 
It is therefore a burning question whether $n^g_\beta$ or $|n^g_\beta|$ are a sensible quantity to 
look at and if so why or under which circumstances precisely\footnote{Why for 
example are virtually all $n_g\in \mathbb{N}$ for one parameter Calabi-Yau and does this
mean that the $n_g^\beta$ are in this case a good  approximation to state count as the 
analysis~\cite{Huang:2007sb} suggest?}. Similar but so far less precise questions about state 
counting arise in the swampland criteria~\cite{Ooguri:2006in}, that attempts to separate 
consistent low energy theories coming from quantum gravity from inconsistent ones. 
In particular the swampland distance conjecture predicts an infinite number of states 
in regions which are at infinite distance in the Weil-Petersson metric from any point in the interior 
of the complex moduli space~\cite{Ooguri:2006in}. In this context it is natural that at the point 
of maximal unipotent monodromy which corresponds to large radius of the mirror the massless 
states are the $D2-D0$ bound states~\cite{Joshi:2019nzi}. Likewise in the various  versions 
of the weak gravity (sub)-lattice conjectures~\cite{Heidenreich:2016aqi}\cite{Lee:2018urn} 
the GVI on compact and non-compact  are used to argue the conjectures hold.  
             
The situations that arise in quantum gravity are markedly different than
those in the local limit. While on the one hand consistent $N=2$ quantum gravity 
theories  forbid any global symmetries like the $U(1)_R$  that is used to define the 5d 
protected index, on the other hand the usual arguments about the decoupling  of vector multiplets-- and 
hyper--multiplets seem to fail in the gravitational sector~\cite{Christensen:1978md}, which 
might be reflected by the fact that the GVN's are actually sensitive to both complex 
structure-- and K\"ahler structure deformations.


\section{The physics of Gopakumar-Vafa invariants and their refinements}\label{sec:gvphys}
\label{sec:Physics}  
The refined Gopakumar-Vafa numbers of a  Calabi-Yau threefold $X$ were originally introduced in \cite{GV2} as 
intermediate  step towards defining the Gopakumar-Vafa invariants as the index as  (\ref{eq:rightsum}). 

In the IIA description, we have a moduli space $\widehat{\mathcal{M}}_\beta$ of bound states 
of D0-D2 branes with charge vector $(\chi,\beta)$. 
$\widehat{\mathcal{M}}_\beta$ has    a projection map to the  moduli space $\mathcal{M}_\beta$ of the curve $C_\beta$ 
obtained by ignoring the gauge fields on the branes.  
It is remarked in \cite{GV2} that the $\mathrm{SU}(2)_R$ action on the component with highest $j_L$ is 
identified with the Lefschetz representation on $H^*(\mathcal{M}_\beta)$, while the action of the diagonally embedded $\mathrm{SU}(2)$ is 
identified with the Lefschetz action on $H^*(\widehat{\mathcal{M}}_\beta)$.  In particular, if $\widehat{\mathcal{M}}_\beta$ is smooth, 
it can be inferred that the GV invariant $n^0_\beta$ is the Euler characteristic of $\widehat{\mathcal{M}}_\beta$ up to sign:
\begin{equation}\label{eq:n0smooth}
n^0_\beta=(-1)^{\dim\widehat{\mathcal{M}}_\beta}e(\widehat{\mathcal{M}}_\beta).
\end{equation}
Similarly, if $\mathcal{M}_\beta$ is smooth and parametrizes genus $g$ curves, we have
\begin{equation}\label{eq:ngsmooth}
n^g_\beta=(-1)^{\dim\mathcal{M}_\beta}e(\mathcal{M}_\beta).
\end{equation}

We give just one example computation here for the generic elliptic fibration over $\mathbb{P}^2$, rephrasing a computation from our earlier 
work~\cite{Huang:2015sta}. Many other explicit computations appear in the physics literature, beginning with \cite{GV2,KKV}.

We let $X\to \mathbb{P}^2$ be a generic Weierstrass elliptic fibration over $\mathbb{P}^2$, which can be obtained by resolving the singularities of a generic weight 18 hypersurface in $\mathbb{P}(1,1,1,6,9)$.
We investigate the fiber class $\beta=f$.  In this case we have
\begin{equation}
\widehat{\mathcal{M}}_\beta=X,\qquad \mathcal{M}_\beta=\mathbb{P}^2,
\end{equation}
and $\widehat{\mathcal{M}}_\beta\to\mathcal{M}_\beta$ is identified with the elliptic fibration itself.
The Hilbert space is just $H^*(X)$, which has Betti numbers $h^i=(1,0,2,546,2,0,1)$ for $i=0,\ldots,6$.

Since the fibers have genus~1, it follows from ref. \cite{GV2}  that the refined GV 
numbers $N^{j_L,j_R}_f$ vanish for $j_L>1/2$, while the $j_R$-content for $j_L=1/2$ 
is given by the Lefschetz representation on $\mathcal{M}_\beta=\mathbb{P}^2$.  
Since the Lefschetz action on $\mathbb{P}^2$ yields just the representation $[1]$, we infer 
that as $SL(2)_L\times SL(2)_R$ representations we have
\begin{equation}\label{eq:refinedx}
H^*(X)=[\frac12,1]\oplus [0,R]
\end{equation}
for some representation $R$ of $SL(2)$.

We also know from \cite{GV2} that the diagonal $SU(2)$ agree with the Lefschetz action on $\widehat{\mathcal{M}}_\beta$, which is $X$ in our case.  From the Hodge numbers $(h^{1,1},h^{2,1})=(2,272)$ we see that the Lefschetz representation $X$ is $[3/2]+[1/2]+546[0]$.  Since the restriction of (\ref{eq:refinedx}) to the diagonal $SU(2)$ is $[1/2][1]+R=[3/2]+[1/2]+R$, we see that $R=546[0]$ and conclude that
\begin{equation}
H^*(X)=[\frac12,1]\oplus 546[0,0].
\end{equation}
Since $Tr_{[j_R]\in H^*(X)}(-1)^{F_R}=3[1/2]+546[0]=3([1/2]+2[0])+540[0]$, we get for the unrefined invariants
\begin{equation}\label{eq:GVfphysics}
n^0_f=540,\ n^1_f=3.
\end{equation}
We will re-derive these results later by more complicated geometric methods, showing that it comes down in essence to this same calculation.  The complication arises from the attempt to define the GV invariants in complete generality, particularly when the spaces $\widehat{\mathcal{M}}_\beta$ and $\mathcal{M}_\beta$ are not smooth.

The refined GV numbers can always be defined in M-theory.  However, refined numbers have proven to be very difficult to compute for compact Calabi-Yau threefolds.  In the case of a local Calabi-Yau, we can turn on an $\Omega$ background and use various successful strategies  of computation as reviewed in the introduction. In particular in the local case one can compute  the refined PT 
invariants by localization, and use these to compute the refined GV invariants \cite{Choi:2012jz}.  However all of the above approaches, with the possible exception of 
the modified refined holomorphic anomaly equation proposed below,   are not applicable to the compact case.

The unrefined GV invariants are true invariants in the sense that they are independent of complex structure deformations.  The GVN's can depend on 
the K\"ahler parameters, as can be seen when a K\"ahler deformation cause the Calabi-Yau to undergo a flop.  
Moreover as we emphasized the GVN's depend on the complex structure of $X$ in a discontinuous manner. 
In the case of a local del Pezzo surface, there are no complex structure deformations and the GVN's  are 
truly invariants by the 5d twisted index.  

\section{The geometry of Gopakumar-Vafa invariants and their refinements} \label{sec:geometry}
While the physical definition of Gopakumar-Vafa invariants (GVI's) and their refinements via M-theory is clear, there is still no direct mathematical definition of the GVI's 
from the M2-brane or D2-brane moduli spaces in spite of the efforts of many mathematicians over a period of more than 20 years, 
although there is at least a definition of the unrefined invariants which depend on a conjecture.  The best that can be done in 
complete generality is to infer/define the GV invariants indirectly in terms of the generating functions of GW invariants, 
DT invariants, or PT invariants \cite{K,PTBPS}.  Nevertheless, the GV invariants can be directly defined in many cases, 
and sometimes the refined invariants can be defined directly.  In this section, we will review the current state of affairs, 
beginning with the moduli spaces.

\subsection{Geometric moduli spaces} 
\label{subsec:m}
The space $\mathcal{M}_\beta$ is geometrically defined as the Chow variety $\mathrm{Chow}(\beta)$ of curves of 
class $\beta$.   We will also refer to this space as $\mathcal{M}_\beta$ in keeping with the notation of 
\cite{GV2} while indicating the dependence on $\beta$.  A definition of the Chow variety 
appears for example in~\cite{Kollar:1996rc}.

More precisely, the points of $\mathcal{M}_\beta$ correspond to \emph{cycles}, i.e.\ finite formal sums $\sum_i n_i C_i$ of curves with multiplicities,
where $C_i$ are irreducible curves, $n_i$ are positive integers, and $\sum n_i[C_i]=\beta$.   The point of the construction is that
 $\mathcal{M}_\beta$ has a natural structure as a complex algebraic variety (not necessarily smooth).

\subsubsection{Instructive examples}
\label{subsubsec:examples}

If $S$ is a del Pezzo surface, $X$ is the corresponding local  Calabi-Yau manifold constructed as the total space 
of ${\cal O}(-K_S)\rightarrow S$, and $\beta\in H_2(S,\mathbb{Z})\simeq H_2(X,\mathbb{Z})$, then we can be more explicit.  Since $\beta$ is now a divisor class on $S$, we get a line bundle $\mathcal{O}_S(\beta)$ on $S$, and the curves of class $\beta$ are precisely the divisors of zeroes of the non-vanishing holomorphic sections of $\mathcal{O}_S(\beta)$.  Since the divisor of zeroes is unchanged after multiplying a section by a scalar, we infer that  
\begin{eqnarray}
\mathcal{M}_\beta = \mathbb{P}\left(H^0\left(S,\mathcal{O}\left(\beta\right)\right)
\right),
\end{eqnarray}
which is a projective space.  

For example, if $S$ is $\mathbb{P}^2$ and $\beta$ is the degree d class, we get $\mathcal{M}_d=\mathbb{P}^{d(d+3)/2}$, the parameter space of degree $d$ plane curves.   In particular, $\mathcal{M}_2=\mathbb{P}^5$ is the parameter space of degree 2 plane curves.  In this case, the points of $\mathcal{M}_2$ correspond either to smooth degree 2 curves $C$, pairs of distinct lines $L_1+L_2$, or double lines $2L$.  These are precisely the curves $\sum n_i C_i$ with 
$\sum n_i \mathrm{deg}(C_i)=2$.  The case of general $d$ is analogous. The space  $\mathcal{M}_d$ is the union of strata parametrized by tuples of pairs $\{(n_i,d_i)\}$, corresponding to cycles $\sum n_iC_i$ with the $C_i$ irreducible (but not necessarily smooth) curves of degree $d_i$.  The methods of \cite{KKV} give $n^0_2=-6$, the Euler characteristic of $\mathbb{P}^5$ multiplied by a sign because the dimension is odd.  The stratification above is not relevant to this calculation at all.  We only include this information to clarify the definition of the Chow variety.

Returning to compact Calabi-Yau 3-folds  $X$, we show by example that $\mathcal{M}_\beta$ can depend on the complex structure of $X$.  Suppose that $X$ contains a ruled surface $S$ fibered over a smooth genus $g$ curve $C$, with all fibers isomorphic to $\mathbb{P}^1$.  If $f$ is the fiber class, then clearly $\mathcal{M}_f$ is identified with $C$.

As we will illustrate by an example below, it can be shown that $X$ admits a deformation of complex structure where the ruled surface $S$ is destroyed and exactly $2g-2$ $\mathbb{P}^1$'s remain.  After the deformation, $\mathcal{M}_\beta$ consists of $2g-2$ points instead of the curve $C$.    A moduli space parametrizing $2g-2$ isolated $\mathbb{P}^1$'s clearly has $n^0_\beta=2g-2$. Since the unrefined invariants are deformation invariant, we see that the situation where $\mathcal{M}_f\simeq C$ must also contribute $2g-2$ to $n^0_\beta$, in agreement with~(\ref{eq:ngsmooth}) with $g=0$.  We will explain (\ref{eq:ngsmooth}) in a different way when we return to $n^0_\beta$ below.

Examples of this geometry appeared in the literature in \cite{Candelas:1994hw,Hosono:1993qy} and the general case was considered in \cite{Katz:1996ht}.  An explicit example is the two-parameter Calabi-Yau given by resolving the orbifold singularities of a weight~8 hypersurface in the weighted projective space $\mathbb{P}(1,1,2,2,2)$.  Suppose the equation of the hypersurface is $x_1^8+x_2^8+x_3^4+x_4^4+x_5^4=0$.  This Calabi-Yau has orbifold singularities along the curve $x_1=x_2=x_3^4+x_4^4+x_5^4=0$, a smooth plane curve of degree 4 and genus 3.  The ruled surface $S$ arises as the exceptional divisor of the blowup of this curve.  

The deformations which replace $S$ with $2g-2=4$ $\mathbb{P}^1$'s were called non-polynomial deformations in \cite{Batyrev:1994hm}. We can make these deformations explicit following \cite{Candelas:1994hw}.  The weighted projective space $\mathbb{P}(1,1,2,2,2)$ is isomorphic to a singular degree 2 hypersurface in $\mathbb{P}^5$ by the map
\begin{eqnarray}
f:\mathbb{P}(1,1,2,2,2)\to\mathbb{P}^5, \qquad f(x_1,x_2,x_3,x_4,x_5)=(x_1^2,x_1x_2,x_2^2,x_3,x_4,x_5).
\end{eqnarray}
Letting $(y_1,\ldots,y_6)$ be homogeneous coordinates of $\mathbb{P}^5$, we see that $f$ is an isomorphism onto the degree~2 hypersurface with equation $y_1y_3=y_2^2$.  Furthermore,
$f$ maps the hypersurface $x_1^8+x_2^8+x_3^4+x_4^4+x_5^4=0$ isomorphically to the complete intersection $y_1y_3=y_2^2, y_1^4+y_3^4+y_4^4+y_5^4+y_6^4=0$.  In this model, the singularity is at $y_1=y_2=y_3=0$, a plane curve of degree 4 as before, and blowing up gives a ruled surface.  

If more generally we take a generic weight 8 hypersurface, we still get orbifold singularities along a smooth plane curve $C$ of degree 4 and genus 3.  We blow up $C$ to get a ruled surface, and $M_f\simeq C$.    Since different hypersurfaces lead to different curves $C$, the variety $\mathcal{M}_f$ 
can depend on the complex structure, but at least its structure as a topological space is invariant, so that its Euler number is invariant as well.

Now instead deform $y_1y_3=y_2^2$ to a rank~4 quadric, say $\ell_1(y)\ell_2(y)=\ell_3(y)\ell_4(y)$, where the $\ell_i(y)$ are four linearly independent linear forms in $y_1,\ldots, y_6$.  Then 
$\ell_1(y)\ell_2(y)=\ell_3(y)\ell_4(y)$ is singular along the line $\ell_1(y)=\ell_2(y)=\ell_3(y)=\ell_4(y)=0$.  This line meets $y_1^4+y_3^4+y_4^4+y_5^4+y_6^4=0$ in 4 points instead of a plane curves of degree~4, so we get a Calabi-Yau with 4 conifolds.  Performing a small resolution gives 4 $\mathbb{P}^1$s as claimed.

To summarize: From  the viewpoint of physics, M2-branes wrapping the $\mathbb{P}^1$ fibers  of the  ruled surface over a curve of genus $g$, that is present for special complex moduli values\footnote{Frozen to that values for 
special embeddings of $X$ into the  toric ambient spaces.} , 
have Hilbert space $2g [0,0]+[0,1/2]$, while for generic complex structure the Hilbert space   of $(2g-2)$ M2 branes wrapping isolated $\mathbb{P}^1$'s contains just $(2g-2)$ multiples of the trivial representation $[0,0]$.    
Both geometries lead to the same GVI's namely  $n^0_{\mathbb{P}^1s}=2g-2$ because 
\begin{equation}
{\rm Tr}_{[j_R]\subset 2 g[0,0]+ [0,1/2]}(-1)^{F_R}={\rm Tr}_{[j_R]\subset (2g-2)[0,0]}(-1)^{F_R}=(2 g-2) [L_0]\ .
\end{equation} 

\smallskip

We next consider an example of a singular $\mathcal{M}_\beta$.  Consider the moduli space $\mathcal{M}_1$ of lines on a quintic threefold.  If the quintic is general, then $\mathcal{M}_1$ consists of 2875 points, corresponding to the 2875 lines on the quintic.  But if  $X$ is the Fermat quintic threefold $\sum_{i=1}^5 x_i^5=0$, we get that $\mathcal{M}_1$ is the union of 50 curves of genus~6 which collectively intersect in 375 points \cite{Albano:1991fqt}.  These curves are all isomorphic to Fermat plane curves $a^5+b^5+c^5=0$, of genus 6.  A typical component of $\mathcal{M}_1$ corresponds to the family of lines on $X$ given parametrically by
\begin{eqnarray}\label{fermatlines}
(s,t)\mapsto (s,-\zeta s,at,bt,ct),
\end{eqnarray}
where $(s,t)$ are homogeneous coordinates on $\mathbb{P}^1$ and $\zeta^5=1$.   The other components of $\mathcal{M}_1$ are all obtained from this one by permuting the five coordinates of $\mathbb{P}^4$ and multiplying the coordinates by fifth roots of unity. A typical line in the intersection of two components is given parametrically by $(s,t)\mapsto (s,-\zeta s, t, \xi t,0)$, where $\xi^5=1$ as well.  Furthermore, it is shown in \cite{Albano:1991fqt} that each of these 50 curves appear with multiplicity 2, and so each curve contributes 2(2g-2)=20 to $n^0_1$.  We will make this calculation more precise below in the context of a more general theory.

The component corresponding to (\ref{fermatlines}) intersects the component corresponding to the family of parametrized lines $(s,t)\mapsto (at,bt,ct,s,-s)$ in the line given parametrically by $(s,t)\mapsto (s,-\zeta s,0,t,-t)$.  There are 375 such lines, all obtained from this one by permuting the coordinates and multiplying by fifth roots of unity.  It is shown in \cite{Albano:1991fqt} that these curves appear with multiplicity 5.   Putting the contributions together, we get $n^0=(50\times 20)+(375 \times 5)=2875$, agreeing with the number of lines on a generic quintic threefold.

For later use, we note here that each of the 50 Fermat plane curves contains 15 of the 375 points, by choosing one of the three coordinate lines $a=0,\ b=0$, or $c=0$ together with a fifth root of unity. 

Our main running example is a Calabi-Yau threefold $X$ which is elliptically fibered over a base $B$, $\pi:X\to B$ as discussed  already for $B=\mathbb{P}^2$ in Section \ref{sec:Physics}.  
Let $f$ be the fiber class.  Then $\mathcal{M}_f\simeq B$, the point $b\in B$ corresponding to the fiber $f_b=\pi^{-1}(b)$. Furthermore, for any $k\ge 1$ we 
have the $\mathcal{M}_{kf}$ is isomorphic to $B$ as well.  In this isomorphism, the point $b\in B$ corresponds to the cycle $k\cdot f_b \in \mathrm{Chow}(kf)$.  

We will return to these examples later by placing them in a more general context.


\subsubsection{Stable sheaves }
\label{subsubsec:stablesheafs}

Consider for simplicity a point $C\in\mathcal{M}_\beta$ corresponding to a smooth and irreducible curve $C$ of genus $g$ with $[C]=\beta$.  
Then the fiber of $\widehat{\mathcal{M}}_\beta\to \mathcal{M}_\beta$ over $C$ is identified with the Jacobian $J(C)$ of $C$.  If however $C$ is not both smooth and irreducible, a more precise description is needed.  

The mathematical formulation of a stable D2-D0 brane is a stable coherent sheaf $F$ on $X$ whose support has pure dimension~1.  Let's unpack the terminology to understand what this means.
The dimension of a sheaf is defined as the dimension of its support, so we are considering sheaves $F$ supported on a curve $C$.  Then for the Chern character of $F$ we 
have $\mathrm{ch}(F)\in H^4(X)\oplus H^6(X)$.  We write $\mathrm{ch}(F)=\beta+\chi$, where $\beta\in H_2(X)\simeq H^4(X)$ is a curve class measuring the 
D2-brane charge and $\chi\in H^6(X)\simeq\mathbb{Z}$ measures the D0-brane charge.  Note that $\beta$ need not equal the class of the support of $F$ due to the possibility of multiplicities.  For example, if $E\subset X$ is an elliptic curve and $F=\mathcal{O}_E^2$ (interpreted as a sheaf on $X$ supported on $E$), we have $\beta=2[E]$.

The condition that $F$ has pure dimension~1 means that $F$ has dimension 1 and contains no zero-dimensional subsheaves.  If $C$ is a curve and $p$ is a point, then $F=\mathcal{O}_C\oplus \mathcal{O}_p$ has dimension~1 but is not pure, since $\mathcal{O}_p\subset F$ is a zero-dimensional subsheaf.  

Finally, we have to describe stability.  For simplicity of exposition we assume that
the Neveu-Schwarz 2-form field $B=0$ and identify the K\"ahler class with a real 
K\"ahler class $\omega\in H^2(X)$\footnote{The case $B\ne0$ can be handled as in~\cite{MR3121850}.}  If $F$ has pure dimension~1, we say that $F$ is stable if for all proper nonzero subsheaves $F'\subset F$ we have
\begin{eqnarray}\label{stable}
\frac{\chi(F')}{\mathrm{ch}_1(F')\cdot\omega} <\frac{\chi(F)}{\mathrm{ch}_1(F)\cdot\omega}.
\end{eqnarray}
If $<$ is replaced by $\le$ in (\ref{stable}) we say that $F$ is semistable.  For simplicity, we assume that all semistable sheaves are stable.  This can be enforced by taking\footnote{By making this restriction, we are ignoring some fundamental points.   In general, $\widehat{\mathcal{M}}$ might only exist as a stack rather than as a scheme.  An alternative approach is to approximate the moduli problem by a scheme called the coarse moduli space, with some loss of information.  Beyond these comments, suffice it to say that it is conjectured that there is a sense in which refined invariants do not depend on $\chi$, so we are free to make this simplifying assumption.} $\chi=1$.

For fixed $\chi$ we  let $\widehat{\mathcal{M}}_\beta$ be the moduli space of stable sheaves $F$ of pure dimension~1 with $\mathrm{ch}(F)=\beta+\chi$.  With 
the caveat about semistability, $\widehat{\mathcal{M}}_\beta$ exists and is projective \cite{Simpson:1994}.  In general, $\widehat{\mathcal{M}}_\beta$ 
is a scheme but not necessarily a variety.
It is apparent that $\widehat{\mathcal{M}}_\beta$ depends on both the complex structure and the K\"ahler class of $X$.  
The map 
$\widehat{\mathcal{M}}_{\beta}\to \mathcal{M}_{\beta}$ is defined by sending a sheaf $F$ to its support, including multiplicities.

For compact Calabi-Yau threefolds, both the moduli spaces of stable sheaves and the moduli spaces of stable pairs, proposed by Pandharipande and Thomas (PT) in~\cite{PT1,PTBPS} 
to define GVI's in certain situations, can be difficult to describe.   In fact, the moduli spaces of sheaves tend to be more complicated.  However, the decisive advantage in using sheaves over pairs is that we only need to specify one value of the D0-brane charge for sheaves in order to extract the GVI's or of GVN's, while for PT pairs we need to do computations for several 
D0-brane charges, producing more obstacles to completing these typically difficult computations.  By contrast, in the local  toric case, the computation of the GVI's and refined GVN's are facilitated by localization and the virtual Bialynicki-Birula decomposition \cite{Choi:2012jz}.  For this reason, the simpler description of the PT moduli spaces makes PT the preferred method in the local case.

We now continue with the examples from Section~\ref{subsubsec:examples}.  We start with local $\mathbb{P}^2$ and $d=1$.  If $F\in\widehat{\mathcal{M}}_{1}$, then by $\chi(F)=h^0(F)-h^1(F)=1$, we have $h^0(F)>0$ and therefore $F$ has a nonzero section $s$.  This section gives rise a map $f:\mathcal{O}_X\to F$ sending a function $g$ to the section $gs$ of $F$.  Since $F$ has dimension 1, $f$ must vanish on a curve $C$, giving an injective map $\mathcal{O}_C\to F$ which we also denote by $f$.  Since $d=1$ and $F$ is pure, $C$ must be a line.  Since $\chi(\mathcal{O}_C)=1$, we infer that $f$ is an isomorphism.  In other words, $\widehat{\mathcal{M}}_1\simeq \mathcal{M}_1\simeq\mathbb{P}^2$.

We similarly see that $\widehat{\mathcal{M}}_1\simeq \mathcal{M}_1$ for all examples of lines on a quintic threefold.

Returning to local $\mathbb{P}^2$, we consider $d=2$ and $F\in \widehat{\mathcal{M}}_{2}$.  As before, we get an injective map $\mathcal{O}_C\to F$.  There are now two cases: either $C$ has degree 1, or $C$ has degree 2.  If $C$ has degree 1, the resulting inclusion $\mathcal{O}_C\hookrightarrow F$ contradicts stability.  If $C$ has degree 2, we have $\chi(\mathcal{O}_C)=1$ and $\mathcal{O}_C\hookrightarrow F$ is an isomorphism, so $\widehat{\mathcal{M}}_2\simeq \mathcal{M}_2\simeq\mathbb{P}^5$.

The case $d=3$ is more interesting.  If $F\in \widehat{\mathcal{M}}_3$, as before we get an inclusion $\mathcal{O}_C\hookrightarrow F$.  The degree of $C$ cannot be 1 or 2 by stability, so $C$ has degree 3.
Since $C$ has (arithmetic) genus $g=1$, we see that $\chi(\mathcal{O}_C)=1-g=0$.  Since $\chi(F)=1$, we infer an exact sequence
\begin{eqnarray}\label{eq:fses}
0\to \mathcal{O}_C \to F \to \mathcal{O}_p\to 0
\end{eqnarray}
for some point $p\in C$.  

Conversely, it can be shown that given $p\in C$, we can find a unique $F$ fitting into an exact sequence (\ref{eq:fses}).\footnote{By applying $\underline{Ext}^*_{\mathcal{O}_{\mathbb{P}^2}}(\cdot,\mathcal{O}_{\mathbb{P}^2}(-3))$ to (\ref{eq:fses}) it can be computed that
$F=I_{p,C}^\vee=\underline{Ext}^1_{\mathcal{O}_{\mathbb{P}^2}}(I_{p,C},\mathcal{O}_{\mathbb{P}^2}(-3))$,
where $I_{p,C}\subset\mathcal{O}_C$ is the ideal sheaf of functions on $C$ which vanish at $p$.} Thus $\widehat{\mathcal{M}}$ is isomorphic to the universal plane curve of degree 3.

Next, we consider $\widehat{\mathcal{M}}_f$ for the fiber class of an elliptically fibered Calabi-Yau threefold $\pi:X\to B$.  
If $F$ is a stable sheaf with class $f$ and $\chi(F)=1$, our previous argument shows that $F$ has a section which necessary induces an injection $\mathcal{O}_{f_b}\hookrightarrow F$ for some fiber $f_b$.  From $\chi(F)=1$ we see that the cokernel of this injection is a skyscraper sheaf $\mathcal{O}_p$ for some $p\in f_b$.  It follows that we have a short exact sequence
\begin{equation}
0\to \mathcal{O}_{f_b}\to F\to \mathcal{O}_p\to 0,
\end{equation}
from which we deduce that 
\begin{equation}
F=I_{p,f_b}^\vee = \underline{Ext}^2(I_{p,f_b},\mathcal{O}_X),
\end{equation}
where $I_{p,f_b}$ is the ideal sheaf of holomorphic functions on $f_b$ which vanish at $p$.
This shows that $F$ is completely determined by a point $p\in X$, since the fiber $f_b$ is then necessarily $f_{\pi(p)}$.  Said differently, we have $\widehat{\mathcal{M}}_f\simeq X$.  

Furthermore, our descriptions of $\widehat{\mathcal{M}}_f$ and ${\mathcal{M}}_f$ show that $\widehat{\mathcal{M}}_f\to {\mathcal{M}}_f$ is identified with the elliptic fibration $\pi:X\to B$ itself.

Similarly, for any $k\ge 1$ we have $\widehat{\mathcal{M}}_{kf}\simeq X$.  To a point $p\in X$, we put $b=\pi(f)$ and consider the rank $k$ degree 1 Atiyah bundle  $E_{k,p}$ on $f_b$ defined 
inductively\footnote{If $p$ is a singular point of $f_b$, then $\mathcal{O}_{f_b}(p)$ is not well-defined, but in any case we can define $E_{1,p}$ as a dual of the ideal sheaf of $p$ in $f_b$.}
by $E_{1,p}=\mathcal{O}_{f_b}(p)$, see \cite{Atiyah:1957,Friedman:1999vbef}
\begin{equation}\label{eq:atiyahseq}
0\to \mathcal{O}_{f_b}\to E_{k,p}\to E_{k-1,p}\to 0.
\end{equation}
It is shown in\cite{Atiyah:1957} that these are the only stable sheaves on the elliptic fiber $f_b$ itself.   
Now, to each point $p\in X$ we associate $E_{k,p}$, viewed as a sheaf on $X$ supported on the 
fiber $f_{\pi(p)}$. For fixed $k$, the sheaves $\{E_{k,p}\}$ only depend on 
the point $p$, hence are parametrized by $X$.

To complete the argument that $\widehat{\mathcal{M}}_{kf}\simeq X$, we have to show that any stable sheaf $F$ of class $kf$on $X$  with $\chi(F)=1$ must be one of the $E_{k,p}$.  First we show that all such sheaves are supported on a single fiber.  Suppose that $F$ were supported on $\ell$ fibers, with $\ell >1$.  Then $F=F_1\oplus \cdots\oplus F_\ell$, where the $F_i$ are supported on distinct fibers.
From $1=\chi(F)=\sum \chi(F_i)$ we see that $\chi(F_i)\ge 1$ for some $i$, in which case the subsheaf $F_i\subset F$ would destabilize $F$.

We finish the argument by induction on $k$, the case $k=1$ being already proven.  Since $F$ must have a section by $\chi(F)=1$, we obtain an exact sequence
\begin{equation}
0\to \mathcal{O}_{\ell {f_b}}\to F \to F'\to 0.
\end{equation}
Here ${\ell f_b}$ refers to a scheme of multiplicity $\ell$ supported on the fiber $f_b$ where $F$ is supported.  We must have $\chi( \mathcal{O}_{\ell f_b})=0$.  
The minimal $\chi$ is obtained by pulling back via $\pi$ a multiplicity $\ell$ structure on $b\in\mathbb{P}$, while if $\chi( \mathcal{O}_{\ell f_b})>0$, 
then $ \mathcal{O}_{\ell f_b}$ would destabilize $F$.  So $F'$ has class $(k-\ell)f$ and $\chi(F')=1$.  By induction we conclude that we have an exact sequence
\begin{equation}\label{eq:newextension}
0\to \mathcal{O}_{\ell f_b}\to F \to E_{(k-\ell),p}\to 0.
\end{equation}
It can be shown that the exact sequence (\ref{eq:atiyahseq}) with $k$ replaced by $k-\ell+1$ is a subsequence of (\ref{eq:newextension}).  So if $\ell>1$, then $E_{k-\ell+1,p}\subset F$ would destabilize $F$.  We conclude that $\ell=1$, which completes the argument since we have recovered (\ref{eq:atiyahseq}).  We omit the details. 
As in the case $k=1$, the map $\widehat{\mathcal{M}}_{kf}\to {\mathcal{M}}_{kf}$ is identified with the elliptic fibration $\pi:X\to B$ itself.  This will allow us to conclude that the refined invariants are independent of the degree $k$!

\smallskip
Returning to the general case, it follows from \cite{Pantev:2013sss,Joyce:2015dc} that  locally $\widehat{\mathcal{M}}_\beta$ is the critical point locus of a superpotential defined on a manifold.  More precisely, for every point $p\in \widehat{\mathcal{M}}_\beta$ we can find a manifold $U$, a holomorphic function $W$ on $U$, and an open set $R\subset \widehat{\mathcal{M}}_\beta$ containing $p$ so that $R$ is analytically isomorphic 
to the critical point locus of $W$ via an inclusion $\iota:R\hookrightarrow U$.  
The data $(R,U,W,\iota)$ is called a \emph{critical chart} in \cite{Joyce:2015dc}.  

\subsection{Genus 0 Gopakumar-Vafa invariants}  It was stated in \cite{GV2} that $n^0_\beta$ is equal to the Euler characteristic of $\widehat{\mathcal{M}}_\beta$ up to a sign.  This is literally true only if 
$\widehat{\mathcal{M}}_\beta$ is smooth.  However, $n^0_\beta$ can be computed as a weighted Euler characteristic in complete generality, as we now describe.

First of all, $n^0_\beta$ can be mathematically defined as a Donaldson-Thomas invariant of $\widehat{\mathcal{M}}_\beta$ \cite{Katz:2008g0gv} associated with a  \emph{symmetric obstruction theory} on $\widehat{\mathcal{M}}_\beta$.  This implies that $n^0_\beta$ is a weighted Euler characteristic \cite{Behrend:2009dt}.  More precisely, for any scheme $S$ supporting a symmetric obstruction theory, there is an integer-valued constructible function
\begin{equation}
\nu_{S}:S\to\mathbb{Z}
\end{equation}
such that
\begin{equation}\label{eq:wteuler}
n_\beta^0=\sum_{m\in\mathbb{Z}}m\,e\left(
\left\{
p\in S\mid \nu_{X}(p)=m
\right\}
\right),
\end{equation}
where $e$ denotes the topological Euler characteristic.  The function $\nu_S$ is called the Behrend function. A key point is that $\nu_S$ depends only on the scheme structure of $S$ and not on the particulars of the symmetric obstruction theory.   In particular, if $S$ is a smooth $n$-dimensional manifold at $p\in S$, then $\nu_S(p)=(-1)^n$, as we will expand on a bit below.  Putting $S=\widehat{\mathcal{M}}_\beta$ and supposing that $\widehat{\mathcal{M}}_\beta$ of dimension $n$, we conclude that $n^0_\beta=(-1)^ne(\widehat{\mathcal{M}}_\beta)$, in complete agreement with \cite{GV2}. 

The physical refinement has the property that the diagonal $SL(2)$ corresponds to the Lefschetz action on $\widehat{\mathcal{M}}_\beta$.  This Lefschetz action is related to the Betti numbers of $\widehat{\mathcal{M}}_\beta$.  So we expect that the mathematical notion of a refinement should be related to the refinement of the Euler characteristic of $\widehat{\mathcal{M}}_\beta$ by its Betti numbers.  We will come back to this point later.

An explicit way to compute $\nu$ is via a superpotential.  Choose a critical chart $(R,U,W,\iota)$ of $\widehat{\mathcal{M}}_\beta$ as above with $p\in R$.  We recall the 
notion of the Milnor fiber $MF(W,p)$ of $W$ at $p$ \cite{Milnor:1968}.  We take a small $(2n-1)$-sphere containing $p$, where $n$ is the dimension of $U$.   Then we have a fibration
\begin{equation}
S^{2n-1}-\{x\in S^{2n-1}\mid W(x)=0\}\to S^1,\qquad x\mapsto \frac{W(x)}{|W(x)|}.
\end{equation}
The fiber of this fibration is called the Milnor fiber.  In terms of the Milnor fiber we have
\begin{equation}\label{eq:bmf}
\nu_{R}(p)=(-1)^{\dim U}\left(1-e\left(MF(W,p)
\right)\right)
\end{equation}
More generally, the critical point scheme $R$ of a holomorphic function $W$ on any complex manifold $U$ supports a symmetric obstruction theory, and its Behrend function is also given by (\ref{eq:bmf}). 

As our first illustration of (\ref{eq:bmf}), suppose that $\widehat{\mathcal{M}}_\beta$ is smooth of dimension $n$.  We can take $U=\widehat{\mathcal{M}}_\beta$ and $W\equiv0$.  In this case the Milnor fiber is empty, and (\ref{eq:bmf}) gives $\nu\equiv (-1)^n$ as  asserted earlier.  It follows immediately from (\ref{eq:wteuler}) that $n^0_\beta=(-1)^{\dim\widehat{\mathcal{M}}_\beta}e(\widehat{\mathcal{M}}_\beta)$.

As another example, consider an isolated curve $C\simeq\mathbb{P}^1$ in a Calabi-Yau $X$, with normal bundle $\mathcal{O}\oplus\mathcal{O}(-2)$.  As above we have $\widehat{\mathcal{M}_\beta}\simeq \mathcal{M}_\beta$, where $\beta=[C]$.  Since $C$ is isolated, $\widehat{\mathcal{M}}_\beta$ is a point.\footnote{We ignore any other curves that may be in the same class $\beta$.}  Since $H^0(\mathcal{O}\oplus\mathcal{O}(-2))$ is 1-dimensional, $\widehat{\mathcal{M}}_\beta$ has a 1-dimensional tangent space.  Thus $\widehat{\mathcal{M}}_\beta$ is defined by an equation $x^n=0$ for some $n>1$ (or $\widehat{\mathcal{M}}=\mathrm{Spec}(\mathbb{C}[x]/(x^n))$), where $n$ is the multiplicity of the curve as a point of its moduli space.  Then $\widehat{\mathcal{M}}_\beta$ can be defined by the superpotential $W=x^{n+1}$ on $\mathbb{C}$, as the equation $dW=0$ matches the defining equation $x^n=0$. Then $MF(W,0)$ consists of $n+1$ points in this case, so $e(MF(W,0))=n+1$, and $\nu_{\widehat{\mathcal{M}}_\beta}(0)=n$.  Thus $n^0_\beta=n$ in this case, as would be expected from an isolated $\mathbb{P}^1$ with multiplicity $n$.

As our final example, we return to the lines on the Fermat quintic threefold.  In this case, the lines have normal bundle $\mathcal{O}(1)\oplus \mathcal{O}(-3)$ which has a two-dimensional space of sections.   So the moduli space is locally planar.  Away from the 375 special lines, the reduced moduli space is a smooth plane curve, with multiplicity 2.  We can choose local analytic coordinates $(x,y)$ near any such point of the moduli space so that the moduli space is defined by $x^2=0$, which can be deduced from a superpotential $W=x^3$.  The Milnor fiber of $W=x^3$ is the disjoint union of three contractible spaces, hence has Euler characteristic 3.  By (\ref{eq:bmf}) we conclude that $\nu=-2$ on this locus.

Near a point of the moduli space corresponding to the 375 lines, the reduced moduli space consists of two intersecting smooth curves.  We choose coordinate $(x,y)$ so that the two curves are given by $x=0$ and $y=0$.  We have already observed that these curves have multiplicity 2 away from the origin.  We also noted that the multiplicity is 5 at the origin.  We deduce that the moduli space has local equations $x^2y^3=x^3y^2=0$, which can be derived from a superpotential $W=x^3y^3$.   Since the singularity of $xy=0$ is a node, whose Milnor fiber has the homotopy type of a circle, we conclude that the Milnor fiber of $W=x^3y^3$ has the homotopy type of the disjoint union of three circles, which has Euler characteristic zero.  By (\ref{eq:bmf}) we conclude that $\nu=1$ at these 375 points of moduli.

We can now calculate $n^0_1$ from the lines on the Fermat quintic threefold, obtaining 2875 as expected.  Let $Z\subset \mathcal{M}_1\simeq\widehat{\mathcal{M}}_1$ be the 375 points corresponding to the lines identified above.  Then $\mathcal{M}-Z$ is the union of 50 non-compact curves $C_i$, each isomorphic to the $g=6$ degree 5 Fermat plane curve with 15 points deleted.  Each of these curves has Euler characteristic $2-2g-15=-25$.  Since $\nu=1$ on $Z$ and $\nu=-2$ on $\mathcal{M}_1-Z$,we can apply (\ref{eq:wteuler}) to get
\begin{equation}
n_1^0=50 \cdot (-2)\cdot (-25)+1\cdot 375=2875,
\end{equation}
as expected.

The definition of $n^0_\beta$ and its calculation by (\ref{eq:wteuler}),(\ref{eq:bmf}) is completely general.  To define $n^g_\beta$ and the refined invariants, we need more concepts. 
 First, we have to explain a compatibility condition on the locally defined superpotentials.  This is explained in terms of the notion of a d-critical locus.  
 Next, we need to spread out the pointwise condition on the topology of the Milnor fibers to a perverse sheaf, which is actually a complex whose cohomologies contain the 
 information of the cohomologies of the Milnor fiber.  To accomplish this, we need the notion of an orientation.  Finally, we use the decomposition theorem \cite{BBD} for
 the map $\widehat{\mathcal{M}}_\beta\to  {\mathcal{M}}_\beta$ together with hard Lefschetz to make  geometrically precise the notions of Lefschetz on the base and Lefschetz on the 
 fibers from \cite{GV2}. The caveat is that orientations need not be unique, and the definitions of the $n^g_\beta$ for $g>0$ and the refined invariants depend on the conjectural 
existence of orientations with certain good properties.   We take up each of these matters in turn.

\subsection{D-critical loci}\label{subsec:dcrit}

In \cite{Pantev:2013sss}, the authors showed that a  derived version of $\widehat{\mathcal{M}}_\beta$ has a $-1$-shifted symplectic structure.  Joyce found a classical truncation of this $-1$-shifted symplectic structure in \cite{Joyce:2015dc} and calls it a \emph{d-critical locus}.   In a rough sense, the structure of a d-critical locus remembers an important piece of the information obtained from choices of various critical charts 
 at distinct points of $\widehat{\mathcal{M}}_\beta$. Later, we will use perverse sheaves to connect this notion to a globalization of the Milnor fiber $MF(W,p)$ en route to refining the invariants and describing $n^g_\beta$.  

Joyce constructs a sheaf of complex vector spaces $S_Y$\footnote{Actually a sheaf of $\mathbb{C}$-algebras, but we don't need this additional structure.} 
on any scheme $Y$ as follows.  First consider the special case where $Y$ is a closed subscheme of a manifold $U$.  We let $i:Y\hookrightarrow U$ be the inclusion and we also let
$I_{Y,U}\subset\mathcal{O}_U$ be the ideal sheaf of holomorphic functions on $U$ which vanish on $Y$.  Then we define
\begin{equation}\label{eq:sdef}
S_Y=\ker\left(\frac{i^{-1}(\mathcal{O}_U)}{I_{Y,U}^2}\stackrel{d}{\to} \frac{i^{-1}(\Omega^1_U)}{I_{Y,U}\cdot i^{-1}(\Omega^1_U)},
\right)
\end{equation}
with the map $d$ being the exterior derivative.  It is shown that $S_Y$ is independent of the choice of manifold $U$ containing $Y$.  For an arbitrary $Y$ we choose an open covering $\{V_\alpha\}$ of $Y$ with embeddings of the $V_\alpha$ into manifolds $U_\alpha$, then the sheaves $S_{V_\alpha}$ constructed from the embeddings $V_\alpha\subset U_\alpha$ agree on the intersections $V_\alpha\cap V_\beta$ and therefore glue to give a well-defined sheaf $S_Y$ on all of $Y$.

\smallskip\noindent
{\bf Example.} Suppose $Y$ is smooth.  Then $S_Y=\mathbb{C}$.  We can take $U=Y$ so that $I_{Y,U}=0$, and the kernel of $d:\mathcal{O}_Y\to \Omega^1_Y$ is the constant sheaf $\mathbb{C}$.

\smallskip\noindent
{\bf Example.} Suppose $Y=\mathrm{Spec}(\mathbb{C}[x]/(x^n))$ is the scheme defined by $x^n=0$ (non-reduced if $n>1$).  Taking $U=\mathbb{C}$, we look at polynomials in $x$ mod $x^{2n}$ whose derivatives are divisible by $x^n$.   This leads to
\begin{equation}
S_Y=\mathrm{span}\left\{[1],
[x^{n+1}],[x^{n+2}],\ldots [x^{2n-1}]
\right\}
\end{equation}
where $[\ldots]$ denotes the equivalence class modulo $x^{2n}$.  

Now suppose  that we have an embedding of $Y$ into a manifold $U$ and we have a superpotential $W$ on $U$ with $Y=\mathrm{Crit}(W)$.   Then $I_{Y,U}$ is the ideal generated by the partial derivatives of $W$ (in any system of coordinates), so that $dW\in I_{Y,U}\cdot i^{-1}(\Omega^1_U)$. It follows immediately from (\ref{eq:sdef}) that $W$ determines a section of $S_Y$.

Since constant functions have vanishing differential, we have an inclusion $\mathbb{C}_Y\subset S_Y$ of the constant sheaf on $Y$ in $S_Y$.  If we let $S^0_Y\subset S_Y$ be the subsheaf represented by functions vanishing on $Y$, we have the direct sum decomposition $ S_Y= S^0_Y\oplus \mathbb{C}$.  

Returning to the example above with $Y=\mathrm{Spec}(\mathbb{C}[x]/(x^n))$, we have 
\[S^0_Y=
\mathrm{span}\left\{[x^{n+1}],[x^{n+2}],\ldots [x^{2n-1}]\right\}.
\] 

We can now define a d-critical locus.

\smallskip\noindent
{\bf Definition.} A \emph{d-critical locus} is a pair $(Y,s)$ with $Y$ a scheme and $s\in H^0(Y,S^0_Y)$, such that for every $p\in Y$ we can find a critical chart $(R,U,W,\iota)$ with $p\in R$ such that the section of $S^0_R$ determined by $W|_R$ as explained above is equal to $s|_R$.

\smallskip
Note that $W$ is not part of the data defining a d-critical locus.  The only requirement is that locally such a $W$ exists which is compatible with the globally defined $s$.

Returning again to $Y=\mathrm{Crit}(W)$ for $W=x^{n+1}$ on $U=\mathbb{C}$, we have that $(Y,[x^{n+1}])$ is a d-critical locus.  Similar reasoning shows that $(Y,[\sum_{i=n+1}^{2n-1}a_ix^i])$ is a d-critical locus if and only if $a_{n+1}\ne0$.  We simply take $W=\sum_{i=n+1}^{2n-1}a_ix^i$ on any open set $U\subset \mathbb{C}$ which contains 0 but does not contain any of the non-zero roots of $W$.

To a d-critical locus $(Y,s)$ we can associate a line bundle $K_{Y,s}$ on $Y$ with its reduced structure\footnote{$Y^{\mathrm{red}}$ is the reduced structure on $Y$, the same topological space with the nilpotent functions set to zero.  In this way, all components of $Y^{\mathrm{red}}$ have multiplicity 1.} $Y^{\mathrm{red}}$. It has the property that for any critical chart $(R,U,W,\iota)$ with $R\subset Y$, we have a canonical isomorphism 
\begin{equation}
(K_{Y,s})|_{R^\mathrm{red}} \simeq (K_U^{\otimes2})|_{R^\mathrm{red}}.
\end{equation}
In particular, if $Y$ is smooth, we have $K_{Y,0}\simeq K_Y^{\otimes 2}$.

\subsection{Perverse sheaves and D-modules}
In preparation for the mathematical definition of the invariants, we give a quick introduction to perverse sheaves.  Part of this material was borrowed from a project of the second author and Wati Taylor, who we thank for permitting us to use the material here.  Perverse sheaves are a topological notion and can be defined for any topological space.  We will be using perverse sheaves on both $\widehat{\mathcal{M}}_\beta$ and $\mathcal{M}_\beta $.  
Perverse sheaves provide, among other things, an object which in a sense globalizes the data provided by the Milnor fibers.   We refer the reader to \cite{decataldo:2009dec} for more details about the ideas in this 
section, including IC sheaves, to be introduced later.

We start by considering complexes of sheaves of vector spaces $\mathcal{S}^\bullet=\cdots \to\mathcal{S}^i\stackrel{d_i}{\to}\mathcal{S}^{i+1}\to\cdots$ on $S$.\footnote{For our purposes, we consider complex vector spaces.  In applications to Hodge theory, rational vector spaces are typically considered.}  We have the cohomology sheaves $\mathcal{H}^i(\mathcal{S}^\bullet)=\ker(d^i)/\mathrm{coker}(d^{i-1})$, a sheaf of vector spaces.  We require the cohomology sheaves $\mathcal{H}^i(\mathcal{S}^\bullet)$ to be \emph{constructible}, i.e.\ each is a sheaf which is a local system on each stratum of a stratification by locally closed subsets.  We do not require the $\mathcal{S}^i$ themselves to be finite dimensional.

We form a bounded derived category from these complexes by the usual procedures, restricting to bounded complexes with constructible cohomology, identifying morphisms of complexes which agree up to homotopy, and inverting quasi-isomorphisms.  The resulting category is called the derived category of constructible complexes.  We will denote this category by $D^b_{\mathrm{con}}(S)$.  

The category $\mathrm{Per}(X)$ of perverse sheaves on $X$ is a subcategory $D^b_{\mathrm{con}}(S)$.  
The category of perverse sheaves is abelian, which means that we have notions of perverse subsheaves, perverse quotients, short exact sequences of perverse sheaves, etc.  It would take us too far afield to define perverse sheaves carefully.  We content ourselves with listing some common examples of perverse sheaves below.   We also digress by including a section on D-modules and the Riemann-Hilbert correspondence, which is another route to perverse sheaves which is based on concepts which are probably more familiar to physicists.

Next, we expand a bit on the shift functor in the derived category.  For any 
$n\in\mathbf{Z}$ and $\mathcal{S}^\bullet\in D^b_{\mathrm{con}}(S)$, we define $\mathcal{S}[n]^\bullet\in D^b_{\mathrm{con}}(S)$ by $\mathcal{S}[n]^i=\mathcal{S}^{i+n}$, so that the complex $\mathcal{S}^\bullet$ gets shifted $n$ places to the left.  We have $\mathcal{H}^i(\mathcal{S}[n]^\bullet)=\mathcal{H}^{i+n}(\mathcal{S}^\bullet)$.

An important special case is when $\mathcal{S}^\bullet$ is a single constructible sheaf $F$ in degree~0 with all other terms of the complex vanishing.  We continue to use the same symbol $F$ to describe this complex.  Then $F[n]\in D^b_{\mathrm{con}}(S)$ is the complex with the sheaf $F$ in degree $-n$ and all other terms zero.

We also have hypercohomology groups $\mathbf{H}^i(K^\bullet)$ which generalize the cohomology of a sheaf.  This leads to the Poincar\'e polynomial
\begin{equation}
P_y(K^\bullet)=\sum_i \dim \mathbf{H}^i(K^\bullet) y^i.
\end{equation}
In the cases which interest us, $K^\bullet$ satisfies a form of Poincar\'e duality,\footnote{In fact, these $K^\bullet$ are self-dual under \emph{Verdier duality}. We will not discuss Verdier duality further in this paper.} and so $P_y(K^\bullet)$ is invariant under $y\mapsto y^{-1}$.  We can therefore identify $P_y(K^\bullet)$ with the character of an $SL(2)$ representation.

If $E^\bullet$ is a single sheaf $E$ in degree~0, then $\mathbf{H}^i(E^\bullet)$ is simply $H^i(E)$.  It follows that
$\mathbf{H}^i(E^\bullet[k])=\mathbf{H}^{i+k}(E^\bullet)$.

Before giving our first examples of perverse sheaves, we recall the notion of a local system, which is a locally free sheaf of finite dimensional vector spaces (or equivalently, the sheaf of flat sections of a flat vector bundle).  Here are some examples of perverse sheaves.

\begin{itemize}
\item The shifted constant sheaf $\mathbb{C}[\dim X]$ on $X$, if $X$ is smooth.  
\item More generally, $F[\dim X]$ for any local system $F$ on $X$, still assumed smooth.
\item The shifted constant sheaf $\mathbb{C}_Z[\dim Z]$ on a smooth submanifold $Z\subset X$, where $\mathbb{C}_Z$ is the constructible sheaf which is the constant sheaf $\mathbb{C}$ on $Z$ and the constant sheaf $0$ on $X-Z$.
\item More generally, if $F$ is a local system on a smooth submanifold $Z$, We consider $F$ as a constructible sheaf on $X$ by putting $F|_{X-Z}=0$.  Then $F[\dim Z]$ is a perverse sheaf on $X$.
\end{itemize}

The fundamental building blocks of perverse sheaves are the $IC$ sheaves \footnote{So named because of a connection to intersection cohomology.}.  Let $Z\subset X$ be an irreducible closed subvariety, not necessarily smooth, and let $U\subset Z$ be a smooth Zariski open subset of $Z$, necessarily dense.  Let $F$ be a local system on $U$.  Then we have a canonically defined perverse sheaf $IC(F)$ on $X$ with the properties $IC(F)|_U\simeq F[\dim Z]$ and $IC(F)$ vanishes on $X-Z$.  If desired, $IC(F)$ can be constructed using a stratification $U=U_0\subset U_1\subset \cdots\subset U_k=Z$, starting with $F[\dim Z]$ on $U_0$ and inductively extending from $U_k$ to $U_{k+1}$.

If $F$ is simple (i.e.\ there are no nontrivial proper sub-local systems), then the objects $IC(F)$ are simple objects in the category of perverse sheaves.  This means that the only perverse subsheaves of $IC(F)$ are $IC(F)$ and 0.

Rather than define perverse sheaves directly, we digress from the main development by defining them from regular holonomic $D$-modules via the Riemann Hilbert correspondence.  While $D$-modules are not necessary for understanding the geometry of the (refined) GV invariants, we expect that concepts underlying $D$-modules will be more familiar to physicists than the concepts underlying perverse sheaves and we hope that the following section will have expository value.

\smallskip
If $X$ is a smooth variety, we denote by $\mathcal{D}_X$ the sheaf of differential operators on $X$.  This is a sheaf of noncommutative rings on $X$, so we have distinguish between multiplication on the right and on the left.  When $X$ is clear from context, we sometimes simply write $\mathcal{D}$ instead of $\mathcal{D}_X$.  A $D$-module is a quasicoherent sheaf $\mathcal{F}$ on $X$ which is a left $\mathcal{D}_X$-module, that is, $\mathcal{D}_X$ acts on $\mathcal{F}$ from the left and satisfies the associative and distributive laws, much as is familiar for $\mathcal{O}_X$ modules.

We refer to \cite{HTT} for a detailed exposition of D-modules and their relationships to perverse sheaves.

\smallskip\noindent
{\bf Examples.} 

\begin{itemize}
\item $\mathcal{O}_X$ is a $D$-module via the usual action of differential operators acting on functions.
\item The sheaf of sections of a flat vector bundle becomes a $D$-module using the covariant derivative with respect to the flat connection.
\item For $X=\mathbb{C}$, the (left) ideal $\mathcal{D}_X(xd_x-k)$ is a $\mathcal{D}$-module.  It follows that the quotient module $\mathcal{F}_k=\mathcal{D}/\mathcal{D}\cdot\left(xd_x - k\right)$ is also a $\mathcal{D}$-module.
\item More generally, let $X$ be arbitrary and let $D_1,\ldots,D_n$ be differential operators on $X$.  Then the $D_i$ generate an ideal $\mathcal{D}_X(D_1,\ldots,D_n)$, and the quotient $\mathcal{D}_X/\mathcal{D}_X(D_1,\ldots,D_n)$ is also a $\mathcal{D}$-module.
\end{itemize}
Another familiar example is the Picard-Fuchs $\mathcal{D}$-module on the complex structure moduli space of a Calabi-Yau threefold $X$, which is associated to the Picard-Fuchs equations on $H^3(X)$. 

\smallskip
Now let $\mathcal{F}$ be a $\mathcal{D}$-module.  The \emph{solution sheaf} of $\mathcal{F}$ is defined as
\begin{equation}
\mathrm{Sol}(\mathcal{F})=\mathrm{Hom}_{\mathcal{D}_X}(\mathcal{F},\mathcal{O}_X),
\end{equation}
the sheaf of $\mathcal{D}_X$-module homomorphisms of $\mathcal{F}$ to $\mathcal{O}_X$ using the complex analytic topology.

The terminology will become clear after some examples.  Since $\mathcal{F}_k=\mathcal{D}/\mathcal{D}\cdot\left(xd_x - k\right)$ is generated as a module by the identity differential operator 1, 
a homomorphism $h:\mathcal{F}_k\to \mathcal{O}_X$ is determined by its value on $1$.  Let $f(x)=h(1)$, so that for any differential operator $D$ we must have $h(D)=h(D\cdot 1)=Dh(1)=Df$.  For the homomorphism to be well-defined modulo $\left(xd_x - k\right)$, it is necessary and sufficient that $f(x)$ satisfy the differential equation $xf'=kf$.  More generally, the solution sheaf of $\mathcal{D}_X/\mathcal{D}_X(D_1,\ldots,D_n)$ is the sheaf of solutions to the system of differential equations $D_1f=\cdots=D_nf=0$.  The solution sheaf of $\mathcal{D}/(\mathcal{D}d_x)$ is the constant sheaf $\mathbb{C}$.  More generally, if $M$ is the $D$-module associated to a flat vector bundle, then $\mathrm{Sol}(M)$ is the associated local system of flat sections.  In the case of the Picard-Fuchs $\mathcal{D}$-module, we get the sheaf of flat sections as the solution sheaf.

Now $\mathrm{Sol}(\mathcal{F}_k)$ is a simple object of a topological nature.  Away from $x=0$, $\mathrm{Sol}(\mathcal{F}_k)$ is a locally free sheaf of 1-dimensional vector spaces, a local system of rank~1.  Local systems of rank $r$ are locally free sheaves of $r$-dimensional vector spaces.  Local systems are characterized by their monodromies.  In the case of $\mathrm{Sol}(\mathcal{F}_k)$, the monodromy is $e^{2\pi ik}$.   Let $L_\lambda$ be the rank~1 local system on $\mathbb{C}^*$    whose monodromy is multiplication by $\lambda$.  The sections of $\mathrm{Sol}(\mathcal{F}_k)$ are identified with the functions $\{cx^k\mid c\in\mathbb{C}\}$, and we have just explained that
$\mathrm{Sol}(\mathcal{F}_k)|_{\mathbb{C}^*}\simeq L_{\mathrm{exp}(2\pi ik)}$.

We have for the stalk at 0
\begin{equation}
\mathrm{Sol}(\mathcal{F}_k)|_0=\left\{
\begin{array}{cl}
\mathbb{C}&k=0\\
0&{\rm otherwise}
\end{array}.
\right.
\end{equation}
If $k>0$, the stalk at the origin is zero because the solutions all vanish at $x=0$, while if $k<0$, the nonzero solutions are singular at $x=0$ and so the only solution which is well-defined at the origin is $f(x)\equiv0$ and the stalk vanishes at the origin in this case as well. 

Notice that we have a stratification of $\mathbb{C}$ with strata $\mathbb{C}^*$ and $\{0\}$ such that the sheaf $\mathrm{Sol}(\mathcal{F}_k)$ is locally constant on each stratum, so that  $\mathrm{Sol}(\mathcal{F}_k)$ is constructible.

Now suppose that we have a $\mathcal{D}$-module $M$ on a smooth $n$-dimensional space $X$.  We have the \emph{characteristic variety} $\mathrm{Ch}(M)\subset T^*(X)$ (also called the singular support), see e.g.\ \cite[Section 2.2]{HTT}.  If $M$ is nonzero and coherent, we have $\dim \mathrm{Ch}(M)\ge n$.  The system is \emph{regular holonomic} if $\mathrm{Ch}(M)$ has dimension $n$ and has no multiplicities (see e.g.\ \cite[Chapter 6]{HTT}).  These modules have especially nice properties.  In particular, the solution sheaf of a regular holonomic $\mathcal{D}$-module is constructible.  From the viewpoint of differential equations, the holonomic systems of differential equations in several variables have finite-dimensional spaces of solutions.  The Picard-Fuchs equations are a good example.  

In general, we lose information in passing from a $\mathcal{D}$-module to its solution sheaf, even for  regular holonomic $\mathcal{D}$-modules.  We compensate by taking the derived functor $R\mathrm{Hom}$ and consider
\begin{equation}R\mathrm{Hom}_{\mathcal{D}_X}(\mathcal{F},\mathcal{O}_X).
\end{equation}
We return to our example of $\mathcal{F}_k$ on $\mathbb{C}$ to get a sense of what this means.  In this case, $\mathcal{F}_k$ has a resolution by free $\mathcal{D}$-modules
\begin{equation}\label{eq:resolution}
0\to\mathcal{D}\to\mathcal{D}\to\mathcal{F}_k\to0,
\end{equation}
where the first nonzero mapping sends a differential operator $D$ to the differential operator $D(xd_x - k)$.  Passing to the derived category of $\mathcal{D}$-modules, we see that $\mathcal{F}_k$ is quasiisomorphic to the two-term complex $\mathcal{D}\to\mathcal{D}$, which can then be used to compute $R\mathrm{Hom}$.\footnote{Warning: this computational method works because $\mathbb{C}$ is affine, analogous to the simplified computation of Ext groups in the affine case.  The computation for general $\mathcal{D}$-modules is more complicated.}  So $R\mathrm{Hom}_{\mathcal{D}}(\mathcal{F}_k,\mathcal{O}_{\mathbb{C}})$ is represented by the complex
\begin{equation}\label{eq:rhomcomp}
\mathrm{Hom}(\mathcal{D},\mathcal{O}_{\mathbb{C}})\to \mathrm{Hom}(\mathcal{D},\mathcal{O}_{\mathbb{C}}).
\end{equation}
Since $\mathrm{Hom}(\mathcal{D},\mathcal{O}_{\mathbb{C}})$ is isomorphic to $\mathcal{O}_{\mathbb{C}}$, (\ref{eq:rhomcomp}) simplifies to the complex
\begin{equation}\label{eq:solncx}
\mathcal{O}_{\mathbb{C}}\to \mathcal{O}_{\mathbb{C}},
\end{equation}
where the map takes $f$ to $xf'-kf$.  We call the complex (\ref{eq:solncx}) the \emph{solution complex} of $\mathcal{F}_k$.  Notice that the kernel is just the solution sheaf described above.  We also can have a cokernel.  Given any function $g\in\mathcal{O}_{\mathbb{C}}$,
we can always solve $xf'-kf=g$ for $f$ away from 0.  So the map in (\ref{eq:solncx}) is surjective on $\mathbb{C}^*$, and so the cokernel, if nonzero, is a skyscraper sheaf supported at 0. 
In particular, the cokernel of (\ref{eq:solncx}) is constructible, hence the solution complex has constructible cohomology.  It  is easy to see that the cokernel of (\ref{eq:solncx}) is zero if $k\ne0$, and is the skyscraper sheaf $\mathbb{C}$ at zero for $k=0$.

We adopt the convention of shifting (\ref{eq:solncx}) one place to the left, so that the first term is in degree $-1$ and the second term is in degree 0.  In other words, we consider $R\mathrm{Hom}_{\mathcal{D}}(\mathcal{F}_k,\mathcal{O}_{\mathbb{C}})[1]$.   In general, we shift $R\mathrm{Hom}_{\mathcal{D}_X}(\mathcal{F},\mathcal{O}_X)$ by $\dim(X)$ places to the left.  

The category of \emph{perverse sheaves} on $X$ is the category formed by the 
\[R\mathrm{Hom}_{\mathcal{D}_X}(\mathcal{F},\mathcal{O}_X)[\dim(X)]
\] 
as $\mathcal{F}$ varies over the category of regular holonomic $\mathcal{D}$-modules.  Usually, perverse sheaves have an independent definition and this equivalence, the \emph{Riemann-Hilbert correspondence}, is a deep theorem.  However, we take this as the definition of perverse sheaves for our purposes.  

We are suppressing a crucial point here.
The functor $\mathcal{F}\mapsto R{\rm Hom}_{\mathcal{D}_X}(\mathcal{F},\mathcal{O}_X)[\dim(X)]$ is contravariant, which is not a natural way to identify categories.  To get a covariant functor and the usual formulation of the Riemann-Hilbert correspondence, we would use $R{\rm Hom}_{\mathcal{D}_X}(\mathcal{O}_X,\mathcal{F})$ instead. Since our goal in giving the contravariant functor was simply to give additional insight into perverse sheaves, we will not pursue the covariant functor in more detail as it would take us too far afield.  But we do state a couple of results which will give us some insights into perverse sheaves.

\smallskip\noindent
1. There is an equivalence of categories $DR:(D-mod)_{rh})(X) \to Perv(X)$ from regular holonomic D-modules on $X$ to perverse sheaves on $X$.

\smallskip
Furthermore the de Rham functor $DR$ can also be applied to a derived category of complexes of D-modules whose cohomologies (necessarily D-modules) are regular holonomic.  The result is a complex of vector spaces with constructible cohomology.  This gives us an equivalence of derived categories, also denoted by DR.

\smallskip\noindent
2. There is an equivalence of categories $DR:D^b_{rh}(D-mod)(X) \to Const(X)$ from the derived category of D-modules on $X$ with regular holonomic cohomology modules to the derived category of complexes of sheaves of vector spaces on $X$ with constructible cohomology.

\smallskip
We emphasize that for a complex $M^\bullet$ of D-modules as above, the complex of sheaves $DR(M^\bullet)$ will not be perverse unless $M^\bullet$ was a single regular holonomic D-module placed in degree 0.

Examples.  

The local system $L_\lambda$ on $\mathbb{C}^*$ is simple, so $IC(L_\lambda)$ is a simple perverse sheaf.  For stalks, we clearly have $IC(L_\lambda)_z=(L_\lambda)_z[1]$ if $z\ne 0$,
 since $IC(L_\lambda)=(L_\lambda)[1]$ on $\mathbb{C}^*$.  Furthermore, it is the case that
\begin{equation}\label{eq:icex}
IC(L_\lambda)_0=\left\{
\begin{array}{cl}
\mathbb{C}[1]&\lambda=1\\
0&{\rm otherwise}
\end{array}
\right. .
\end{equation}

Let's construct perverse sheaves $S_\lambda$ satisfying the conditions required of $IC(L_\lambda)$: $(S_\lambda)|_{\mathbb{C}^*}=L_\lambda[1]$ and (\ref{eq:icex}).  We construct these by the Riemann-Hilbert correspondence.\footnote{This discussion is of course incomplete, since we have not actually defined what an IC sheaf is.}  For $\lambda=1$, we note that $\mathbb{C}[1]$ is the (shifted) solution complex of $\mathcal{D}/(\mathcal{D}d_x)$, so is perverse by definition.  It turns out that $IC(L_1)=\mathbb{C}[1]$, which is immediately checked to be consistent with the restriction to $\mathbb{C}^*$ and with (\ref{eq:icex}).

Similarly, for $\lambda\ne 1$, we pick $k$ with $\lambda=\mathrm{exp}(2\pi i k)$ and then $IC(L_\lambda)$ is the (shifted) solution complex of $\mathcal{F}_k$, which we also check is consistent with the restriction to $\mathbb{C}^*$ and with (\ref{eq:icex}).

\smallskip
We now return to our main development.  Suppose $W$ is a holomorphic function on a smooth $U$, and $Y=\mathrm{Crit}(W)$.  To this data is associated a perverse sheaf on $Y$, 
called the perverse sheaf of vanishing cycles and denoted by $\Phi_W$.  

\smallskip\noindent
{\bf Example.}  Suppose $Y$ is smooth and take $U=Y$, $W=0$.  Then $\Phi_W=\mathbb{C}[\dim Y]$, the constant sheaf viewed as the degree $(-\dim Y)$ term of a complex with all other terms zero.

\smallskip

For any $p\in Y$, the stalk $(\Phi_W)_p$ is a complex of vector spaces with cohomology groups $H^i((\Phi_W)_p)$.  This gives a pointwise Euler characteristic 
\begin{equation}
\chi(\Phi_W)(p):=\sum_i(-1)^i\dim H^i(Y,(\Phi_W)_p)=\left(-1\right)^{\dim U}\left(1-e(MF(W,p)\right) .
\end{equation}
So we have
\begin{equation}
\nu_Y(p)=(-1)^{\dim U-1}\chi(\Phi_W)(p).
\end{equation}
However, since the stalks $(\Phi_W)_p$ are complexes of vector spaces, they have more information than their Euler characteristics.  In fact, the cohomologies $H^i((\Phi_W)_p)$ coincide up to a shift with the reduced cohomologies of $MF(W,p)$, which is the same as its cohomology except for $H^0$, which is reduced in dimension by 1.  So the Euler characteristic of the reduced cohomology of $MF(W,p)$ is equal to $e(MF(W,p))-1$.  We conclude that up to a sign, $\nu_Y(p)$ is equal to the  Euler characteristic of the reduced cohomology of $MF(W,p)$.  Our refinement comes from using all of the reduced cohomology of $MF(W,p)$ instead of forgetting information by taking the Euler characteristic.

As elegant as this construction is, in application to GV invariants the construction only works locally.  The next step is to glue the locally defined perverse sheaves of vanishing cycles in order to get a globally defined perverse sheaf.  To do this, we will introduce the notion of an orientation.

\subsection{Orientations}\label{subsec:orientation}
Let $(Y,s)$ be a d-critical locus, with canonical bundle  $K_{Y,s}$.  An \emph{orientation} of $(Y,s)$ is a choice of square root $(K_{Y,s})^{1/2}$ of $K_{Y,s}$ on $Y^{\mathrm{red}}$, together with an isomorphism 
\begin{equation}\label{eq:orientiso}
\left(\left(K_{Y,s}\right)^{1/2}\right)^{\otimes2}\stackrel{\sim}{\to}K_{Y,s}
\end{equation}
on $Y^{\mathrm{red}}$.  An orientation determines a perverse sheaf $\Phi$ on $Y$ as follows \cite{Joyce:2015dc}.

Given a critical chart $(R,U,W,i)$, (\ref{eq:orientiso}) induces an isomorphism
\begin{equation}\label{eq:square}
\left(\left(K_{Y,s}\right)^{1/2}|_{R^{\mathrm{red}}}\right)^{\otimes2}\simeq \left(K_U^{\otimes2}\right)|_{R^{\mathrm{red}}}.
\end{equation}
Let $\sigma:\tilde{R}\to R^{\mathrm{red}}$ be the principal $\mathbb{Z}_2$-bundle of square roots of the isomorphism (\ref{eq:square}).  Let $L$ be the rank 1 local system on $R^{\mathrm{red}}$ defined by $\sigma_*(\mathbb{Q})\simeq
\mathbb{Q}\oplus L$.  Then the perverse sheaf $\Phi$ is determined by natural isomorphisms
\begin{equation}
\Phi|_R\simeq \Phi_W\otimes L.
\end{equation}

\smallskip\noindent
{\bf Example.} Returning the case of $Y$ smooth, $U=Y$, $W=0$, we have $K_{Y,0}=K_Y^2$.  If we choose $K_Y$ as an orientation, then we get $\Phi=\Phi_0=\mathbb{C}[\dim Y]$.  But if there is a 2-torsion line bundle $L$ on $Y$, then $K_Y\otimes L$ is another orientation.  The resulting perverse sheaf is then $L[\dim Y]$, viewing $L$ as a local system.

\smallskip
A canonical orientation has been recently described in \cite{ju}.  While its properties have not yet been elucidated, it is natural to hope that this orientation is the right one for physics. To avoid this uncertainty, whenever a moduli space $Y$ is smooth, we choose $K_Y$ as orientation, so that the corresponding perverse sheaf is $\mathbb{C}[\dim Y]$.  We make this choice to match the physics literature.

\smallskip
Consider the situation where $\mathcal{M}$ is a point, corresponding to a curve $C$ of genus $g$.  Then $\widehat{\mathcal{M}}=J(C)$.  Since the canonical bundle of $J=J(C)$ is trivial, we can choose $\mathcal{O}_J$ as an orientation on  $\widehat{\mathcal{M}}$ by the above discussion, giving the perverse sheaf $\mathbb{C}[g]$ on $\widehat{\mathcal{M}}$.  In this case, there are certainly non-trivial 2-torsion line bundles $L$.  If we choose such an $L$ as an orientation, we get $L[g]$ as a perverse sheaf.  If $g=1$, we have for the Poincar\'e polynomial $P_y(\mathbb{C}[1])=y^{-1/2}+2+y$ which corresponds to $n^1=1, n^0=0$.  But if we take $L[1]$ instead, all cohomology vanishes and we get $n^1=n^0=0$, which is not correct.

So we have to restrict the orientations to get the correct GV invariants.  In \cite{MaulikToda}, the notion of a \emph{Calabi-Yau orientation} is introduced.  These have the property that they are trivial on the Jacobians of curves in the fibers of $\widehat{\mathcal{M}}_\beta\to \mathcal{M}_\beta$.\footnote{A fine point: this need only hold on the fibers of a Stein factorization of  $\widehat{\mathcal{M}}_\beta\to \mathcal{M}_\beta$ \cite{MaulikToda}.} Furthermore, the inferred GV invariants are proven to be independent of the chosen Calabi-Yau orientation.  In the rest of this section, we will restrict ourselves to Calabi-Yau orientations.  It is conjectured in \cite{MaulikToda} that a Calabi-Yau orientation always exists.

As of this writing, it is not known which class of orientations produces the correct refined invariants.

\subsection{Higher genus Gopakumar-Vafa invariants}

At last we can define the unrefined invariants, at least up to a conjecture.  Consider $\pi:\widehat{\mathcal{M}}_\beta\to \mathcal{M}_\beta$.  Conjecturally, there exists a Calabi-Yau orientation on $\widehat{\mathcal{M}}_\beta$ with good properties.  Let $\Phi$ be the associated perverse sheaf on $\widehat{\mathcal{M}}_\beta$.  Now consider the derived pushforward $R\pi_*\Phi$, which is in the constructible derived category of $\mathcal{M}$.  There is no reason for $R\pi_*\Phi$ to be perverse, much as there is no reason for a complex of sheaves to be a sheaf.  But much as a complex of sheaves has cohomology sheaves, $R\pi_*\Phi$ has \emph{perverse cohomology}, which are perverse sheaves.  The $i$th perverse cohomology is denoted by $^pR^i\pi_*\Phi$.  Perverse cohomology can be understood by the Riemann-Hilbert correspondence.  To a perverse sheaf, the (inverse of the) Riemann-Hilbert correspondence associates a regular holonomic $D$-module.  But to a general complex of sheaves with constructible cohomology such as $R\pi_*\Phi$, we can only associate a complex $M^\bullet$ of $D$-modules whose cohomologies are regular holonomic.  Let $\mathcal{H}^i(M^\bullet)$ be the $i$th cohomology sheaf of $M^\bullet$, which is a regular holonomic $D$-module.  Then $^pR^i\pi_*\Phi$ is the perverse sheaf associated to $\mathcal{H}^i(M^\bullet)$ by the Riemann-Hilbert correspondence.  Or more formally albeit slightly imprecisely
\begin{equation}\label{eq:perversecoh}
^pR^i\pi_*\Phi=DR\left(\mathcal{H}^i\left(DR^{-1}\left(R^i\pi_*\Phi\right)\right)
\right),
\end{equation}
where DR is the de Rham isomorphism describing the Riemann-Hilbert correspondence on $\mathbb{P}^2$, and $\mathcal{H}^i\left(DR^{-1}\left(R^i\pi_*\Phi\right)\right)$ in keeping with our notation is the $i$th cohomology D-module of the complex of D-modules
$DR^{-1}\left(R^i\pi_*\Phi\right)$.  So $\mathcal{H}^i\left(DR^{-1}\left(R^i\pi_*\Phi\right)\right)$ is a perverse sheaf.

Then at last we can define the GV invariants $n^g_\beta$ by
\begin{equation}\label{eq:gvdef}
\sum_i \chi(^pR^i\pi_*\Phi)y^i=\sum n^g_\beta(y^{1/2}+y^{-1/2})^{2g}.
\end{equation}
This definition makes sense because the left-hand side of (\ref{eq:gvdef}) is invariant under $y\mapsto y^{-1}$, by a variant of Poincar\'e duality in this situation which is a consequence of the self-duality of $\Phi$ with respect to Verdier duality \cite[Theorem 6.9]{BBDJS}.

Comparing (\ref{eq:gvdef}) to the physics definition in \cite{GV2} recalled at the beginning of Section~\ref{sec:gvphys}, we see that the $\chi$ on the left hand side implements $Tr(-1)^{F_R}$ while on the right hand side $(y^{1/2}+y^{-1/2})^{2g}$ is the character of the $I_g$ representation.

It was shown in \cite{MaulikToda} that the $n^g_\beta$ so defined are independent of the choice of Calabi-Yau orientation.

\smallskip\noindent
Example.  We return to the elliptic fibration $\pi:X\to \mathbb{P}^2$ and consider $n^g_{kf}$.  We have already seen that $\widehat{\mathcal{M}}_{kf}\to {\mathcal{M}}_{kf}$ is identified with $\pi:X\to B$ itself, so the answer will be independent of $k$.  Since $X$ is smooth, its associated d-critical locus has canonical bundle $K_{X,0}\simeq K_X^{\otimes2}\simeq \mathcal{O}_X$.  There is an obvious orientation $K_{X,0}^{1/2}=\mathcal{O}_X$, which is the only possible orientation since $X$ is simply connected.  Clearly this is a Calabi-Yau orientation.  The perverse sheaf of vanishing cycles in this case is $\Phi_0=\mathbb{C}[3]$. 

Let $\Delta\subset \mathbb{P}^2$ be the discriminant curve, so that the elliptic fibers $E_b$ are all smooth for $b\in B^0:=\mathbb{P}^2-\Delta$.  The cohomologies $H^i(E_b,\mathbb{C})$ form the fibers of local systems on $B^0$, of ranks 1,2,1 for $i=0,1,2$ respectively.  These local systems are denoted $(R^i\pi_*\mathbb{C})|_{B^0}$ in the language of algebraic geometry.  Note that $(R^0\pi_*\mathbb{C})|_{B^0}$ and $(R^2\pi_*\mathbb{C})|_{B^0}$ are both the trivial local system $\mathbb{C}|_{B^0}$, since both $H^0(E_b,\mathbb{C})$ and $H^2(E_b,\mathbb{C})$ are canonically isomorphic to $\mathbb{C}$.

The local systems are interpolated by an object of the constructible derived category of $B^0$ denoted by $(R\pi_*\mathbb{C})|_{B^0}$.  This object can be represented by a complex of sheaves of vector spaces whose cohomology sheaves are the local systems $(R^i\pi_*\mathbb{C})|_{B^0}$.

Consider the complex
\begin{equation}
\mathbb{C}|_{B_0}\oplus (R^1\pi_*\mathbb{C})|_{B^0}[-1] \oplus \mathbb{C}|_{B_0}[-2],
\end{equation}
a complex with terms $\mathbb{C}|_{B_0}$ in degree 0, $(R^1\pi_*\mathbb{C})|_{B^0}$ in degree 1, and $\mathbb{C}|_{B_0}$ in degree 2, with all differentials vanishing.  So the nonzero terms are the same as the cohomologies, and we have exhibited an object of the derived category with the same cohomologies as $(R\pi_*\mathbb{C})|_{B^0}$.  A theorem of Deligne \cite{Deligne:1968lef} says that they are actually equal:
\begin{equation}\label{eq:deligne}
(R\pi_*\mathbb{C})|_{B^0}=\mathbb{C}|_{B_0}\oplus (R^1\pi_*\mathbb{C})|_{B^0}[-1] \oplus \mathbb{C}|_{B_0}[-2].
\end{equation}
This equality arises from the Hard Lefschetz theorem for the map $\pi$.  In this formulation, the Lefschetz raising operators are maps
\begin{equation}
R^i\pi_*\mathbb{C}|_{B^0}\to R^{i+2}\pi_*\mathbb{C}|_{B_0}
\end{equation}
arising from cup product with a K\"ahler class on the base (``Lefschetz on the base," in the terminology of \cite{GV2}).  For $i=0$, this is the Hard Lefschetz isomorphism which identifies $H^0$ of the fiber with $H^2$ of the fiber.

We shift (\ref{eq:deligne}) 3 places to the left so as to have a perverse sheaf on the left hand side: 
\begin{equation}\label{eq:decompb0}
(R\pi_*\mathbb{C}[3])|_{B^0}=\mathbb{C}[3]|_{B_0}\oplus (R^1\pi_*\mathbb{C}[2])|_{B^0}\oplus \mathbb{C}[1]|_{B_0}.
\end{equation}
Since $\dim B^0=2$ and $R^1\pi_*\mathbb{C}$ is a local system on $B^0$, the summand $(R^1\pi_*\mathbb{C}[2])|_{B^0}$ in the right hand side of (\ref{eq:decompb0}) is a perverse sheaf.  The other two terms are shifts of the perverse sheaf $\mathbb{C}[2]$ on $B^0$.  This is a special case of the \emph{decomposition theorem} \cite{BBD}, which guarantees that $R\pi_*$ of a perverse sheaf decomposes into a direct sum of perverse sheaves and their shifts.   Another useful reference for the decomposition theorem is (\cite{decataldo:2009dec}).

In this context Hard Lefschetz corresponds to maps of perverse cohomologies
\begin{equation}
{}^pR^i\pi_*\Phi\to {}^pR^{i+2}\pi_*\Phi,
\end{equation}
which play the role of ``Lefschetz on the base" in the phraseology of \cite{GV2}.  This is another reason why we need perverse sheaves: in order for hard Lefschetz to hold, we must use perverse sheaves instead of sheaves.

The decomposition theorem extends (\ref{eq:decompb0}) from $B^0$ to $\mathbb{P}^2$ in a natural way.
We have
\begin{equation}\label{eq:pushdown}
R\pi_*\left(\mathbb{C}[3]\right)=\mathbb{C}[3]_{\mathbb{P}^2}\oplus IC\left(R^1\pi_*\mathbb{C}|_{B^0}\right)\oplus\mathbb{C}[1]_{\mathbb{P}^2}.
\end{equation}
Since $P=\mathbb{C}[2]$ is perverse on $B$ while $\mathbb{C}[3]=P[1]$ and $\mathbb{C}[1]=P[-1]$, we conclude that
\begin{equation}\label{eq:decomp}
{}^pR^{-1}\pi_*(\mathbb{C}[3])=\mathbb{C}[2],\ {}^pR^{0}\pi_*(\mathbb{C}[3])=IC\left(R^1\pi_*\mathbb{C}|_{B^0}\right),\ {}^pR^{1}\pi_*(\mathbb{C}[3])=\mathbb{C}[2].
\end{equation}

We have $\chi(\mathbb{C}[2])$ is just the Euler characteristic 3 of $\mathbb{P}^2$ (there is no sign change since the shift is even).  We next set out to compute $\chi(IC\left(R^1\pi_*\mathbb{C}|_{B^0}\right)$.  

For a sheaf $F$ on $X$ we have the functorial isomorphism $H^0(X,F)\simeq H^0(\mathbb{P}^2,\pi_*F)$.  This implies the isomorphism of derived functors
\begin{equation}\label{eq:Leray}
\mathbb{H}^*(X,F^\bullet) = \mathbb{H}^*(\mathbb{P}^2,R\pi_*F^\bullet)
\end{equation}
for any $F^\bullet$ in the derived category.   Applying (\ref{eq:Leray}) to $F^\bullet=\mathbb{C}$, we have
\begin{equation}\label{eq:lerayc}
\mathbb{H}^*(\mathbb{P}^2,R\pi_*\mathbb{C})=H^*(X,\mathbb{C}).
\end{equation}
Taking Euler characteristics in (\ref{eq:lerayc}) we get
\begin{equation}\label{eq:eulerellp2}
\chi(R\pi_*\mathbb{C}) = e(X)=-540,
\end{equation}
Also the Euler characteristic of the constant sheaf on $\mathbb{P}^2$ is just the topological Euler characteristic 3.  Finally, odd shifts change the sign of the Euler characteristic.  So taking Euler characteristics in (\ref{eq:pushdown}) we get
\begin{equation}\label{eq:iccalc}
540=-3+\chi(IC\left(R^1\pi_*\mathbb{C}|_{B^0}\right)-3,
\end{equation}
so that $\chi(IC\left(R^1\pi_*\mathbb{C}|_{B^0}\right)=546$.
Then (\ref{eq:gvdef}) becomes 
\begin{equation}\label{eq:gvgen}
3y^{-1}+546+3y^1=\sum n_{kf}^g(y^{1/2}+y^{-1/2})^{2g},
\end{equation}
so that $n^0_{kf}=540,\ n^1_{kf}=3$, and $n^g_{kf}=0$ for $k>1$, in agreement with (\ref{eq:GVfphysics}) as calculated using the ideas of \cite{GV2}.  Not only do the results agree, but the arithmetic calculations of these numbers agree.

We can repeat this calculation for the elliptic fibration over $\mathbb{F}_0$ or $\mathbb{F}_1$, both of which have Hodge numbers $(h^{11},h^{21})=(3,243)$.  In this context, (\ref{eq:pushdown})--(\ref{eq:lerayc}) are unchanged, while (\ref{eq:eulerellp2}) is replaced by $\chi(R\pi_*\mathbb{C})=-480$.  Replacing the left hand side of (\ref{eq:iccalc}) with 480 and replacing each 3 on the RHS with the Euler characteristic 4 of a Hirzebruch surface leads to $\chi(IC\left(R^1\pi_*\mathbb{C}|_{B^0}\right)=488$.  Then replacing the left hand side of (\ref{eq:gvgen}) with $4y^{-1}+488+4y^1$, we arrive at the familiar result $n^0_{kf}=480,\ n^1_{kf}=4$, and $n^g_{kf}=0$ for $k>1$.

While the calculation using the methods of \cite{GV2} is much simpler, we emphasize that the methods explained in this section apply in principle to the more common situation where $\widehat{\mathcal{M}}_\beta$ is singular.

\subsection{Refined Gopakumar-Vafa numbers}
We would like to refine the GV invariants by replacing the Euler characteristics of the perverse sheaves in the decomposition theorem by its cohomologies.   We outline the ideas here, referring to \cite{KiemLi} for more details.

We try to refine (\ref{eq:gvdef}) in the spin basis by the formula
\begin{equation}\label{eq:refgvdef}
\sum_{i,j} \mathbb{H}^j(^pR^i\pi_*\Phi)u^jy^i=\sum N^{j_L,j_R}_\beta(u^{-2j_L}+u^{-2j_L+2}+\ldots +u^{2j_L})(y^{-2j_R}+y^{-2j_R+2}+\ldots +y^{2j_R}).
\end{equation}
The refined numbers in the genus basis are defined similarly.

However, for (\ref{eq:refgvdef}) to make sense, its left hand side must be invariant under $y\mapsto y^{-1}$ and $u\mapsto u^{-1}$.   Unfortunately, these identities do not hold for a general $\Phi$.    A sufficient condition is for $\Phi$ to underly a pure Hodge module, a concept which we will say more about below.  Suffice it to say that for a smooth space $X$ of dimension $d$, the perverse sheaf $\mathbb{C}[d]$ underlies a pure Hodge module.

In our example of the elliptic fibration over $\mathbb{P}^2$, we have
\begin{equation}\label{eq:cohp2}
\dim \mathbb{H}^i(\mathbb{P}^2,\mathbb{C}[2])=H^{i+2}(\mathbb{P}^2,\mathbb{C})=\left\{
\begin{array}{cl}
1 & i =-2, 0, 2\\
0 & {\rm otherwise}
\end{array}
\right.
\end{equation}
Computing (hyper)cohomology rather than Euler characteristics in (\ref{eq:pushdown}) together with (\ref{eq:lerayc}) we arrive at
\begin{equation}
\dim \mathbb{H}^i(\mathbb{P}^2,IC\left(R^1\pi_*\mathbb{C}|_{B^0}\right))=\left\{
\begin{array}{cc}
546&i=0\\
0& {\rm otherwise}
\end{array}
\right.
\end{equation}
Forming a generating function with $y$ keeping track of the cohomology of degree and $u$ keeping track of the perverse cohomology degree, we get the generating function
\begin{equation}
\left(u^{-1}+u\right)\left(y^{-2}+1+y^2\right)+546
\end{equation}
which is the character of the representation
\begin{equation}
[\frac{1}{2},1]\oplus 546[0,0].
\end{equation}
In Section~\ref{basedeg0}, we will directly connect this calculation with the calculational methods of \cite{GV2,KKV}.

\smallskip
Making the obvious changes analogous to what we already did for the unrefined GV invariants in addition to replacing (\ref{eq:cohp2}) by 
\begin{equation}\label{eq:cohfk}
\dim \mathbb{H}^i(\mathbb{F}_k,\mathbb{C}[2])=H^{i+2}(\mathbb{F}_k,\mathbb{C})=\left\{
\begin{array}{cl}
1 & i =-2, 2\\
2 & i = 0\\
0 & {\rm otherwise}
\end{array},
\right.
\end{equation} 
we get
\begin{equation}
[\frac{1}{2},1]\oplus [\frac12,0]\oplus 488[0,0]
\end{equation}
for the elliptic fibration over $\mathbb{F}_0$ or $\mathbb{F}_1$.

\smallskip
More generally, the decomposition theorem holds for perverse sheaves underlying pure Hodge modules. 

Roughly, a \emph{Hodge module} is the data of a D-module $M$, a perverse sheaf $\Phi$ of \emph{rational} vector spaces, and an isomorphism
\begin{equation}
DR(M)\simeq \Phi\otimes_{\mathbb{Q}}\mathbb{C},
\end{equation}  
together with a good filtration of $M$ satisfying some properties analogous to Griffiths transversality and other conditions which don't concern us.  

As an already familiar example of a pure Hodge module, consider a family of Calabi-Yau threefolds $\pi:\mathcal{X}\to B$ over a smooth base $B$, with fiber $X_b$ over $b\in B$.  Then the families of cohomologies $H^3(X_b,\mathbb{Q})$ are a local system $L$ of rational vector spaces, and $\Phi=L[\dim B]$ is a perverse sheaf on $B$.  The sheaf of holomorphic sections of the associated vector bundle is a D-module $M$ via differentiation with respect to the Gauss-Manin connection.  We have $DR(M)\simeq \Phi\otimes_{\mathbb{Q}}\mathbb{C}$, and the filtration is the usual Hodge filtration.  

An even simpler example of a pure Hodge module over a smooth $B$ is the D-module $\mathcal{O}_B$, together with the perverse sheaf $\mathbb{Q}[\dim B]$ on $B$.  We have $DR(\mathcal{O}_B)=\mathbb{C}[\dim B]=\mathbb{Q}[\dim B]\otimes\mathbb{C}$.  The filtration is simply the filtration with $F^0\mathcal{O}_B=\mathcal{O}_B$ and $F^1\mathcal{O}_B=0$.

Unfortunately this method is not as general as could be hoped for, even when $\Phi$ is a pure Hodge module, for several reasons.

As pointed out in \cite{MaulikToda},  see also~\cite{MR3504535}, the refined invariants can depend on the choice of orientation.  A simple example the situation considered in Section~\ref{subsec:m} where the Calabi-Yau contains a ruled surface over a genus $g$ curve $C$.  In this case, $\widehat{\mathcal{M}}_\beta=\mathcal{M}_\beta = C$, and so any orientation is a Calabi-Yau orientation and produces the correct refined invariants.  Let $g=1$ for simplicity.  By the discussion in Section~\ref{subsec:orientation}, we see that in this situation from an appropriate orientation we can get the perverse sheaf $L[1]$ where $L$ is a nontrivial 2-torsion local system.  Now $\mathbb{C}[1]$ has Poincar\'e polynomial $y^{-1}+2+y$, while if $L$ is nontrivial, then $L[1]$ has vanishing Poincar\'e polynomial.  In the former case we get refined invariant $[0,1/2]+2[0,0]$, while in the latter case we get zero, so the results do not agree.  Only the first agrees with \cite{GV2}.  The unrefined invariant are identically zero in both cases.  We make the ansatz that whenever a moduli space $M$ is smooth, we must choose the natural orientation $K_{M,0}^{1/2}=K_M$.

In \cite{KiemLi}, it was attempted to avoid the issue of purity by using the associated graded Hodge module of a mixed Hodge module.
But an example is given in \cite{MaulikToda} where it is shown that no orientation exists which can give the correct answer using the ansatz of \cite{KiemLi}; it even gives the wrong unrefined invariants!

In conclusion, the geometric theory of refined GV numbers still needs to be refined! 


\section{Basic properties of elliptic fibred Calabi-Yau 3-fold} 
\label{ellipticfibrations} 
To describe the solution of  the topological string on 
elliptically fibred Calabi-Yau spaces we denote by $M$ the elliptically 
Calabi-Yau manifold, by $B$ its base and  by ${\cal E}$ its fibre  
\be
\begin{array}{ccc}
{\cal E}&\to&M \\
&&\phantom{\pi}\downarrow\pi \\ 
&&B
\end{array}
.
\ee 
In the following we will assume that for generic 
complex moduli of $M$, the fiber ${\cal E}$ degenerates  only with Kodaira type $I_1$ or $II$ in codimension two over 
the base $B$.  We also assume that $M$ is smooth as usual and denote its mirror manifold as $W$. We further assume that $M$ has a simple section 
in the elliptic fibration, which we sometimes identify with $B$. 
This corresponds in the $F$-theory setting to type IIB 
compactifications with no generic gauge group.

\subsection{Classical topological properties}  \label{subsecTopoproper}

We start by discussing the topological properties of elliptically
fibered Calabi-Yau 3-folds in this setting and set out notation.  First we pick convenient  
sets of generators for  $H_2(M)$ and  $H^2(M)$, beginning
with bases for the Mori and K\"ahler cones.  

Let  $\{[\tilde {\cal{C}}^k]\}$, $k=1,\ldots,h_{11}(B)=h_{11}(M)-1$ be generators for the Mori cone of $B$, and
$\{[D_k']\}$ the dual basis for the K\"ahler cone of $B$.  Identifying $B$ with the section of the elliptic fibration, the $[\tilde {\cal{C}}^k]$ may also
identified with curve classes on $M$.  Letting $[\tilde {\cal{C}}^e]$ be the class of the
elliptic fiber, we have that $\{[\tilde{\cal{C}}^e],[\tilde {\cal{C}}^k]\}$ are
generators for the Mori cone of $M$.

The K\"ahler cone of $M$ is then generated by the dual basis $\{[\tilde D_e],[\tilde D_k]\}$.
We have
\be
[\tilde D_k]=\pi^*[D_k'],\qquad [\tilde D_e]=[E]+\pi^*c_1(B), 
\label{basicdivisors}
\ee
where $[E]$ is the divisor class of the section.

The complexified K\"ahler areas of the curves in the base are
\be  
{\tilde T}^k=\int_{\tilde{\cal C}^k} {\cal B}+i\omega \ , 
\label{Kahlerareabase} 
\ee
where $\omega$ is the K\"ahler class of $M$ and ${\cal B}$ is the Neveu-Schwarz ${\cal B}$-field. 
Similarly we denote by 
\be 
{\tilde \tau}=\int_{\tilde {\cal C}^e}  {\cal B}+i\omega
\label{Kahlerareafibre}
\ee
the complexified K\"ahler area  of the fibre. We use here the $\tilde{}$ for the 
divisors, curves and variables corresponding to the K\"ahler and Mori cones, 
because later 
we need to define closely related shifted variables without the tilde on which 
the ${SL}(2,\mathbb{Z})$ act naturally.  

The enumerative results will be reported in the K\"ahler cone basis. These are obtained in the 
large volume limit in which ${\tilde T}^k$ ($T^k$) and ${\tilde \tau}$ ($\tau$) 
have large imaginary parts.  According to the base and fibre decomposition 
we split the degrees $[\kappa]\in H_2(M,\mathbb{Z})$ into 
\be 
[\kappa]=[f,\beta]\ ,
\ee
where $f$ denotes a multiple class of the elliptic fibre and 
$[\beta]\in H_2(B,\mathbb{Z})$ denotes a class in the base.

We assume that the anticanonical class $c_1(B)=-K_B$ of $B$ is 
semi-positive. The simplest case is that $B$ is a del Pezzo surface, but the 
semi-positivity of $c_1(B)$ includes more general cases as for example 
the toric spaces $\mathbb{P}_\Delta$ associated to two dimensional reflexive 
integer lattice polyhedra $(\Delta,\mathbb{Z}^2)$.      

The key point of these fibrations  is that due to the Leray-Serre spectral sequence    
all topological information of the total space comes from the base. One can easily 
calculate the following classical intersections on $M$ \cite{Klemm:2012}\cite{Klemm:1996ts}
\begin{equation}
\tilde D_e^3= \int_{B} c_1(B)^2, \qquad \tilde D_e^2 \cdot \tilde D_k= a_k, 
\qquad \tilde D_e \cdot \tilde D_i\cdot \tilde D_j= c_{ij}\ .       
\label{class1}
\end{equation}
In (\ref{class1}) we defined
\be 
a_k= c_1(B) \cdot  D_k', 
\qquad  c_{ij}=D_i' \cdot D_j', 
\label{ak}
\ee 
where the divisor classes $D_k'\in H^2(B,\mathbb{Z})$  were defined in (\ref{basicdivisors}).
Note that due to the elliptic fibration structure there are no triple intersections 
between the vertical divisors.   

Furthermore, one gets the the intersections of the second Chern 
class with divisors and the  Euler number $\chi=\int_M c_3(M)$  of $M$  as 
\begin{equation}
c_2(M)\cdot \tilde D_e = \int_{B} 11 c_1(B)^2+ c_2(B), \ \ c_2(M)\cdot \tilde D_k = 12 a_k, \ \  \chi= -60 \int_{B} c_1(B)^2 \  .       
\label{class2}
\end{equation}
Since the arithmetic genus of del Pezzo surfaces is one, we have further from  
Noether's formula for $\chi(\mathcal{O}_B)$ that $12=\int_B c_1(B)^2+ c_2(B)$.   

It is useful to make a change of basis on the curves 
\begin{equation}
{\cal C}^e =\tilde {\cal C}^e, \qquad {\cal C}^k=\tilde {\cal C}^k + \frac{a^k}{2} \tilde {\cal C}^e \ ,
\label{curvechange}
\end{equation}
where we have now put
\be
a^k= c_1(B) \cdot \tilde {\cal C}^k\ .
\label{dualak}
\ee
This changes the complexified K\"ahler parameter to
\be 
\tilde \tau=\tau, \qquad   T^k =\tilde T^k + \frac{a^k}{2} \tau\ .
\label{parameterchange} 
\ee

The dual basis to this basis of curves is
\begin{equation}
  \label{changedivisor}
  D_e=\tilde{D}_e-\frac12\pi^*c_1(B)=E+\frac12\pi^*c_1(B), \qquad D_k=\tilde{D}_k.
\end{equation}

The intersections in the basis (\ref{changedivisor}) become
\begin{equation}
D_e^3=\frac{1}{4} \int_B c_1^2, 
\quad D_e^2 \cdot D_k=0, 
\quad  D_e \cdot  D_i\cdot D_j= c_{ij}\ ,
\label{class3}
\end{equation}
so that the quadratic intersections of $D_e$ with the base classes vanish. 
In these computations we used 
\be
a^ka_k=\int_B c_1^2,
\label{dualas}
\ee 
which follows immediately from $c_1=a^k D_k'=a_k\tilde{C}^k$.

We note that in this basis there is a subgroup of the monodromy group of the mirror $W$ of 
$M$, which after identification with the mirror map generates a 
$PSL(2,\mathbb{Z})$ action on $\tau$ and does not act on the modified base classes 
$S^k$, $k=1,\ldots, b_2(B)$. This was first noted 
for the genus zero contribution for the base $B=\mathbb{P}^2$ in~\cite{Candelas:1994}.

To separate the instanton contribution it is also convenient to introduce 
exponentiated variables   
\be 
q=\exp(2 \pi i {\tau}), \quad  {\rm and }\quad  Q_k=\exp(2 \pi i  {T}^k)\ ,
\ee
which are invariant under integer shifts of the Neveu-Schwarz $B$-field ${\cal B}$, 
a trivial invariance of string theory. We use similar expressions for the variable 
with the $\tilde{} $ and abbreviate in both cases  
\be
Q^\beta=\prod_{k=1}^{b_2(B)} Q_k^{\beta_k}, \qquad \tilde Q^\kappa=\tilde q^f {\tilde Q}^{\tilde \beta} \quad {\rm etc.} 
\ee

\begin{figure}
\center
\includegraphics[width=10cm]{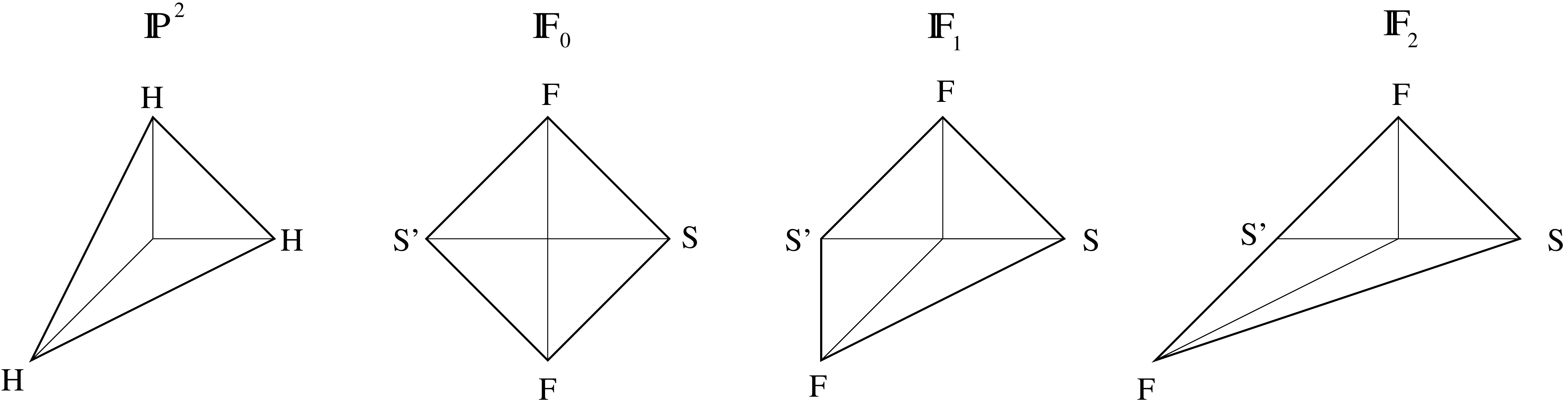}
\caption{The reflexive polyhedra $\Delta^{B*}$ associated to the four bases $B$ of the elliptically
fibred Calabi-Yau spaces discussed in this section.}
\label{mon}
\end{figure}

Let us finish this section with  the data for the bases that we discuss as key examples in the following. 
\begin{itemize}
\item For $B=\mathbb{P}^2$ the only K\"ahler generator is the hyperplane class $D_1'=H$ a 
$\mathbb{P}^1$ with $H\cdot H=1$. $H$ also represents the dual curve 
$[\tilde {\cal C}^1]$. The anticanonical class is $c_1(B)=3 H$ and $\int_{B} c_1(B)^2=9$. 
Hence $a^1=H\cdot (3 H)=3$. A general curve class can be abbreviated as 
$\beta=[\tilde {\cal C}^\beta]=[b H]$, 
so that $\beta \cdot (\beta-c_1(B))=b(b-3)$. 
\item The Hirzebruch surfaces $B=\mathbb{F}_k$, lead to smooth fibrations\footnote{The  Hirzebruch surface $\mathbb{F}_2$ is not a del Pezzo surface, but comes from a 
2d  reflexive polyhedron.}    
      for $k=0,1,2$. They are ruled surfaces, i.e $\mathbb{P}^1$ fibrations 
      over $\mathbb{P}^1$. We denote the fibre $\mathbb{P}^1$ by $D'_1=F$, hence $F\cdot F=0$. 
      Further there is a $-k$ genus $0$ curve $S'$ representing one section, i.e.  $S'\cdot F=1$ 
      and  $S'$ has self intersection $(S')^2=-k$, and another disjoint section 
      $D'_2=S=S'+k F$, with $S\cdot F=1$, self intersection $S^2=k$ and $S'\cdot S=0$. The 
      K\"ahler cone of divisors inside $B$ is spanned by $F$ and 
      $S$, the dual curves are $\tilde {\cal C}^1=S'$ and $\tilde {\cal C}^2=F$. 
      The anticanonical class is 
\be 
c_1(B)=2 S + (2-k)F
\ee 
hence $a_1=2,a_2=2+k$, $a^1=2-k,a^2=2$  and $(c_{ij})=\left(\begin{array}{cc} 0 & 1\\1 & k\end{array}\right)$. For a  general curve class $\beta=[b_1 S'+b_2 F]$ 
we calculate   
\be 
\beta\cdot \beta' =b_1 b'_2 + b_1' b_2- k b_1 b_1', \quad   \beta\cdot (\beta-c_1(B))=2 b_2 (b_1-1)-b_1(kb_1+(2-k))\ . 
\label{propertiesFk}
\ee
\end{itemize}

\subsection{The B-model geometry: Mirror construction, complex moduli space and Picard-Fuchs system} 
\label{subsec:bmodel}
In this subsection we describe the B-model geometry of the elliptic fibrations 
using Batyrev's construction~\cite{Batyrev:1994hm} of mirror pairs as 
hypersurfaces in dual toric ambient spaces $\mathbb{P}_{\Delta^*}$ and $\mathbb{P}_\Delta$ 
for pairs of reflexive polyhedra  $(\Delta^*,\Delta)$. We 
follow~\cite{Klemm:2012}\cite{Huang:2013}\cite{Bizet:2014uua} to describe 
elliptic fibrations with one section over toric Fano basis given  by 
$\mathbb{P}_{\Delta^{B*}}$. We  discuss the general features of the moduli 
spaces in particular the local limits, the discriminants, the involution symmetry $I$  
and the Picard-Fuchs systems governing the periods. Using this discrete symmetry, 
the monodromy data reflected by the Picard Fuchs equations, we  can prove 
the occurrence of the modular functions and the holomorphic anomaly equation.

The elliptic fibration of the hypersurface is inherited from the 
ambient space $\mathbb{P}_{\Delta^*}$. The polyhedron $\Delta^*$  
is constructed from a base polyhedron $\Delta^{B*}$ and a fibre 
polyhedron $\Delta^{F*}$  as the convex hull of the points on the left  
\begin{equation} 
  \footnotesize 
  \begin{array}{|ccc|ccc|} 
   \multicolumn{3}{c}{ \nu^*_i\in \Delta^*} &\multicolumn{3}{c}{ \nu_j\in \Delta}  \\ 
    &            &\nu_i^{F*} &             &\nu_j^{F} & \\
    & \Delta^{B*} & \vdots                &s_{ij}\Delta^{B}&\vdots & \\
    &            &\nu_i^{F*} &                   & \nu_j^{F}& \\
    & 0 \ldots 0      &                       & 0\dots 0              &                          & \\
    & \vdots     &\Delta^{*F}             &   \vdots          & \Delta^{F}            & \\
    & 0 \ldots 0      &                       & 0\ldots 0              &                          & \\
  \end{array} \, ,
\label{polyhedrafrombaseandfibre}
\end{equation} 
while the convex hull of the points on the right yield $\Delta$ and $(\Delta^*,\Delta)$ is a 
reflexive pair.  We represent $\Delta^{F*}$ as $\Delta^{F*}={\rm conv}((-1,0),(0,-1),(2,3))$ 
and $\Delta^{*B}$ can be any reflexive $2$ dimensional polyhedron, for example 
the ones given in Figure~1. It follows that  $\Delta^{F}={\rm conv}((1, 1), (1, -1),(-2, 1))$ and  we choose 
$\nu_1^{F*}=(2,3)$ and $\nu_1^{F}=(1,1)$, so that the scale factor 
$s_{11}=\langle \nu_1^F,\nu_1^{F*} \rangle +1=6$. The sections of the canonical bundle in the blowup of 
$\mathbb{P}_{\Delta^*}$  defined by a suitable triangulation of ${\Delta}$, is a family of Calabi-Yau 
hypersurfaces~\cite{Batyrev:1994hm}, compare(\ref{Pstar}), with an elliptic fibration, 
see~\cite{Klemm:2012}. The generic member of this family of elliptic Calabi-Yau spaces has 
a single section and the only singular fibres are of Kodaira types $I_1$ and $II$. 
The case $B=\mathbb{P}^2$ is described in this language in~\cite{Huang:2015sta}. 
Given the toric description, the topological data and the periods 
of the mirror can be obtained by standard methods~\cite{Hosono:1993qy}. 

The mirror manifold $W$ is described similarly by the canonical sections in $\mathbb{P}_{\Delta}$, 
which  in turn is given by the vanishing of the Newton polynomial of 
$\Delta^*$ written in the coordinates $\underline Y$ of $\mathbb{P}_{\Delta}$, which 
are associated to the points $\nu_j\in \Delta$
\be 
P^*=P_{\Delta^*}=\sum_{\nu^*_i \in \Delta^*}a_i \prod_{\nu_k\in\Delta} Y_k^{\langle \nu_i, 
\nu^*_k\rangle+1}=0
\label{Pstar}
\ee  
while the original  manifold $M$ is defined by exchanging ${\Delta^*}$ 
with ${\Delta}$ in (\ref{Pstar}). The complex structure of the 
mirror $W$ is redundantly parametrized by the coefficients $a_i$. 
This redundancy is removed by setting the coefficients $a_i$ 
for points on codimension one faces in $\Delta^*$ to zero and 
considering  scale invariant variables for the rest of the coefficients 
\be 
z^a=(-a_0)^{l_0^{(\alpha)}} \prod_{k=1}^s a_k^{l_k^{(\alpha)}} \quad a=1,\ldots,b_2(M)=h_{21}(W)
\label{defz}
\ee 
Here $\sum_{i} {\overline {\nu_i^*}} l^{(k)}_i$ are relations among points 
in ${\overline {\Delta^*}}={\rm conv}\{{\overline {\nu_i^*}}=(1,\nu_i^*)\}$ 
and if the $l^{(k)}$ are chosen to be the Mori vectors, $z_k=0$ is the maximal 
degeneration point which corresponds to a large radius limit of $M$~\cite{Batyrev:1994hm}.
It will be convenient below to use the scaling just to set some of the coefficients to 
one\footnote{Five of them can be set to $1$ by the toric actions in $\mathbb{P}_{\Delta^*}$ 
as in (\ref{Newtonpolyhedron})}.

Due to combinatorics of (\ref{polyhedrafrombaseandfibre}), the mirror itself 
has Weierstrass-Tate form 
\begin{equation} 
\begin{array}{rl} 
P^*&=y^2+ x^3 + a\,  xyz \prod_{i=1}^r \hat u_i +z^6  P^*_B(\hat {\underline u},b,m_i)\\ [ 2mm]
   &=y^2+ x^3 + a\,  xyz \prod_{i=1}^r \hat u_i +z^6 [ b\, \prod_{i=1}^r \hat u_i^6+ P^{\circ *}_B(\hat {\underline u},m_i)]\ .  
   \label{Newtonpolyhedron} 
\end{array}
\end{equation}
Here $(x,y,z)$ are the coordinates  associated to the points 
\be 
\{(0,0,1, 1), (0,0,1, -1),(0,0,-2, 1)\}\subset \Delta\ .
\ee 
The $\hat u_i$ can be viewed as coordinates for the mirror of the 
base. One can chose coordinates associated to the corners of the 
polyhedron $6\Delta^{B}$ embedded in $\Delta$ according 
to (\ref{polyhedrafrombaseandfibre}).  

The $a,b$ are distinguished complex structure coordinates of the mirror $W$. 
They correspond in $M$ to the canonical class of $\mathbb{P}_{\Delta^*}$ 
restricted to $M$ and the canonical class of the base $B=\mathbb{P}_{\Delta^*}$ 
respectively. The $m_i$, $i=1,\ldots, b_2(B)-1$ are complex moduli of $W$ 
that correspond to further K\"ahler moduli of the base. These are 
called $m_i$, because in the geometric engineering context one combination 
sets the mass scale of the compactification radius from $5d$ to $4d$  
and the others become the masses of the charged matter (quarks) in 
$4d$ $N=2$ QCD.  

Notice that the involution symmetry $I$ with $I^2=1$  on $M$ observed 
for $B=\mathbb{P}^2$ in~\cite{Candelas:1994} depends just on the fibre type 
and extends to all models under discussion. 
$I$  is realized in general on (\ref{Newtonpolyhedron}) by the transformations~\cite{Klemm:2012}
\be 
\begin{array}{rl} 
x &\rightarrow x+ \frac{a^2}{2}  (z \prod_{i=1}^r \hat u_i)^2 \\
y &\rightarrow y+ \frac{i-1}{2}  a x z  \prod_{i=1}^r \hat u_i-\frac{a^3}{12}(z \prod_{i=1}^r \hat u_i)^3\ .
\label{Ireparametrization} 
\end{array}
\ee
This leaves $P^*$ invariant provided $I$ acts on the moduli by
\be 
I:(a,b)\rightarrow \left(i a, b-\frac{a^6}{432}\right)\ . 
\ee 
The holomorphic $(3,0)$ form $\Omega$  for hypersurfaces in $\mathbb{P}_\Delta$ is defined by
\be
\Omega=\int_{\gamma} \frac{a \mu}{P^*}\ ,      
\ee
where $\gamma$ is a cycle encircling $P^*=0$ in $\mathbb{P}_\Delta$ and $\mu$ is a 
suitable measure on $\mathbb{P}_\Delta$, so that $\frac{a \mu}{P^*}$ is invariant 
under the torus actions on $\mathbb{P}_\Delta$. Note that $\mu$ is invariant 
under reparametrization compatible  with the torus actions such as 
(\ref{Ireparametrization}).  Therefore the involution acts on $\Omega$ as   
\be 
I:\Omega \rightarrow i \Omega \ .
\label{IactiononOmega}
\ee

The expression for the discriminants and the involution symmetry  simplifies
if one rescales the $a_i$ to $\alpha_i$, which induces a rescaling from 
$z_\alpha$ to $x_\alpha$. It is particularly convenient to set\footnote{The best rescaling 
of the $m_i\rightarrow \mu_i$ depends on the basis.}         
\be 
a=\frac{\alpha}{6^\frac{2}{3} s^\frac{1}{6}}, \quad b=- \frac{\beta}{3 s }, \quad   x_k:=(-a_0)^{l_0^{(k)}} \prod_{i=1}^s \alpha_k^{l_i^{(k)}},
\ee 
which makes the involution symmetry act as
\be 
I:(\alpha,\beta)\rightarrow \left(i \alpha, \beta+\alpha^6\right)\ \label{involution}
\ee
and leaves a further scaling $s$ which can be used to make the discriminant 
factors as simple as possible.   
Independent of $s$ we have for the fibre parameter  
$x_e=432 z_e$. The involution symmetry acts on $x_e$ and the base parameters $x_k$ variables 
as 
\be 
I:(x_e,x_k)\rightarrow \left(1-x_e, x_k \left(\frac{x_e}{x_e-1}\right)^{a^k}\right)\ . 
\label{involutionx}
\ee

A key observation that allows us to infer many properties of the global mirror geometry 
from the local mirror geometry is that $P^*_B(\hat {\underline u},\mu_i)$ is related simply 
to the geometry of the local mirror described by, i.e. 
\be 
v w = P^*_B({\underline u},b,\mu_i)\ , 
\label{loc}
\ee
by the \'etale map 
\be 
u_i=(\hat u_i)^\frac{1}{6}.
\label{etale}
\ee  
For example the discriminants of $W$ are almost completely determined 
from the discriminants of (\ref{loc}). Generically the discriminant 
is the locus in the moduli space where $P^*=0$ and $d P^*=\frac{\partial P^*}{\partial \chi_i} d \chi_i=0$ 
can both be satisfied for special values of the coordinates generically called $\chi_i$ 
at which $W$ is singular. One solution is to set $x=y=0$ and $z=1$. Then by (\ref{etale}) 
and (\ref{loc}), one component of the discriminant of $M$ is just the 
generic  discriminant of the local  mirror $P^*_B({\underline u},b,\mu_i)=0$, where $P^*_B({\underline u},b,\mu_i)=0$ develops a node where an 
$S^1$ shrinks.  We denote this component by
$\Delta_{loc}(b,\mu_i)=0$.   By (\ref{loc}), 
this corresponds on $W$ to an $S^3$ shrinking, i.e.\ 
$\Delta_{loc}(b,\mu_i)=0$ is a conifold locus in the complex moduli space of $W$. Moreover 
the other solutions of the discriminant condition with $x,y\neq 0$ are obtained 
from $\Delta_{loc}(b,\mu_i)=0$ by applying $I$. In conclusion $W$ has two conifold 
discriminants
\be 
\Delta_1=\Delta_{loc}(b,\mu_i)=0 \quad   {\rm and} \quad \Delta_2=\Delta_{loc}\left( b-\frac{a^6}{432},\mu_i\right)=0\ .
\ee
If the 2d base polyhedron has points on the edges, there are invariant discriminant 
loci $\Delta_{loc}(\mu_i)=0$, which do not depend on $\beta$ and correspond 
to divisors in $M$ contracting to curves.  In this case, the corresponding shrinking $3$-cycles 
in $W$ do not have the topology of $S^3$.       

\begin{figure}
\center
\includegraphics[width=12cm]{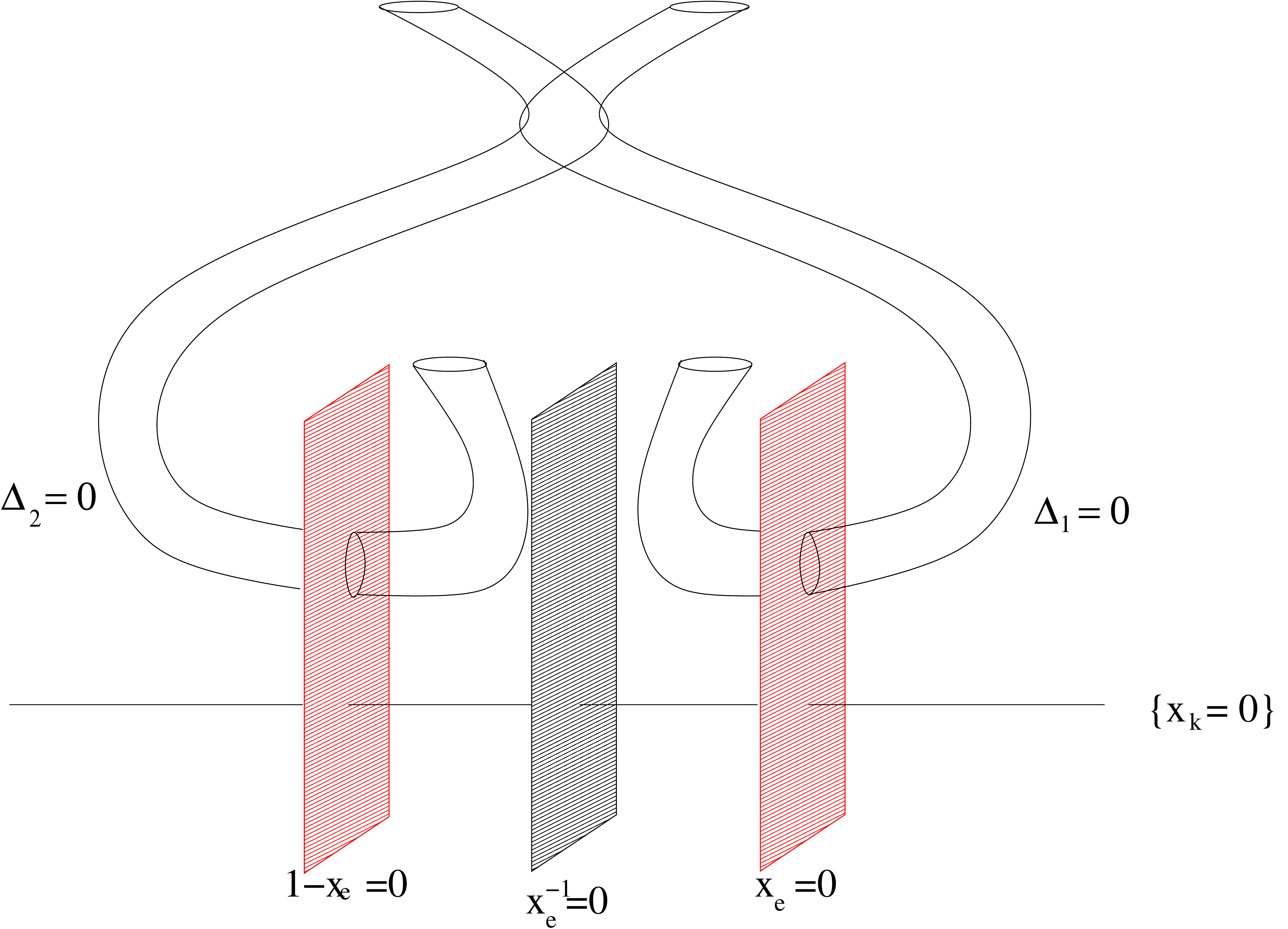}
\caption{The general moduli of the elliptic fibration. The moduli space 
of the local model, which arises for large elliptic fibre, is represented 
by the vertical red planes at $x_e=0$ and $x_e=1$. These are 
two equivalent planes exchanged by the involution symmetry. 
There are two large radius points at the intersection  of the common 
vanishing of the coordinates for the local model 
$\{ x_k=0\}$, which describe the large K\"ahler volume of the base. These are represented by the horizontal line  
with the vertical $x_e=0,1$ planes.}
\end{figure}

The A-periods $X^I$, $I=0, \ldots, h_{21}(W)$ 
are homogeneous coordinates\footnote{In the present section the  index $1$ 
labels the fibre class $e$ and the indices $k,l,m=2,\ldots h_{21}(W)$ 
the base classes for which we use latin characters  from the middle of the alphabet.} on the 
complex moduli space of $W$. Moreover they can be identified at the large 
complex structure point (characterized by its maximal unipotent 
monodromy) with the complexified K\"ahler volumes associated with the 
corresponding curves $\tilde {\cal C}^e$ and $\tilde {\cal C}^k$ on $M$.      

Corresponding to the K\"ahler parameters of $M$, we can henceforth introduce $z_e$  
and  $\tau=\frac{X^e}{X^0}=\frac{1}{2 \pi i} \log(z_e)+ {\cal O}(z)$ 
can  be identified with the complex volume of the  fibre curve as  well as  $z_k$, $k=1,\ldots, b_2(B)$  so that  
$\tilde T^k=\frac{X^k}{X^0}=\frac{1}{2 \pi i} \log(z_k)+ {\cal O}(z)$  are identified with  
the complex volumes of the base curves.

The fibration structure in (\ref{polyhedrafrombaseandfibre}) 
implies that there is in the elliptic phase a Mori vector 
\be 
l^{(e)}=(-6;3,2,1,0,\ldots, 0) \ ,
\label{ellipticmori}
\ee
which corresponds to the curve class $\tilde {\cal C}^e$. The dual complex 
modulus is $z_e=\frac{b}{a^6}$. The K\"ahler moduli of the base and its dual Mori vectors and 
therefore by (\ref{defz}) the moduli $z_k(b,m_i)$ that correspond to the base curve classes 
$\tilde {\cal C}^k$ on $M$ are  obtained from triangulations of $\Delta_B^*$. These have been 
given explicitly in ~\cite{Klemm:2012}.        
   
The periods are annihilated by a ring of differential linear 
operators called Picard-Fuchs operators. For the mirror manifold $W$ 
dual to an elliptic fibration with one section, this  ring 
contains an order two operator, which governs the mirror map of the 
fibre  elliptic curve $\tau=\frac{1}{2 \pi i} \log(z_e)+ {\cal O}(z)$  
in the large base limit ${\rm Im}(T_i)\rightarrow \infty$, ${\rm Im}(\tau)$ 
finite, or $z_i\rightarrow 0$ and $z_e$ finite. It is convenient to 
write the ring elements in terms of the logarithmic derivatives 
$\theta_\alpha=z_a\frac{d}{d z_a}$. The ring element in question
reads for a smooth Weierstrass fibration with a single section\footnote{It 
can be derived in the toric context from the Mori vectors~\cite{Klemm:2012}.}   
\be 
{\cal L}_e=\theta_e(\theta_e-\sum_{i=1}^{r-1} a^i \theta_i) - 12 z_e( 6 \theta_e +1)(6 \theta_e+5) \ .   
\label{Le}
\ee
It is an exercise in differential equations associated to modular function forms~\cite{Zagierbook} 
to see that for $z_i\rightarrow 0$ and $z_e$ finite the two solutions of (\ref{Le}) determine the 
mirror map  as $z_e|_{Q_i=0}=z_e(q)=z_e^-(q)$ with $x_e=432 z_e$,         
\be 
x^\pm_e(q)= \frac{J\pm\sqrt{J^2-J}}{2 J},
\label{zeJ}
\ee
and $J=\frac{j}{1728}$, where  $j$ is the well known modular invariant and 
\be 
X^0= E_4^\frac{1}{4}(q) + \sum_{b_1,\ldots,b_{r-1}} f_{b_1,\ldots,b_{r-1}} (q) \prod_{k=1}^{r-1}\tilde Q_k^{b_{i_k}} \ ,
\label{splittinggauge}
\ee
where the Eisenstein series $E_4$ is well known. We can  also derive 
these  equations directly from geometry starting from (\ref{Newtonpolyhedron})  and taking the large 
base limit. This is defined after reparametrizing $z\mapsto \frac{z}{b^\frac{1}{6} \prod_{i=1}^r \hat u_i}$ 
by sending $b$ in (\ref{Newtonpolyhedron}) to infinity while keeping $z_e=\frac{b}{a^6}=x_e/432$ finite
\be 
{\rm lim}_{b\rightarrow \infty} P^*\left(z,y,\frac{z}{b^\frac{1}{6} \prod_{i=1}^r \hat u_i},\hat {\underline u}\right)=y^2 + x^3 + z^6 +z_e^{-\frac{1}{6}} x y z =0\ .
\ee
The $J$ invariant of this limiting curve is calculated by bringing it into Weierstrass normal form as
\be
J=\frac{1}{1728 z_e (1 - 432 z_e )}=\frac{1}{4 x_e (1 - x_e )}\ 
\label{Jze}
\ee
and yields the solutions (\ref{zeJ}). Note that  the Galois group of (\ref{Jze}) 
exchanges  $z^+_e\leftrightarrow z_e^-$ and corresponds precisely to the 
operation of the involution symmetry $I$ on $z_e$.  In (\ref{splittinggauge}) we have normalized $\Omega$ so that $X_0(q=0)=1$ and have chosen the branch of the 4th root accordingly.  

The complete ring of Picard-Fuchs equations contains further operators which 
can be derived from the Mori cone generators of the base~\cite{Klemm:2012}. 
We give here the ones for the bases $\mathbb{P}^2$ and $\mathbb{F}_n$, $n=0,1,2$ 
whose base polyhedron is shown in Figure 1. For these cases,  the Mori vector for the elliptic fibre (\ref{ellipticmori}) is 
supplemented by the ones of the base
\be
\begin{array}{rl} 
\mathbb{P}^2 \ :& \   \quad l^{(1)}\ \ =(0;0,0,\phantom{2}-3,\phantom{-}1,\phantom{-}1,\phantom{-}1), \\[1 mm] 
\mathbb{F}_n\  : & \quad \begin{array}{rl} l^{(1)}&=(0;0,0,n-2,\phantom{-}1,\phantom{-}1,-n,\phantom{-}0),\\
                                           l^{(2)}&=(0; 0,0,\phantom{2}-2,\phantom{-}0,\phantom{-}0,\phantom{-}1,\phantom{-}1)\ .
\end{array}
\end{array}
\ee
By the standard methods~\cite{Hosono:1993qy} these Mori vectors lead to the following Picard-Fuchs operators  
\be
\begin{array}{rl} 
\mathbb{P}^2 \ :&  \ {\cal  L}_1\ \ =\theta_1^3+z_1 \prod_{k=0}^2( 3 \theta_1+ \theta_e+k) , \\[1 mm] 
\mathbb{F}_n \  :& \begin{array}{rll} {\cal L}_1&=\theta_1^2-z_1\prod_{k=0}^1 (2\theta_1 +2\theta_2-\theta_e+k), \  &  \ n=0,\\
                                            {\cal L}_1&=\theta_1^2-z_1 (\theta_1 +2\theta_2-\theta_e +1)(\theta_1 -\theta_2), \ &  \ n=1,\\
                                            {\cal L}_1&=\theta_1^2-z_1 \prod_{k=0}^1 (2 \theta_1 -\theta_2+k), \  &   \ n=2,\\
                                            {\cal L}_2&=\theta_2(\theta_2-n \theta_1)- z_2 \prod_{k=0}^1 (2\theta_1 +(2-n)\theta_2-\theta_e +k),\ & n=0,1,2\ ,   
\end{array}
\end{array}
\ee
which complete (\ref{Le}) to the ring of differential operators, which fix the periods 
up to linear combinations. Let us provide the local discriminant
\be
\begin{array}{rll} 
\mathbb{P}^2 \ :& \  \quad \Delta_{loc}  =1-\beta^3, \ \   s=1 \\ [1 mm]
\mathbb{F}_n\  : & \quad \begin{array}{rll} \Delta_{loc}&=\prod_{\pm} [(1 \pm \beta)^2+\mu], & n=0, \ \  s = 2/3, \ \ \mu=m \\
                                            \Delta_{loc}&=1 - \beta^3 + 12 \beta \mu - 9 \beta^4 \mu + 24 \beta^2 \mu^2 - 16 \mu^3,&  n=1, \ \  s = 1, \ \ \ \ \mu=3 m \\
                                            \Delta_{loc}&=\prod_{\pm} [1 \pm (2  \beta^2-\mu)][1 \pm \mu] , &   n=2, \ \  s = 4/3, \ \ \mu=2 m \ .  \\
\end{array}
\end{array}
\ee
Note that in the $z_i$ or $x_i$ variables the corresponding discriminants do not 
factorize. Similarly to e.g.\ the local model for $\mathbb{P}^2$, where one can 
either can work in $b$ plane, with three conjugated conifold monodromies at 
$b_k=e^{2 k \pi i /3}$ for $k=0,1,2$, or in the $x=1/b^3$ plane with one conifold 
at $x=1$ and an order three orbifold at $x=0$, we will describe the monodromies in 
a compactification of the $x_i$ coordinates, where this factorization does not occur.

\section{The refined theory on elliptic fibration over $\mathbb{P}^2$ } \label{refinedsec3.7}

Let us recall some basic ingredient of the mirror geometry. For details see \cite{Huang:2015sta}. The mirror geometry has two complex structure parameters $z_1, z_2$. The involution acts on the geometry by 
\begin{eqnarray} \label{involution2.1}
I: ~~~ (z_1, z_2) \rightarrow (x_1, x_2) = (\frac{1}{432}-z_1, -\frac{z_1^3 z_2}{(\frac{1}{432}-z_1)^3}).
\end{eqnarray} 
We will sometimes use the tilde symbol to denote the involution transformation, e.g. $\tilde{z_i}=x_i$. On a related note, an earlier paper \cite{Alim:2013eja} considered also the Fricke involution, which acts on quasi-modular forms of a subgroup of $SL(2, \mathbb{Z})$. There are two discriminant divisors of the complex moduli space, which are exchanged under the involution action 
\begin{eqnarray}
\Delta_1 &=& (1-432 z_1)^3 -27 z_2 (432 z_1)^3, \nonumber \\
\Delta_2 &=& 1+27 z_2. 
\end{eqnarray}
The 3-point couplings can be computed from the PF operators~\cite{Hosono:1993qy} 
\begin{eqnarray} 
\label{3point2.3}
&& C_{111}  = \frac{9}{z_1^3 \Delta_1},   ~~~~~ 
C_{112}  =C_{121}  =C_{211}  = \frac{3\Delta_3 }{z_1^2z_2  \Delta_1},   \nonumber \\ 
&& C_{122}  = C_{212}  =C_{221}  = \frac{\Delta_3^2 }{z_1 z_2^2 \Delta_1},   ~~~
C_{222}  = \frac{ 9(\Delta_3^3 + (432z_1)^3 ) }{z_2^2 \Delta_1 \Delta_2 },   
\end{eqnarray}
where for convenience we can define the factor $\Delta_3 = 1-432z_1$.

The Bershadsky-Cecotti-Ooguri-Vafa (BCOV) propagators $S, S^i, S^{ij}$ are assigned with weights $3,2,1$ respectively, and are defined by anti-holomorphic derivatives 
\begin{eqnarray}
\bar{\partial}_{\bar{i}} S^{jk} = \bar{C}_{\bar{i}}^{jk}, ~~~  \bar{\partial}_{\bar{i}} S^j = G_{\bar{i} k} S^{jk}, ~~~
\bar{\partial}_{\bar{i}}  =  G_{\bar{i} j}S^j,
 \end{eqnarray}
 where $G_{\bar{i} j} = \bar{\partial}_{\bar{i}}  \partial_j K$ is the special Kahler metric of the Calabi-Yau moduli space. It is convenient to shift the propagators \cite{Alim:2007}
 \begin{eqnarray}
S^{ij}\rightarrow S^{ij}, ~~~S^i\rightarrow S^i - S^{ij}K_j, ~~~ S\rightarrow S- S^iK_i +\frac{1}{2} S^{ij}K_iK_j. 
\end{eqnarray}

We shall compute the B-model refined topological string amplitudes $\mathcal{F}^{(n,g)}$, where the two indices $n,g$ are non-negative integers. The case without refinement corresponds to $n=0$ and we may simplify the notation as 
$\mathcal{F}^{(g)}\equiv \mathcal{F}^{(0, g)}$. We will see that the refined amplitude $\mathcal{F}^{(n,g)}$ is a polynomial of degrees $3(n+g)-3$ of the shifted propagators $S, S^i, S^{ij}$ with rational functions of $z_{1,2}$ as coefficients. Hereafter by default we refer the propagators as those after the shifts. One can integrate the defining equations of the propagators and the special geometry relation \cite{Alim:2007}, and find   
\begin{eqnarray} \label{propa2.6}
\Gamma_{ij}^k &=& \delta^{k}_{i}K_j + \delta^k_jK_i -C_{ijl}S^{kl} +s^k_{ij}, \nonumber \\
\partial_iS^{jk} &=& C_{imn} S^{mj} S^{nk} +\delta^j_i S^k +\delta^k_i S^j -s^j_{im} S^{mk} -s^k_{im} S^{mj} +h^{jk}_i,  \nonumber \\
\partial_i S^j &=& C_{imn} S^{mj} S^n +2 \delta^j_i S - s^j_{im} S^m -h_{ik} S^{kj} +h^j_i,  \nonumber \\
\partial_i S &=& \frac{1}{2} C_{imn} S^m S^n -h_{ij} S^j +h_i, \nonumber \\
\partial_i K_j &=& K_iK_j -C_{ijn} S^{mn} K_m +s^m_{ij} K_m -C_{ijk}S^k +h_{ij} .  
\end{eqnarray}
Here the holomorphic ambiguities $s^k_{ij}, h^{jk}_i, h_{ij}, h^j_i, h_i$ are some rational functions from the integration constants of the anti-holomorphic derivatives. There are some redundancies and a gauge choice is presented in \cite{Alim:2012ss}.

In our previous paper \cite{Huang:2015sta} we work out the involution transformations of the propagators. It turns out the propagators $S^{jk}$ transform as a tensor with an extra minus sign, while for the propagators $S, S^i$, there is also an additional shift which can be determined. We postulate that the refined amplitude $ \mathcal{F}^{(n,g)}$ is a section of $\mathcal{L}^{2(n+g)-2}$, where $\mathcal{L}$ is the vacuum line bundle. So under the involution symmetry, it transforms as
\begin{eqnarray}
 \mathcal{F}^{(n,g)} \rightarrow (-1)^{n+g-1}  \mathcal{F}^{(n,g)}.
\end{eqnarray}

According the mirror symmetry, the A-model refined topological string amplitude for a Calabi-Yau three-fold $M$ is related to the B-model formula on the mirror by
\begin{eqnarray}
F_{A-model} ^{(n,g)} =w_0^{2(n+g)-2} \mathcal{F}^{(n,g)} ,
\end{eqnarray}
where $w_0$ is the power series solution of the Picard-Fuchs equation, which is no longer a constant for the compact Calabi-Yau case. In the large volume limit, the A-model amplitude is a constant plus the world-sheet instanton contributions. The generating function for the instanton contributions can be written in terms of the refined Gopakumar-Vafa numbers 
\begin{eqnarray}
F &=& \sum_{n,g=0}^{\infty} (\epsilon_1+\epsilon_2)^{(2n)} (\epsilon_1\epsilon_2)^{g-1} F_{inst}^{(n,g)} (t_i) 
\\ \nonumber 
&=& \sum_{g_L,g_R=0}^{\infty} \sum_{m=1}^{\infty} \sum_{\beta\in H_2(M,\mathbb{Z})}  \frac{ (2 \sinh( \frac{m \epsilon_{-}}{4}))^{2g_L} (2 \sinh( \frac{m \epsilon_{+} }{4}))^{2g_R}}{4 m \sinh(\frac{m \epsilon_1 }{2}) \sinh(\frac{m \epsilon_2 }{2})} (-1)^{g_L+g_R} n^{g_L,g_R }_\beta  e^{m(\beta\cdot t)} , 
\end{eqnarray}
where $t_i$ are the Kahler moduli of the Calabi-Yau manifold, and $\epsilon_{\pm} =\epsilon_1\pm \epsilon_2$. From the above equation,  we see that in order to extract the refined Gopakumar-Vafa numbers $n^{g_L,g_R }_\beta$ in the integer basis, it is necessary and  sufficient to compute the refined topological string amplitudes $F_{inst}^{(n,g)}$ for all  $n\leq g_R, n+g\leq g_L+g_R$. 

Here in the integer basis, the usual Gopakumar-Vafa invariants without refinement correspond to the case of $g_R=0$. The convenient basis for curve counting considerations is the spin basis, and we denote the Gopakumar-Vafa numbers as  $N^{j_L,j_R }_\beta$, where $j_L,j_R$ are non-negative half-integers. Here to avoid confusion we denote the refined Gopakumar-Vafa numbers in the two different basis with small or capital letters n. They are related by a simple transformation according to the rule of decomposition of products of $SU(2)$ representations
\begin{eqnarray} \label{changebasis}
\sum_{j_L,j_R} N^{j_L,j_R }_\beta [j_L]\otimes [j_R] = \sum_{g_L,g_R} n^{g_L,g_R }_\beta ([\frac{1}{2}]+2 [0] )^{g_L} \otimes ([\frac{1}{2}]+2 [0] ) ^{g_R}. 
\end{eqnarray} 
The refined Gopakumar-Vafa numbers $N^{j_L,j_R }_\beta$ in the spin basis have a better geometric meaning, and are closely related to motivic invariants in mathematics. In particular, they are always non-negative and typically have far fewer non-vanishing entries than those in the integer basis.

 \subsection{The genus one amplitude} 
 
The formula for the unrefined case is known in the literature \cite{Alim:2012ss, Klemm:2012, Huang:2015sta} 
 \begin{eqnarray} \label{genusoneamplitude}
\mathcal{F}^{(0, 1)} = \frac{1}{2}(3+h^{1,1} -\frac{\chi}{12}) K + \frac{1}{2} \log\det G^{-1} -\frac{1}{12} \log (\Delta_1\Delta_2)     - \frac{1}{24}\sum_{i=1}^{h^{1,1}} s_i\log(z_i)  , 
\end{eqnarray} 
where the hodge number and Euler character are $h^{1,1}=2$ and $\chi= -540$ for our model, and the numbers $s_1=114, s_2=48$ are related to the second Chern class of the geometry. 

As a first step we  consider the simplest refined topological string amplitude $\mathcal{F}^{(1,0)}$. In the studies of refined topological strings for local Calabi-Yau models, we learn that the  $\mathcal{F}^{(1,0)}$ is holomorphic and has logarithmic cuts at conifold divisors and large volume points \cite{Huang:2010kf, Huang:2011qx, Krefl:2010}. Here for the compact models we should also expect an extra term from the K\"ahler potential, which was gauged away in the local models.  We make the following ansatz 
\begin{eqnarray}  \label{ansatz3.107}
\mathcal{F}^{(1,0)} = \frac{1}{24}[ \log(\Delta_1\Delta_2) - c_1 \log(z_1) - c_2 \log(z_2)] +c_0 K ,
\end{eqnarray}
where the constant $\frac{1}{24}$ is fixed by boundary conditions at the conifold divisors as in the local models. The constants  $c_0, c_1, c_2$  should be related to the classical topological numbers of the Calabi-Yau manifolds. For now the relations are not well understood for refined theory, so we will fix these constants by the fiber modularity constraints.  

We first consider the action of the involution symmetry (\ref{involution2.1}). Since the Kahler potential is invariant, the constant $c_0$ is a free parameter. The invariance of the remaining parts impose a constraint $3+c_1=3c_2$ on the parameters $c_1, c_2$. 
 
We can expand the refined amplitudes in the base degree, 
\begin{eqnarray}  \label{base3.108}
w_0^{2n+2g-2} \mathcal{F}^{(n,g)} = \sum_{k=0}^{\infty} P^{(n,g)}_k(q_E) ~(\frac{q_E}{\eta(q_E)^{24}})^{\frac{3k}{2}} q_B^k,
\end{eqnarray} 
where $q_B$ and $q_E$ are the exponentials of the Kahler parameters for the base and fiber, and behave like $q_B\sim z_2\sim 0, q_E\sim z_1\sim 0$ near the large volume point. Sometimes it is more convenient to make a shift $q_B\rightarrow q_B q_E^{-\frac{3}{2}}$, though here we use the original unshifted parameters. We consider the constraint from base degree zero amplitude. In this case, the genus zero and one amplitudes actually depend also on $\log(q_B)$ due to the perturbative contributions. In the previous paper \cite{Huang:2015sta}, in the calculations for the unrefined amplitudes $P^{(0,0)}_0$ and $P^{(0,1)}_0$, we find that if we replace $\log(q_B)\rightarrow -\frac{3}{2} \log(q_E)$, the amplitudes are modular
\begin{eqnarray} \label{modularbasezero2.11}
\frac{d^3}{d(\log q_E)^3} P^{(0,0)}_0\sim E_4(q_E), ~~~~ \frac{d}{d\log q_E} P^{(0,1)}_0\sim E_2(q_E) . 
\end{eqnarray}  

We should expect the same modularity constraint for the refined amplitude $\mathcal{F}^{(1,0)}$  as well, namely the base degree zero amplitude $ P^{(1,0)}_0\sim \log\eta(q_E)$ up to a constant and a term proportional to $\log(q_B) +\frac{3}{2} \log(q_E)$.    It turns out this condition can fix two of the constants $c_0, c_1, c_2$ in the ansatz  (\ref{ansatz3.107}). We find 
\begin{eqnarray}  \label{F103.110}
\mathcal{F}^{(1,0)} = \frac{1}{24}[ \log(\Delta_1\Delta_2) - (4c_0+3) \log(z_1)- (\frac{4}{3}c_0+2) \log(z_2)  ] + c_0 K. 
\end{eqnarray}
As a consistency check, it is easy to see that the formula (\ref{F103.110}) is invariant under the involution transformation, so here for genus one, the modularity of the base degree zero amplitude is stronger than the constraint of involution invariance.

To further fix the constant we need one more condition from the refined Gopakumar-Vafa numbers. We note that the case of $d_E=0$ corresponds to the local $\mathbb{P}^2$ model and the refined invariants have been computed by the refined topological vertex \cite{IKV} or the B-model method \cite{Huang:2010kf}. However, the corresponding refined invariants for $d_E=0$ extracted from the ansatz (\ref{F103.110}) are independent of the free constant $c_0$ and agree with the previous works.  

So we need to go beyond the local limit with $d_E>0$. For some low degrees, i.e. the cases of $(d_B,d_E) =(0,1), (1,1), (2,1) $, these numbers can be completely determined by motivic curve-counting calculations. For example, for the case of $(d_B,d_E) =(0,1)$, there are two non-vanishing motivic invariants in the spin basis, $N^{\frac{1}{2},1}_{(0,1)}= 1, N^{0,0}_{(0,1)}= 546$. We can change to the integer basis using the formula (\ref{changebasis}). We provide the tables of refined GV numbers in Appendix \ref{AppendixA1}. We use the datum $n^{g_L,g_R}_{(d_B,d_E)} =8 $ for $(g_L,g_R)=(0,1), (d_B,d_E)=(0,1)$ to fix the constant $c_0=-\frac{101}{8}$ in the ansatz (\ref{F103.110}). So we have the first non-trivial refined topological string amplitude 
\begin{eqnarray}  \label{F10exact}
\mathcal{F}^{(1,0)} = \frac{1}{24}[ \log(\Delta_1\Delta_2) + \frac{95}{2} \log(z_1) + \frac{89}{6} \log(z_2)  ]  -\frac{101}{8} K. 
\end{eqnarray}

We then can then compute the refined GV numbers $n^{g_L,g_R}_{(d_B,d_E)}$ for  $(g_L,g_R)=(0,1)$ and all $ (d_B,d_E)$ with the above exact formula (\ref{F10exact}). The results are listed in table \ref{GVtablegenus01} in the Appendix. The results provide non-trivial tests of the formula by comparing the numbers 524 and -10760 for 
$(d_B,d_E)= (1,1), (2,1) $ with the corresponding entries in the tables \ref{GVtabledegree11}, \ref{GVtabledegree21}. 

From the tables  \ref{GVtablegenus00},  \ref{GVtablegenus10},  \ref{GVtablegenus01}, we also see that for $d_B=0$, the refined GV numbers are the same for all fiber degrees $d_E>0$. This suggests that the results of the motivic calculations for all degrees $d_B=0, d_E>0$ are the same, which we confirm in Section~\ref{basedeg0}.  This is not a trivial result from the curve counting considerations.

\subsection{Higher genus amplitudes}

The refinement of the higher genus BCOV holomorphic anomaly equation for Seiberg-Witten theory and local Calabi-Yau was proposed in \cite{Huang:2010kf,Krefl:2010}. However, we shall see later that the equation needs a non-trivial modification in the compact Calabi-Yau case. The equation in \cite{Huang:2010kf,Krefl:2010} is
\begin{eqnarray} \label{noncompactrefine}
\bar{\partial}_{\bar{i}} \mathcal{F}^{(n,g)} = \frac{1}{2} \bar{C}_{\bar{i}}^{jk} [ D_j D_k  \mathcal{F}^{(n,g-1)} +
(\sum_{n_1=0}^n \sum_{g_1=0}^{g})^{\prime} D_j \mathcal{F}^{(n_1,g_1)}  D_k \mathcal{F}^{(n-n_1,g-g_1)} + \cdots], 
\end{eqnarray}
where the prime over the sum denotes the exclusion of the cases of $n_1=0, g_1=0$ and $n_1=n,  g_1=g$, and we have put $\cdots$ in the equation to symbolize the need to some non-trivial extra terms in the compact Calabi-Yau case.

The unrefined case of $(n,g)=(0,2)$ has been studied in previous work \cite{Huang:2015sta}. We consider first the refined case $(n,g)=(1,1)$. It turns out for this case, the refined holomorphic anomaly equation for the non-compact Calabi-Yau is still valid. 
Similar to the unrefined case, the refined holomorphic anomaly equation can be written as partial derivatives with respect to the an-holomorphic propagators. First we compute the derivative of the genus one amplitudes 
\begin{eqnarray} \label{F1i2.13}
\partial_i \mathcal{F}^{(0,1)} &=& \frac{1}{2} C_{ijk} S^{jk} - (\frac{\chi}{24}-1) K_i - \frac{1}{2} s^j_{ij} - \partial_i [ \frac{1}{12}  \log (\Delta_1\Delta_2)+ \frac{19}{4}\log z_1+ 2\log z_2 ],  \nonumber \\ 
\partial_i \mathcal{F}^{(1,0)} &=&  -\frac{101}{8} K_i + \partial_i  \frac{1}{24}[ \log(\Delta_1\Delta_2) + \frac{95}{2} \log(z_1) + \frac{89}{6} \log(z_2)  ]  
\end{eqnarray} 
where the first equation is also available in our previous paper \cite{Huang:2015sta}. The equations for partial derivative can be generalized to the refined case 
\begin{eqnarray} \label{partial2.14}
\frac{\partial \mathcal{F}^{(n,g)} }{\partial S^{ij}} & =& \frac{1}{2} \partial_i(\partial _j^{\prime} \mathcal{F} ^{(n,g-1)}  ) +\frac{1}{2} (C_{ijl }S^{lk} - s^k_{ij} )\partial_k^{\prime}  \mathcal{F} ^{(n,g-1)}   
+ \frac{1}{2}  (C_{ijk}S^k - h_{ij} ) c_{n,g-1}    \nonumber \\
&& + \frac{1}{2} (\sum_{n_1=0}^{n} \sum_{g_1=0}^{g})^{\prime} \partial_i^{\prime}  \mathcal{F}^{(n_1,g_1)}   \partial_j^{\prime}  \mathcal{F}^{(n-n_1,g- g_1)},  \nonumber \\
\frac{\partial \mathcal{F}^{(n,g)} }{\partial S^{i}} & =& (2n+2g-3) \partial_i^{\prime}  \mathcal{F}^{(n,g-1)} 
+ (\sum_{n_1=0}^{n} \sum_{g_1=0}^{g})^{\prime}  c_{n_1,g_1}   \partial_i ^{\prime} \mathcal{F}^{(n-n_1,g- g_1)}, \nonumber \\
\frac{\partial \mathcal{F}^{(n,g)} }{\partial S } & =& (2n+2g-3) c_{n,g-1} +  (\sum_{n_1=0}^{n} \sum_{g_1=0}^{g})^{\prime} c_{n_1,g_1} c_{n-n_1,g-g_1}, 
\end{eqnarray}
where the prime over summation is from the refined holomorphic anomaly equation, and the $c_{n,g}$ is defined as
\begin{eqnarray}  \label{cg3.36}
c_{n,g} &=& \left\{
\begin{array}{cl}
 \frac{\chi}{24} -1,    &   (n,g)=(0,1) ;   \\
 \frac{101}{8},    &   (n,g)=(1,0) ;   \\
 (2g+2n-2) \mathcal{F}^{(n,g)},             &   n+g>1 .   
\end{array}    
\right.
\end{eqnarray}
We have also used the notation $\partial^{\prime} $ to denote 
\begin{eqnarray}  \label{F1extra3.37}
\partial_i^{\prime}  \mathcal{F}^{(n,g)}  &=& \left\{
\begin{array}{cl}
 \partial_i \mathcal{F}^{(n,g)}  +  (\frac{\chi}{24}-1) K_i  ,    &  (n,g)=(0,1)  ;   \\
 \partial_i \mathcal{F}^{(n,g)}  +  \frac{101}{8} K_i  ,    &   (n,g)=(1,0)  ;   \\
 \partial_i \mathcal{F}^{(n,g)} ,             &   n+g>1 ,
\end{array}    
\right.
\end{eqnarray}
i.e. on the right hand in (\ref{partial2.14}), we use the formula (\ref{F1i2.13}) for the refined genus one amplitudes omitting the $K_i $ term.

As in the unrefined case, the right hand side of first equation in (\ref{partial2.14}) needs to be multiplied by an extra factor of 2 to take account of the double contribution by identifying the propagators $S^{ij}=S^{ji}$ for $i\neq j$. By induction, we can now see that the topological string amplitudes $\mathcal{F}^{(n,g)}$ are polynomial of degree $3n+3g-3$ with rational function coefficients, where one assigns degree 1,2,3 respectively to the propagators $S^{ij}, S^{i}, S$.

We integrate the equation (\ref{noncompactrefine}) without the modification and determine the amplitude $\mathcal{F}^{(1,1)}$ up to a holomorphic ambiguity. The holomorphic ambiguity has a pole at the conifold divisors $\Delta_1(z_1,z_2)$ and $\Delta_2(z_1,z_2)$. We use the gap condition as in the non-compact case \cite{Huang:2010kf,Krefl:2010} and also the regularity condition near $(z_1, z_2) \sim (\infty, 0)$ to completely fix the holomorphic ambiguity. The expression for $\mathcal{F}^{(1,1)}$ is already too long to write down here. We extract the refined BPS numbers for $(g_L, g_R)= (1,1)$ from the amplitudes, and list them in table \ref{GVtablegenus11}. We see that the numbers -4, 11, -80 match the corresponding entries in tables \ref{GVtabledegree01}, \ref{GVtabledegree11}, \ref{GVtabledegree21} for degrees $(d_B,d_E) = (0,1), (1,1), (2,1)$ from curve counting considerations. The results provide non-trivial tests of the validity of the  $\mathcal{F}^{(1,1)}$ formula, and further predict many higher degree numbers for $(g_L, g_R)= (1,1)$. 

The study of the case of $(n,g)=(2,0)$ encounters the problem of modification of the refined holomorphic anomaly equation. We find that the naive uses of the equation (\ref{noncompactrefine}) and boundary conditions produce inconsistent non-integral values of refined BPS numbers. To see this point more explicitly, we shall show that if our conjecture that the refined numbers are the same for all $d_B=0, d_E>0$ is true, then the base degree zero amplitudes for genus $n+g\geq 2$ defined by (\ref{base3.108}) can be written in terms of quasi-modular forms with certain appropriately chosen constant contributions, where the modular parameter $q$ is the exponential of the fiber Kahler parameter. In particular, using the refined numbers $N^{\frac{1}{2},1}_{(0,d_E)}= 1, N^{0,0}_{(0,d_E)}= 546$ computed in Section~\ref{basedeg0}, we can write the single cover contribution for the topological string amplitudes and expand for small $\epsilon_{1,2}$ parameters as 
\begin{eqnarray} \label{singleexpansion}
F_{single} &=& (\sum_{d_E=1}^{\infty} q^{d_E} )\frac{546 - (e^{\frac{\epsilon_1-\epsilon_2}{2}} + e^{-\frac{\epsilon_1-\epsilon_2}{2}} )
(e^{\epsilon_1+\epsilon_2}+1+e^{-\epsilon_1-\epsilon_2}) }{(e^{\frac{\epsilon_1}{2}}-e^{-\frac{\epsilon_1}{2}})(e^{\frac{\epsilon_2}{2}}-e^{-\frac{\epsilon_2}{2}}) } \nonumber \\
&=&  (\sum_{d_E=1}^{\infty} q^{d_E} ) [ \frac{540}{\epsilon_1\epsilon_2} + 48 - \frac{101}{4} \frac{(\epsilon_1 + \epsilon_2)^2}{\epsilon_1\epsilon_2}  
+\frac{9}{4}  \epsilon_1 \epsilon_2
-\frac{89}{48} (\epsilon_1 + \epsilon_2)^2    \nonumber \\ && 
+ \frac{65}{192} \frac{(\epsilon_1 + \epsilon_2)^4 }{\epsilon_1\epsilon_2}  + \mathcal{O}(\epsilon^4)  ] . 
\end{eqnarray}
To include the multi-cover contributions, we shall divide the above single cover contribution by a natural number $m$, rescale the parameters $q\rightarrow q^m, \epsilon_{1,2}\rightarrow m\epsilon_{1,2}$, and sum over  $m$. We also recall the well-known formula for Eisenstein series 
\begin{eqnarray}
E_{2k} (q) = 1 -\frac{4k}{B_{2k}} \sum_{m, d=1}^{\infty} m^{2k-1} q^{md}.  
\end{eqnarray} 
So it is easy to see that for genus zero and one, we have the modularity property as in (\ref{modularbasezero2.11}), while for the higher genus $n+g\geq 2$ cases we have 
\begin{eqnarray}
P_0^{(n,g)} \sim E_{2(n+g)-2} (q )  ,
\end{eqnarray} 
with an appropriate chosen constant contribution, which is a refinement of the well known constant map contributions in Gromov-Witten theory \cite{GV1, Marino:1998, Faber}. This is also considered in \cite{Huang:2013} in the case of a reduced geometry of M-strings \cite{Haghighat:2013}. We shall not write them explicitly until they can be confirmed by some independent calculations in Gromov-Witten theory.

Using the Bernoulli number $B_2 = \frac{1}{6}$, we can derive the exact coefficients from the expansion (\ref{singleexpansion}) as 
\begin{eqnarray} \label{basezero}
 P_0^{(0,2)} =  -\frac{3}{32} E_2(q), ~~  P_0^{(1,1)} =  \frac{89}{1152} E_2(q), ~~  P_0^{(2,0)} = -\frac{65}{4608}  E_2(q). 
\end{eqnarray}
In the next subsection \ref{subsecproofgenus01}, we shall show that the coefficients of $E_2(q)$ can be derived from the refined holomorphic anomaly equation.  The coefficients for $P_0^{(0,2)}$ and $P_0^{(1,1)}$ are the same as those derived from the naive refined holomorphic anomaly equation (\ref{noncompactrefine}) without extra modification. However, for the case of $P_0^{(2,0)}$, we would get a coefficient, namely $-\frac{7921}{497664}$, different from the 
 $-\frac{65}{4608}$ in the above equation. Therefore we are forced to the conclusion that the equation (\ref{noncompactrefine}) needs extra corrections in the compact Calabi-Yau case.

We also proceed to calculate the refined amplitude $\mathcal{F}^{(1,2)}$ at genus 3, which only needs lower genus amplitudes $\mathcal{F}^{(n,g)}$ with $n\leq 1, n+g\leq 2$ in the BCOV recursion. We are able to determine the amplitude and extract the refined GV numbers at $(g_L,g_R)=(2,1)$, listed in table \ref{GVtablegenus21}. The number -396 at $(d_B,d_E)=(3,1)$ has been tested by motivic curve counting considerations. So it is clear that the refined holomorphic anomaly equation (\ref{noncompactrefine}) needs modification only for $n\geq 2$. 

\subsection{Derivation of modular anomaly equation} \label{subsecproofgenus01}

The topological string amplitude of elliptic Calabi-Yau manifolds for a given base homology class are quasi-modular forms, i.e. homogeneous polynomials of Eisenstein series $E_2, E_4, E_6$. Here the second Eisenstein series $E_2$ is not modular, and the modular anomaly equation relating the partial derivative of $E_2$ to lower base degree and lower genus amplitudes has been proposed in \cite{Alim:2012ss, Klemm:2012}. This generalizes the earlier works on $\mathcal{N}=4$ topological gauge theory \cite{Minahan:1997}  and topological string theory on half K3 Calabi-Yau space \cite{HST}. 

Here we shall further generalize the modular anomaly equation in \cite{Alim:2012ss, Klemm:2012} to the refined theory. For half K3 case, the refined modular anomaly equation has been proposed in \cite{Huang:2013} and confirmed in \cite{Haghighat:2014} using domain wall blocks from the M-brane picture. We shall start from the refined BCOV holomorphic anomaly equation, e.g. in (\ref{noncompactrefine}), and derive the refined version of the modular anomaly equation.

First we introduce a notation and derive some general formulas. There are two independent complex structure parameters $z_1, z_2$, with the corresponding Kahler parameters $t_1, t_2$ and exponentials $q_1, q_2$ determined by the mirror maps. We will need to be careful when taking partial $E_2(q_1)$ derivative, by specifying the independent variables that are fixed. We can either keep $z_2$ fixed or $t_2$ fixed.  To avoid confusion, we use the notation of the operator $\mathcal{L}_{E_2}$ for this first case and reserve the notation $\partial_{E_2}$ only for the second case, i.e. 
\begin{eqnarray} \label{defineE23.145}
\mathcal{L}_{E_2} f := \partial_{E_2(q_1)} f(q_1, z_2), ~~~~~\partial_{E_2} f := \partial_{E_2(q_1)} f(q_1, q_2). 
\end{eqnarray} Alim:2013eja
Also in this subsection the default argument of $E_2$ is $q_1$ when omitted. The conventional modular anomaly equations in e.g. \cite{Alim:2012ss, Klemm:2012} are written using $\partial_{E_2}$ in our notation, while here we introduce the $\mathcal{L}_{E_2}$ as we shall see that it is more convenient for the derivation from BCOV holomorphic anomaly equation.

It is  obvious that 
\begin{eqnarray} \label{obvious2.23}
\mathcal{L}_{E_2} z_2 =0, ~~~\mathcal{L}_{E_2} t_1 =0, 
\end{eqnarray} 
while the non-trivial results of our previous paper \cite{Huang:2015sta} amount to 
\begin{eqnarray} \label{results3.145}
\mathcal{L}_{E_2} z_1 =0, ~~~\mathcal{L}_{E_2} w_0 =0, ~~~ \mathcal{L}_{E_2} t_2 = \frac{1}{12} \partial_{t_2} P^{(0)},
\end{eqnarray} 
where $w_0$ is the power series solution to the Picard-Fuchs equation, and $P^{(0)}\equiv P^{(0,0)}$ is the generating function of non-zero base degree instanton contributions in the prepotential. 
So for any rational or logarithmic functions $f(z_1, z_2)$ and $g(w_0)$, we also have 
\begin{eqnarray} \label{general3.147}
\mathcal{L}_{E_2} f(z_1,z_2) = 0 , ~~~ \mathcal{L}_{E_2} g(w_0) = 0.
\end{eqnarray} 
For higher genus, we also define the generating function for non-zero base degree contributions 
\begin{eqnarray}
P^{(n,g)} = \sum_{k=1}^{\infty}P^{(n,g)}_k Q^k, 
\end{eqnarray}
with $P^{(n,g)}_k$ as defined in (\ref{base3.108}) by the expansion of A-model topological string amplitude and the parameter  $Q\equiv  q_2q_1^{\frac{3}{2}}/ \eta^{36}$.

From the  chain rule for computing derivatives and the last equation in (\ref{results3.145}), we can relate these two derivatives 
\begin{eqnarray}
\mathcal{L}_{E_2} f &=&   \partial_{E_2} f + [\partial_{t_2} f(t_1,t_2) ] \mathcal{L}_{E_2}(t_2)  
\nonumber \\
&=&  \partial_{E_2} f + \frac{1}{12} (\partial_{t_2} f )  (\partial_{t_2} P^{(0)}), 
\end{eqnarray}

We will  need to compute the action of  the operator $\mathcal{L}_{E_2}$ on partial derivatives $\partial_{t_i}$. It is obvious that $\partial_{E_2}=\partial_{E_2(q_1)}$ commutes with $\partial_{t_2}$ since they are independent variables. For the $\partial_{t_1}$ case, we will need to use the following formulas in the previous papers \cite{Huang:2012, Huang:2015sta}. Suppose $G_k$ is a rational function of quasi-modular form, with modular weight $k$, then the commutation rule is 
\begin{eqnarray} \label{generalformula}
\partial_{E_2}\partial_t^n G_k &=& \partial_t^n  \partial_{E_2}G_k +\frac{n(k+n-1)}{12} \partial_t^{n-1} G_k , \nonumber \\ 
\partial_{E_2}\partial_t \log(G_k) &=& \partial_t  \partial_{E_2} \log(G_k) +\frac{k}{12}. 
\end{eqnarray}
Applying this general formula, since $z_i$ has modular weight zero and $w_0$ have modular weight one, we can compute 
\begin{eqnarray}  \label{E23.149}
\mathcal{L}_{E_2} \partial_{t_i} z_j &=&  \partial_{t_i} (\partial_{E_2} z_j) +\frac{1}{12}(\partial_{t_2} \partial_{t_i} z_j )  (\partial_{t_2} P^{(0)})  \nonumber \\  
&=& - \frac{1}{12}(\partial_{t_2} z_j )  (\partial_{t_2} \partial_{t_i}  P^{(0)}) , \\
\mathcal{L}_{E_2} \partial_{t_i} \log w_0  &=&  \partial_{t_i} (\partial_{E_2} \log w_0 ) + \frac{1 }{12} \delta_{i}^1 +\frac{1}{12}(\partial_{t_2}  \partial_{t_i}  \log w_0  )  (\partial_{t_2} P^{(0)})  \nonumber \\  
&=& \frac{1 }{12}\delta_{i}^1  - \frac{1}{12}(\partial_{t_2}  \log w_0 )  (\partial_{t_2} \partial_{t_i}  P^{(0)}) ,
\label{E23.150}
\end{eqnarray} 
where we have used $\mathcal{L}_{E_2} z_j=0 $ and $\mathcal{L}_{E_2} \log(w_0)=0 $.

The action of $\mathcal{L}_{E_2}$ on second derivative $\partial_{t_i}\partial_{t_j} z_k $ is more complicated. Since $z_k$ can be expanded as power series of $q_2q_1^{\frac{3}{2}}$ with zero modular weight coefficients, we can redefine variables 
\begin{eqnarray}
\tilde{t}_2 = t_2 +\frac{3}{2} t_1 ,~~~~ \tilde{t}_1 = t_1, 
\end{eqnarray} 
so that the derivative $\partial_{\tilde{t}_2} = \partial_{t_2}$ does not change the modular weight, while the derivative $\partial_{\tilde{t}_1} = \partial_{t_1} -\frac{3}{2} \partial_{t_2}$ increases the modular weight by 2. Applying the general formula (\ref{generalformula}) we find 
\begin{eqnarray}
\partial_{E_2} \partial_{\tilde{t}_i}\partial_{\tilde{t}_j} z_k =\partial_{\tilde{t}_i}\partial_{\tilde{t}_j} \partial_{E_2}  z_k
+ \frac{1}{6}\delta^1_i \delta^1_j  \partial_{\tilde{t}_1} z_k. 
\end{eqnarray} 
Therefore the $\mathcal{L}_{E_2}$ action is computed as 
\begin{eqnarray} \label{E2second3.154}
\mathcal{L}_{E_2}( \partial_{t_i}\partial_{t_j} z_k ) &=& \frac{1}{6}\delta^1_i \delta^1_j  (\partial_{t_1} -\frac{3}{2} \partial_{t_2}) z_k  
-\frac{1}{12} (\partial_{t_2} z_k )  (\partial_{t_2} \partial_{t_i}  \partial_{t_j}  P^{(0)})  
\nonumber \\ &&
-\frac{1}{12} (\partial_{t_2} \partial_{t_i} z_k )  (\partial_{t_2} \partial_{t_j}  P^{(0)})  
-\frac{1}{12} (\partial_{t_2} \partial_{t_j} z_k )  (\partial_{t_2} \partial_{t_i}  P^{(0)})
\end{eqnarray}

Now we consider first the genus zero modular anomaly equation. For convenience we defined a variable $\tilde{q}$ by 
\begin{eqnarray}
J(\tilde{q}) = \frac{1}{z_1(1-432z_1)},
\end{eqnarray}
which behaves like $\tilde{q} \sim z_1\sim 0$ in the small $z_1$ limit, but it is different from the physical Kahler parameter $q_1$ which depends on both $z_1$ and $z_2$. A rational function $G_k(q_1)$ of quasi-modular form of weight $k$ can be expanded in terms of power series of quasi-modular forms of $\tilde{q}$. Some useful formulas from the previous paper \cite{Huang:2015sta} are 
\begin{eqnarray} \label{qtilde2.35}
\partial_{E_2(\tilde{q})} G_k(q_1) = \partial_{E_2(q_1)} G_k(q_1) + \frac{k}{12} (t_1-\tilde{t}) G_k(q_1), 
\end{eqnarray}
where $t_1=\log(q_1)$ and $\tilde{t}=\log(\tilde{q})$ are the corresponding flat coordinates. In the previous paper \cite{Huang:2015sta}, we find the formula for $\partial_{t_2} P^{(0)}(\tilde{q}, z_2)$ in terms of the variables $\tilde{q}, z_2$ by solving the Picard-Fuchs system of differential equations. For more details see the previous paper. The upshot is that  the coefficient of $z_2^n$ in the series expansion of $\partial_{t_2} P^{(0)}(\tilde{q}, z_2)$ has modular weight $-2$, and its derivative with respect to $E_2(\tilde{q})$ can be explicitly computed. After some calculations, we find that 
\begin{eqnarray}
\partial_{E_2(\tilde{q})} \partial_{t_2} P^{(0)} (\tilde{q}, z_2) 
= - \frac{1}{6} (t_1-\tilde{t})   \partial_{t_2} P^{(0)} . 
\end{eqnarray}
Using the formula (\ref{qtilde2.35}) relating the derivatives of $E_2(q_1)$ and $E_2(\tilde{q})$, we see actually 
\begin{eqnarray} \label{genuszero3.134}
\mathcal{L}_{E_2(q_1)} \partial_{t_2} P^{(0)} = 0 ,
\end{eqnarray} 
or equivalently 
\begin{eqnarray} 
\partial_{E_2} \partial_{t_2} P^{(0)} = - \frac{1}{12}(\partial_{t_2} P^{(0)}) (\partial_{t_2} ^2 P^{(0)} ) . 
\end{eqnarray} 
Since we do not include the zero base degree contributions, this equation can be integrated and we arrive at the genus zero modular anomaly equation 
\begin{eqnarray}
\partial_{E_2} P^{(0)} = - \frac{1}{24}(\partial_{t_2} P^{(0)})^2.  
\end{eqnarray}

It is also straightforward to show that the genus zero amplitude $P^{(0)}_{k}$ for $k>0$ is a quasi-modular form, i.e. without negative or fractional powers, using relevant expansion formulas in the previous paper \cite{Huang:2015sta}. 

For genus $g\geq 1$, we can rewrite the modular anomaly equation in \cite{Alim:2012ss, Klemm:2012} for the unrefined case as 
\begin{eqnarray} \label{modular3.154}
\mathcal{L}_{E_2} P^{(0,g)} = -\frac{1}{24} \sum_{n=1}^{g-1} (\partial_{t_2} P^{(0,n)} ) (\partial_{t_2} P^{(0,g-n)} ) 
+ \frac{1 }{8} \partial_{t_2} P^{(0,g-1)} -  \frac{1}{24} \partial_{t_2}^2  P^{(0,g-1)} , 
\end{eqnarray} 
where we have moved the genus zero and genus $g$ amplitudes to the left hand side, and use the 
$\mathcal{L}_{E_2} $ notation. We should reproduce this equation from BCOV holomorphic anomaly equation and further generalize to the refined case. 

Now we consider the genus one modular anomaly equation. The formulas are available in the previous subsection in equations (\ref{genusoneamplitude}, \ref{F10exact}). From the discussions of in the beginning of this subsection, it is clear that only the determinant of the special Kahler metric has possible non-vanishing contribution under $\mathcal{L}_{E_2}$ action. so we have 
\begin{eqnarray} \label{E2F13.158}
&&  \mathcal{L}_{E_2} \mathcal{F}^{(1,0)} = 0 ,  \\ \nonumber 
&&  \mathcal{L}_{E_2} \mathcal{F}^{(0,1)} = \frac{1}{2}\mathcal{L}_{E_2} \log\det(G^{-1})  
 = \frac{1}{2}\mathcal{L}_{E_2} \log [ (\partial_{t_1} z_1)( \partial_{t_2} z_2) - (\partial_{t_2} z_1)( \partial_{t_1} z_2)]. 
 \end{eqnarray} 
We can compute the derivatives using the formulas (\ref{E23.149}).  We find 
\begin{eqnarray}
 \mathcal{L}_{E_2} (\frac{\partial z_1}{\partial t_1} \frac{\partial z_2}{\partial t_2}) &=&
 -\frac{1}{12}
(\partial_{t_2} z_2) [(\partial_{t_1} z_1) (\partial_{t_2}^2 P^{(0)} ) + (\partial_{t_2} z_1) (\partial_{t_1}\partial_{t_2} P^{(0)} )], \nonumber \\ 
\mathcal{L}_{E_2} (\frac{\partial z_2}{\partial t_1} \frac{\partial z_1}{\partial t_2}) &=&
 -\frac{1}{12}
(\partial_{t_2} z_1) [(\partial_{t_1} z_2) (\partial_{t_2}^2 P^{(0)} ) + (\partial_{t_2} z_2) (\partial_{t_1}\partial_{t_2} P^{(0)} )]. 
\end{eqnarray}
Subtracting the two equations and using equation (\ref{E2F13.158}) we have 
\begin{eqnarray}
\mathcal{L}_{E_2}   \mathcal{F}^{(0,1)}  =     -  \frac{1}{24} \partial_{t_2}^2  P^{(0)}. 
\end{eqnarray}
It is convenient to separate the base degree zero contribution, which is the usual convention for modular anomaly equations in the previous literature \cite{HST, Alim:2012ss, Klemm:2012}. Around the large volume point, it is easy to calculate for example $\det(G^{-1})\sim q_1q_2\sim z_1z_2$. So we have 
\begin{eqnarray} \label{genusone2.51}
 \mathcal{F}^{(1,0)} &=& \frac{89}{144} (t_2+\frac{3}{2}t_1)  + \frac{101}{4} \log\eta(q_1)+ P^{(1,0)},  \nonumber \\
 ~~~  \mathcal{F}^{(0,1)} &=& -\frac{3}{2}  (t_2+\frac{3}{2}t_1) -48 \log\eta(q_1)+ P^{(0,1)}. 
\end{eqnarray}
According to (\ref{obvious2.23}, \ref{results3.145}), we can write the modular anomaly equation for the positive base degree contributions as
\begin{eqnarray} 
\mathcal{L}_{E_2}  P^{(1,0)} =  -\frac{89}{1728}  \partial_{t_2} P^{(0)} , ~~~
\mathcal{L}_{E_2}  P^{(0,1)} =  \frac{1}{8}  \partial_{t_2} P^{(0)} -  \frac{1}{24} \partial_{t_2}^2  P^{(0)}. 
\end{eqnarray} 
The equation for $P^{(0,1)}$ agrees with that of \cite{Alim:2012ss, Klemm:2012}, while the equation for refined case $P^{(1,0)}$ is somewhat different.

We should also show that the coefficients $P^{(0,1)}_k(q_1)$ and  $P^{(1,0)}_k(q_1)$  for $k\geq 1$ are quasi-modular forms.  We can rearrange the genus one amplitude a little 
\begin{eqnarray}
\mathcal{F}^{(0,1)} &=& -\frac{1}{2}(3+h^{1,1} -\frac{\chi}{12})\log(w_0)  + \frac{1}{2} \log[(1-432z_1)^3\det G^{-1}]  -\frac{1}{12} \log (\frac{\Delta_1\Delta_2}{1-432z_1)^3}) \nonumber \\ && - [ \frac{19}{4} \log (z_1( 1-432z_1)) +2 \log(\frac{z_2}{(1-432z_1)^3} )] .
\end{eqnarray} 
According to the results in our previous paper \cite{Huang:2015sta}, except for the term $\frac{1}{2} \log[(1-432z_1)^3\det G^{-1}] $, all other terms can be expanded as power series of $Q\equiv  q_2q_1^{\frac{3}{2}}/ \eta^{36}$, the coefficients are quasi-modular forms. Furthermore, we can also  expand 
\begin{eqnarray} \label{expandz13.166}
 z_1 &=& \frac{1}{864E_4(q_1) ^{\frac{3}{2}}} \big[ E_4(q_1)  ^{\frac{3}{2}} -E_6(q_1)  +  \sum_{n=1}^{\infty} \frac{f_{18n-6} (q_1)} {\eta(q_1) ^{36n-24} } (q_2  q_1^{\frac{3}{2}})^n \big], 
\nonumber \\ 
 \frac{z_2}{(1-432z_1)^3}   &=& E_4(q_1) ^{\frac{9}{2}}\big[  \sum_{n=1}^{\infty} \frac{g_{18n} (q_1)} {\eta(q_1) ^{36n} } (q_2  q_1^{\frac{3}{2}})^n \big],
\end{eqnarray}
where $f_{18n-6}$ and $g_{18n}$ are quasi-modular forms of weight $18n-6$ and $18n$. Using the expansions we can  compute 
\begin{eqnarray}
&&(1-432z_1)^3 \det(G^{-1})  \\
&=& (\partial_{t_1} z_1)( \partial_{t_2} \frac{z_2}{(1-432z_1)^3}) - (\partial_{t_2} z_1)( \partial_{t_1} \frac{z_2}{(1-432z_1)^3} ) \nonumber \\ 
&=& E_4^2 \eta^{24} \{ \big[ 1+ \frac{E_4 ^{\frac{5}{2}}}{\eta^{24}} \sum_{n=1}^{\infty} \partial_{t_1} (\frac{f_{18n-6} (q_1)} {E_4 ^{\frac{3}{2}} \eta ^{36n-24} } )(q_2  q_1^{\frac{3}{2}})^n
\big] \big[  \sum_{n=1}^{\infty} \frac{g_{18n} (q_1)} {\eta ^{36n} } n (q_2  q_1^{\frac{3}{2}})^n \big]
\nonumber \\ && 
- \big[   \sum_{n=1}^{\infty} \frac{f_{18n-6} (q_1)} {864 \eta ^{36n} } n (q_2  q_1^{\frac{3}{2}})^n \big] \big[  E_4 ^{-\frac{7}{2}}  \sum_{n=1}^{\infty}\partial_{t_1}( \frac{E_4 ^{\frac{9}{2}} g_{18n} (q_1)} {\eta ^{36n} }) (q_2  q_1^{\frac{3}{2}})^n \big] 
\}. 
\end{eqnarray} 
It is now straightforward to check each term of the above expression, and verify its logarithm can be expanded as power series of $Q$ with quasi-modular forms of weight $18n$ as coefficients of $Q^n$ for $n\geq 1$, i.e. without negative or fractional powers. The arguments for the case of refined amplitude $\mathcal{F}^{(1,0)}$ are similar. 

Next we consider the higher genus $n+g\geq 2$ case, and derive the refined modular anomaly equation for $n+g\geq 2$ from the BCOV holomorphic anomaly equation. Due to the general formula in (\ref{general3.147}), the holomorphic ambiguity vanishes under the $\mathcal{L}_{E_2}$ action, so does not affect the modular anomaly equation. Since the refined topological string amplitude $\mathcal{F}^{(n,g)}$ is a polynomial of the propagators with rational function coefficients, we can compute the $\mathcal{L}_{E_2}$ action
\begin{eqnarray} \label{E2action3.169}
\mathcal{L}_{E_2} \mathcal{F}^{(n,g)} = \frac{\partial \mathcal{F}^{(n,g)}} {\partial S^{ij}} (\mathcal{L}_{E_2} S^{ij}) 
+\frac{\partial \mathcal{F}^{(n,g)}} {\partial S^{i}} (\mathcal{L}_{E_2} S^{i}) +
\frac{\partial \mathcal{F}^{(n,g)}} {\partial S} (\mathcal{L}_{E_2} S) . 
\end{eqnarray} 
  
It is clear that we need to compute the action of $\mathcal{L}_{E_2}$ on the propagators. It turns out there are simple formulas 
\begin{eqnarray}
\mathcal{L}_{E_2} S^{ij} &=&  -\frac{1}{12w_0^2} (\partial_{t_2} z_i)  (\partial_{t_2} z_j),  \label{E2Sij}  \\
\mathcal{L}_{E_2} S^{i} &=&  -\frac{1}{12w_0^2} (\partial_{t_2} z_i)  (\partial_{t_2} \log w_0),  \label{E2Si}  \\
\mathcal{L}_{E_2} S &=&  -\frac{1}{24w_0^2}  (\partial_{t_2} \log w_0)^2 .  \label{E2S} 
\end{eqnarray} 
Here the appearance of the base flat coordinate  $t_2$ derivatives in these formulas signals that only $t_2$ derivative should appear in the modular anomaly equation.

We should provide some details on the derivations of the above formulas (\ref{E2Sij}, \ref{E2Si}, \ref{E2S}). The propagators are determined by the equations (\ref{propa2.6}), with the Kahler potential and the Christoffel connection in the holomorphic limit  given by
\begin{eqnarray} \label{holoformulas3.173}
K=-\log(w_0),  ~~~~  \Gamma_{ij}^k = (\partial_{z_i} \partial_{z_j } t_m )(\partial_{t_m} z_k) = - ( \partial_{z_i} t_m) ( \partial_{z_j} t_n)  (\partial_{t_m} \partial_{t_n} z_k) .
\end{eqnarray} 
For the first formula (\ref{E2Sij}), we multiply both sides of the second equation in (\ref{propa2.6}) by $( \partial_{t_m} z_i) ( \partial_{t_n} z_j)$, sum over $i,j$ and act with the $\mathcal{L}_{E_2}$ operator. Utilizing the formulas (\ref{E23.149}, \ref{E23.150}, \ref{E2second3.154}), we find 
\begin{eqnarray} \label{firstline3.174}
( \partial_{t_m} z_i) ( \partial_{t_n} z_j) C_{ijl} (\mathcal{L}_{E_2} S^{kl}) &=& - \frac{1}{12} ( \partial_{t_m} z_k)\delta_n^1 - \frac{1}{12} ( \partial_{t_n} z_k)\delta_m^1 + \frac{1}{6} \delta_m^1 \delta_n^1 (\partial_{t_1} -\frac{3}{2}\partial_{t_2})z_k  \nonumber \\
&&   -\frac{1}{12} (\partial_{t_2} z_k )  (\partial_{t_2} \partial_{t_m}  \partial_{t_n}  P^{(0)}) 
\\ &=& -\frac{1}{12}  (\partial_{t_2} z_k ) [\partial_{t_2} \partial_{t_m}  \partial_{t_n}  (\mathcal{F}^{(0)}_{classical}+P^{(0)})] ,
\end{eqnarray} 
where we see that  the first line in (\ref{firstline3.174}) provides exactly  the derivative of  the classical contributions to the prepotential from intersection numbers, given in e.g. \cite{Huang:2015sta}. This equation overdetermines $\mathcal{L}_{E_2} S^{kl}$. We can now easily check the formula (\ref{E2Sij}) satisfies the equation, so it is the correct formula. 

Similarly, we can multiply both sides of the third and fourth equations in (\ref{propa2.6}) by $\partial_{t_m} z_i$, and act with the $\mathcal{L}_{E_2}$ operator. After some calculations we confirm the formulas (\ref{E2Si}, \ref{E2S}).  

From the first and fifth equations in (\ref{propa2.6}), we can also obtain 
\begin{eqnarray}
\partial_i K_j -K_k \Gamma_{ij}^k +K_iK_j = -C_{ijk} S^k +h_{ij},
\end{eqnarray}
and multiply both sides by $( \partial_{t_m} z_i) ( \partial_{t_n} z_j)$ we get 
\begin{eqnarray} \label{Ceq3.177}
( \partial_{t_m} z_i) ( \partial_{t_n} z_j) (C_{ijk} S^k - h_{ij}) = -\partial_{t_m}\partial_{t_n} K - (\partial_{t_m} K)  (\partial_{t_n} K).
\end{eqnarray} 

Next we should compute the propagator derivatives of $\mathcal{F}^{(n,g)}$ in equations (\ref{partial2.14}) in terms of the flat coordinator derivatives. For simplicity first we consider the case of $n+g> 2$. We find 
\begin{eqnarray} \label{Fderivative3.178}
\frac{\partial \mathcal{F}^{(n,g)}} {\partial S^{ij}} &=& \frac{1}{2} (\partial_{z_i} t_m) (\partial_{z_j} t_n) (\partial_{t_m}\partial_{t_n} \mathcal{F}^{(n,g-1)} ) +\frac{\chi}{48} (\partial_{z_i} K ) (\partial_{z_j}   \mathcal{F}^{(n,g-1)} )
 +\frac{\chi}{48} (\partial_{z_j} K ) (\partial_{z_i}   \mathcal{F}^{(n,g-1)} ) \nonumber \\
 && +\frac{1}{2} (C_{ijk}S^k - h_{ij}) (2n+2g-4)\mathcal{F}^{(n,g-1)}  +\frac{101}{16} (\partial_{z_i} K ) (\partial_{z_j}   \mathcal{F}^{(n-1,g)} )  \nonumber \\ &&
 +\frac{101}{16} (\partial_{z_j} K ) (\partial_{z_i}   \mathcal{F}^{(n-1,g)} )   
 +\frac{1}{2}  (\sum_{n_1=0}^{n} \sum_{g_1=0}^{g})^{\prime}  (\partial_{z_i}   \mathcal{F}^{(n_1, g_1 )} ) (\partial_{z_j}   \mathcal{F}^{(n-n_1,g-g_1)} ) ,
\end{eqnarray}
where we have used the equations in (\ref{propa2.6}) and take into account the extra genus one contributions in the derivatives from (\ref{F1extra3.37}). 

Similarly we can rewrite the other derivatives in convenient forms (for $n+g>2$)
\begin{eqnarray} \label{Fderivative3.179}
\frac{\partial \mathcal{F}^{(n,g)} }{\partial S^{i}} & =& (2n+2g-4+\frac{\chi}{24}) \partial_{z_i}  \mathcal{F}^{(n,g-1)} 
 + (\frac{\chi}{24}-1) (\partial_{z_i} K) (2n+2g-4)\mathcal{F}^{(n,g-1)}   
 \nonumber \\ && +\frac{101}{8} \partial_{z_i}  \mathcal{F}^{(n-1,g)}  
 + \frac{101}{8}  (\partial_{z_i} K) (2n+2g-4)\mathcal{F}^{(n-1,g)}   
 \nonumber \\ &&  + (\sum_{n_1=0}^{n} \sum_{g_1=0}^{g})^{\prime} (2n_1+2g_1-2) \mathcal{F}^{(n_1,g_1)}   \partial_{z_i} \mathcal{F}^{(n-n_1,g-g_1)} ,   \\
\frac{\partial \mathcal{F}^{(n,g)} }{\partial S } & =& (\sum_{n_1=0}^{n} \sum_{g_1=0}^{g})^{\prime}   (2n_1+2g_1-2)[2(n-n_1)+2(g-g_1)-2] \mathcal{F}^{(n_1,g_1)}  \mathcal{F}^{(n-n_1,g-g_1)}  \nonumber \\ &&
+ (2n+2g-5+\frac{\chi}{12}) (2n+2g-4)\mathcal{F}^{(n,g-1)}   + \frac{101}{4}  (2n+2g-4)\mathcal{F}^{(n-1,g)}  ,  \nonumber 
\end{eqnarray}

Now, using the equations (\ref{E2Sij}, \ref{E2Si}, \ref{E2S}), (\ref{Ceq3.177}), and (\ref{Fderivative3.178}, \ref{Fderivative3.179}), we can straightforwardly compute the $\mathcal{L}_{E_2}$ action in (\ref{E2action3.169}). We find that the terms with Euler number $\chi$ and the number $101$ in the derivatives (\ref{Fderivative3.178}, \ref{Fderivative3.179}) cancel out and do not appear in the modular anomaly equation. We find the following refined modular anomaly equation 
\begin{eqnarray}  \label{refinedmodular}
 w_0^{2n+2g-2} \mathcal{L}_{E_2} \mathcal{F}^{(n,g)} 
&=& -\frac{1}{24} (\sum_{n_1=0}^{n} \sum_{g_1=0}^{g})^{\prime}  [\partial_{t_2} w_0^{2n_1+2g_1-2} \mathcal{F}^{(n_1,g_1)} ][ \partial_{t_2} w_0^{2n+2g-2n_1-2g_1-2} \mathcal{F}^{(n-n_1,g-g_1)}  ]
 \nonumber  \\ &&  -  \frac{1}{24} \partial_{t_2}^2  w_0^{2n+2g-4} \mathcal{F}^{(n,g-1)}  . 
\end{eqnarray}

We should also check the special case $n+g=2$. We consider only the extra contributions from (\ref{cg3.36}) and  (\ref{F1extra3.37}) which is independent of the genus one amplitudes. For the case of $ \mathcal{F}^{(0,2)}$ we find 
\begin{eqnarray}
\frac{\partial \mathcal{F}^{(0,2)}} {\partial S^{ij}} &=& \cdots +\frac{1}{2} (\frac{\chi}{24}-1) [\partial_i K_j +(C_{ijl}S^{kl} -s^k_{ij} )K_k + C_{ijk}S^k-h_{ij}] +\frac{1}{2} (\frac{\chi}{24}-1)^2 K_i K_j \nonumber \\
&=&  \cdots +\frac{1}{2} K_i K_j [(\frac{\chi}{24}-1) + (\frac{\chi}{24}-1)^2 ] ,  \nonumber \\
\frac{\partial \mathcal{F}^{(0,2)}} {\partial S^{i}} &=&  \cdots + K_i [(\frac{\chi}{24}-1) + (\frac{\chi}{24}-1)^2 ] , 
\nonumber \\
\frac{\partial \mathcal{F}^{(0,2)}} {\partial S } &=&  \cdots + (\frac{\chi}{24}-1) + (\frac{\chi}{24}-1)^2  , 
\end{eqnarray}
where the $\cdots$'s denote $\mathcal{F}^{(0,1)}$ dependent parts already calculated in (\ref{Fderivative3.178}, \ref{Fderivative3.179}). In the calculations of the first equation, we have use the fifth equation in (\ref{propa2.6}). It is now easy to check that these extra contributions cancel out in the total contribution to the modular anomaly in (\ref{E2action3.169}), so do not affect the derivation of the modular anomaly equation. 

The cases of $ \mathcal{F}^{(1,1)}$ and $ \mathcal{F}^{(2,0)}$ are similar. We straightforwardly check that the extra contributions cancel out, therefore the equation (\ref{refinedmodular}) is valid for $n+g\geq 2$. As before, we separate the A-model amplitude into zero and positive base degrees as $ w_0^{2n+2g-2} \mathcal{F}^{(n,g)} = P^{(n,g)}_0+P^{(n,g)}$. For  $n+g\geq 2$,  the dependence of base Kahler modulus only appears as world-sheet instanton contributions in the form of positive integer powers of $e^{t_2}$, so the degree zero amplitude $P^{(n,g)}$ is independent of $t_2$. While for the case of  $n+g=1 $, there is a linear term of $t_2$ as in (\ref{genusone2.51}) due to classical contributions. Therefore from (\ref{refinedmodular}), we find the modular anomaly of the base degree zero amplitude 
\begin{eqnarray} 
&&\partial_{E_2} P^{(0,2)}_0= -\frac{3}{32}, ~~ \partial_{E_2} P^{(1,1)}_0 = \frac{89}{1152},
 ~~ \partial_{E_2} P^{(2,0)}_0 = -\frac{1}{24}(\frac{89}{144})^2 = -\frac{7921}{497664},  \nonumber \\ &&
\partial_{E_2} P^{(n,g)}_0 =0 , ~~~~\textrm{for} ~~n+g>2. 
\end{eqnarray} 
We see that the genus two anomaly agree with the results (\ref{basezero}) from curve counting considerations, except for the case of $P^{(2,0)}_0$, suggesting that the refined anomaly equation (\ref{noncompactrefine}) needs some correction. 

For higher base degree amplitudes, we obtain the modular anomaly equation 
\begin{eqnarray} 
  \mathcal{L}_{E_2} P^{(n,g)} 
&=& -\frac{1}{24} (\sum_{n_1=0}^{n} \sum_{g_1=0}^{g})^{\prime}  [\partial_{t_2} P^{(n_1,g_1)} ][ \partial_{t_2} P^{(n-n_1,g-g_1)}  ] -  \frac{89}{1728} \partial_{t_2} P^{(n-1,g)}
 \nonumber  \\ &&  + \frac{1}{8} \partial_{t_2} P^{(n,g-1)} -  \frac{1}{24} \partial_{t_2}^2 P^{(n,g-1)}  . 
\end{eqnarray} 
In the component basis, the equation for degree $k>0$ can be written as 
\begin{eqnarray}
\partial_{E_2} P^{(n,g)}_k 
&=& -\frac{1}{24} \sum_{s=1}^{k-1} \sum_{n_1=0}^{n} \sum_{g_1=0}^{g}  s(k-s) P^{(n_1,g_1)}_s P^{(n-n_1,g-g_1)}_{k-s} -  \frac{89k }{1728} P^{(n-1,g)}_k
 \nonumber  \\ &&  + \frac{k(3-k)}{24} P^{(n,g-1)}_k , 
\end{eqnarray} 
where we include the genus zero contribution in the sum and change to the $\partial_{E_2}$ notation. This equation refines the anomaly equation in \cite{Alim:2012ss, Klemm:2012} with the extra term from $P^{(n-1,g)}$.

In the above derivation, we work entirely in the holomorphic limit, but the formulas are not exactly modular invariant due to the $E_2$ dependence. One may wonder we can alternatively keep the anti-holomorphic dependence, so that the formulas are modular invariant. In fact, the BCOV holomorphic anomaly equation is entirely  covariant and does not depend on the coordinate system, so naively it seems puzzling why  there is a difference between $\mathcal{L}_{E_2}$ and $\partial_{E_2}$ operators in (\ref{defineE23.145}).  Of course this is because the $E_2$ Eisenstein series also appears in the base flat coordinate $t_2$. From the derivation we see that it is  the  $\mathcal{L}_{E_2}$ operator that really appears in the modular anomaly equation. So we can ask how the anti-holomorphic dependence appears in the topological string amplitudes in the coordinate system $t_1, z_2$.  

To understand the issue, we should recall the simpler case of Seiberg-Witten theory, which is essentially a local topological string model. In this case one can compute the Christoffel connection $\Gamma_{uu}^u$ in Coulomb modulus coordinate $u$, keeping the anti-holomorphic dependence. One find that indeed the anti-holomorphic dependence in the connection $\Gamma_{uu}^u$ comes entirely from the shifted Eisenstein series $\hat{E}_2=  E_2 -\frac{6i}{\pi (\tau-\bar{\tau})} $  \cite{HK:2009}. We should note that $E_2$ also appears in the formula for flat coordinate $a$, but we should regard the flat coordinate as holomorphic and the $E_2$ here is not shifted. So the appropriate propagator for constructing the higher genus topological amplitudes, i.e. the gravitational couplings in the Seiberg-Witten theory is holomorphic except for the shifted $\hat{E}_2$. In this case the anti-holomorphic derivative is simply related to the $E_2$  derivative.  

For the compact Calabi-Yau model, the situation is much more complicated. Here the Kahler potential  is 
\begin{eqnarray} \label{Kahler02032013}
K= -\log(-i \Pi^{\dagger} \Sigma \Pi), 
\end{eqnarray} 
where $\Pi$ is the period vector and $\Sigma$ is the matrix, 
\begin{eqnarray}
 \Pi = w_0 \begin{pmatrix}
  1  \\
  t_i \\
  \partial_{t_i} \mathcal{F}^{(0)}  \\
  2\mathcal{F}^{(0)} - t_k \partial_{t_k} \mathcal{F}^{(0)} 
 \end{pmatrix},
 ~~~~~
  \Sigma = \begin{pmatrix}
  0 & 0 & 0 & 1 \\
  0 & 0 & I & 0 \\
  0  & -I & 0 &0 \\
  -1 & 0 & 0 & 0 
 \end{pmatrix}. 
\end{eqnarray} 
The Kahler metric is $G_{i\bar{j}} = \partial_{z_i} \bar{\partial}_{\bar{z}_j} K$, and the connection is computed accordingly with the well known formula $\Gamma_{z_iz_j}^{z_k} = G^{k \bar{l}} \partial_j G_{i\bar{l}} $. In the holomorphic limit we find the well known formulas (\ref{holoformulas3.173}). However it is difficult to see how the anti-holomorphic dependence could all fit into the shifted $\hat{E}_2$ Eisenstein series in the connections and the propagators $S^{ij}, S^i, S$. It would be interesting to understand the relation between anti-holomorphic derivative and the $E_2$  derivative in this case.

Finally, we discuss the quasi-modularity of  the amplitude $P^{(n,g)}_k(q_1)$ for higher genus $n+g\geq 2$. For a generic holomorphic ambiguity at genus $(n,g)$, the amplitude $P^{(n,g)}_k(q_1)$ could be a rational function of quasi-modular forms and may also contains half integer powers of  $E_4(q_1)$. We shall show that  the amplitudes $P^{(n,g)}_k$ are indeed  quasi-modular forms for  the space of holomorphic ambiguities such that the total topological string amplitudes  is involution symmetric, i.e. $\tilde{\mathcal{F}}^{(n,g)} = (-1)^{n+g-1} \mathcal{F}^{(n,g)}$, and $w_0^{2n+2g-2} \mathcal{F}^{(n,g)}$ is regular near the small fiber limit $z_1\sim \infty$. Our results in the previous paper \cite{Huang:2015sta},  say that if the amplitudes $P^{(n,g)}_k$ are quasi-modular for any one of the holomorphic ambiguities at genus $(n,g)$, then they would be so for and only for such homographic ambiguities.   

We have shown that the quasi-modularity of $P^{(n,g)}_k(q_1)$ for the case of genus $n+g=1$ explicitly, but the arguments here will work for the case of $n+g\geq 1$ as well. We note that in terms of the coordinates $t_1$ and $Z= \frac{z_2}{(1-432z_1)^3 E_4^{\frac{9}{2}}}$, the involution transformation  is equivalent to switching the sign of $E_4^{\frac{1}{2}} $ and keeping $E_2$, $E_6$ and $Z$ invariant
 \begin{eqnarray} \label{invo3.183}
E_4^{\frac{1}{2}} \rightarrow - E_4^{\frac{1}{2}}, ~~~E_2\rightarrow E_2,
~~~ E_6\rightarrow E_6, ~~ Z\rightarrow Z . 
\end{eqnarray} 
This can be easily checked using the expansion formulas of $1-864z_1$ in the previous paper \cite{Huang:2015sta}. The parameter $Q= q_2q_1^{\frac{3}{2}}/\eta(q_1)^{24}$ is also invariant under the involution transformation according to the expansion formula of $Z$.

The above transformation (\ref{invo3.183}) should be thought of as a procedure to realize the involution transformation for rational functions of $z_1, z_2$, and it is not necessary to find the underlying transformation of parameter $q_1, q_2$. Since the relevant amplitudes can be expanded as power series of $Q$ with coefficients as rational function of $E_4^{\frac{1}{2}}$, $E_2$ and $E_6$, the transformation can always be consistently applied. For example, the involution transformation of the propagators  can be obtained by expanding them as power series of $Q$ and applying the transformation (\ref{invo3.183}).

We should expand the higher genus refined topological string amplitudes as power series of  $z_1,z_2$, which is possible since the higher genus amplitudes are regular near the large volume point $(z_1,z_2) \sim (0,0)$. Then we can expand further using the expansions formulas in our previous paper \cite{Huang:2015sta} in terms of power series of $Q$
\begin{eqnarray}  \label{series3.184}
w_0^{2n+2g-2} \mathcal{F}^{(n,g)} = \sum_{k=0}^{\infty} P^{(n,g)}_k(q_1) Q^k , 
\end{eqnarray} 
The coefficients of $P^{(n,g)}_k(q_1) $ would be rational functions of $E_4^{\frac{1}{2}}$ and $E_6$. Furthermore, we note that the denominator of $P^{(n,g)}_k(q_1) $ can only be a power of $E_4$ due to e.g.  the first expansion formula  in (\ref{expandz13.166}).

We  apply the involution transformation (\ref{invo3.183}) to the above series expansion (\ref{series3.184}). According to the expansion formula of $w_0$ in \cite{Huang:2015sta}, the involution transformation of $w_0^{2n+2g-2}$ contributes a factor of $(-1)^{n+g-1}$. So the A-model amplitude $w_0^{2g-2} \mathcal{F}^{(n,g)}$ and the coefficients $P^{(n,g)}_k(q_1)$ are invariant under the involution transformation. We infer that $P^{(n,g)}_k(q_1)$ must have no half integer power, i.e. only integer powers of $E_4$. 

The small fiber limit $z_1\rightarrow \infty$ is the same as the $E_4\rightarrow 0$ limit according to e.g. the expansion formula in \cite{Huang:2015sta}. The regularity of A-model amplitude $w_0^{2n+2g-2} \mathcal{F}^{(n,g)}$ near small fiber limit then excludes the possible power of $E_4$ in the denominator of $P^{(n,g)}_k(q_1) $. Therefore, the amplitudes $P^{(n,g)}_k(q_1) $ are quasi-modular forms. The modular weight of $18k+2(n+g)-2$ can also be worked out using the expansion formulas in our previous paper \cite{Huang:2015sta}.

\section{More on elliptic fibrations over Fano basis}
\label{refinedfano}

We generalize to analysis for elliptic fibration over $\mathbb{P}^2$ to other examples. Particularly interesting is the  elliptic fibration over $\mathbb{F}_1$. This is a three-parameter model studied e.g. in \cite{Klemm:1996}, and it reduces to the half K3 model when one of the two Kahler modulus parameters in the base becomes large. In this section we consider the main example of $\mathbb{F}_1$ model and then also discuss the $\mathbb{F}_0$ and $\mathbb{F}_2$ cases in the last subsection. 
 
Th complex moduli space of the mirror Calabi-Yau space are parametrized by three parameters $z_1, z_2, z_3$. Here the large volume point in the A-model correspond to  $(z_1, z_2, z_3)=(0,0,0)$. The generators of the Mori cone and the Picard-Fuchs operators are known in the literature 
\begin{eqnarray} \label{PF4.1}
\mathcal{L}_1 &=& \theta_1(\theta_1- \theta_2-2\theta_3) -12z_1 (6\theta_1+5)(6\theta_1+1),
 \nonumber \\
\mathcal{L}_2 &=& \theta_2^2 -z_2 (\theta_2-\theta_3) (\theta_2+2\theta_3-\theta_1), \nonumber \\ 
\mathcal{L}_3 &=& \theta_3(\theta_3-\theta_2) -z_3 (2\theta_3+\theta_2-\theta_1)(2\theta_3+\theta_2-\theta_1+1), 
\end{eqnarray} 
where $\theta_i := z_i \frac{\partial }{\partial z_i}$.  Here the limit  $z_1\sim 0$ corresponds to large elliptic fiber, while $z_3\sim 0$ limit gives the half K3 model. 
 
There are two conifold discriminants in the complex structure moduli space
\begin{eqnarray} \label{coni4.2} 
\Delta_1 &=& (\frac{1}{432}-z_1)^3(\frac{1}{432}-z_1+z_1z_2) -z_1^2 z_3 [8(\frac{1}{432}-z_1)^2 -16z_1^2 z_3 \nonumber \\
&& +36(\frac{1}{432}-z_1) z_1z_2 +27z_1^2 z_2^2 ],  \nonumber \\ 
\Delta_2 &=& (1-4z_3)^2-z_2 +36z_2z_3 -27z_2^2 z_3. 
\end{eqnarray}
The involution transformation which exchanges the two conifold divisors are
\begin{eqnarray} \label{invo4.3}
I: ~~~ (z_1,z_2,z_3)\rightarrow (x_1,x_2,x_3) = (\frac{1}{432}-z_1, -\frac{432z_1z_2}{1-432z_1},\frac{(432z_1)^2 z_3}{(1-432z_1)^2}). 
\end{eqnarray}     
We see that the transformation of conifold divisors are 
\begin{eqnarray}
I(\Delta_2) = \frac{\Delta_1}{(\frac{1}{432}-z_1)^4},~~~~  I(\Delta_1) =  z_1^4 \Delta_2
\end{eqnarray}
 
We can solve the Picard-Fuchs equations. Around the large volume point, there are one power series solution, 3 logarithmic solutions, 3 double logarithmic solutions, and one triple logarithmic solution. We provide some low order terms of the power series and single logarithmic solutions 
\begin{eqnarray} \label{period4.5}
w_0 &=& 1+60 z_1+13860 z_1^2 +9240 z_1^2 (442 z_1+3 z_3) +3063060 z_1^3 (437 z_1+8 z_3) \nonumber \\
&& + 12252240 z_1^3 (38019 z_1^2+1311 z_1 z_3+2 z_2 z_3)+\mathcal{O}(z^6) , \nonumber \\ 
w_{z_1} &=& w_0\log(z_1) + (312 z_1-z_3) +(77652 z_1^2+60 z_1 z_3+2 z_2 z_3-\frac{3}{2} z_3^2 )  +[23485136 z_1^3 \nonumber \\  
&& +196884 z_1^2 z_3+ \frac{2 z_3^2 }{3} (18 z_2-5 z_3)+30 z_1 z_3 (-2 z_2+z_3)]  +\mathcal{O}(z^4) , \\ 
w_{z_2} &=& w_0\log(z_2) +(60 z_1+z_3) +[20790 z_1^2-2 z_2 z_3+ \frac{3}{2} z_3^2 -60 z_1 (z_2+z_3)] + [ 7487480 z_1^3 \nonumber \\  
&&   -27720 z_1^2 (z_2-z_3)+30 z_1 (2 z_2-z_3) z_3+\frac{2}{3} z_3^2 (-18 z_2+5 z_3) ] +\mathcal{O}(z^4)    , \nonumber \\ 
w_{z_3} &=& w_0\log(z_3) +2 (60 z_1+z_3) +[41580 z_1^2+60 z_1 (z_2-2 z_3)+z_3 (-4 z_2+3 z_3)] + \frac{4}{3}[11231220 z_1^3  \nonumber \\  
&&  +20790 z_1^2 (z_2-2 z_3)+45 z_1 (2 z_2-z_3) z_3+z_3^2 (-18 z_2+5 z_3)] +\mathcal{O}(z^4) .  \nonumber 
\end{eqnarray}
The flat coordinates are given by the ratios $t_i = \frac{w_{z_i}}{w_0}$. We denote the exponentials of the flat coordinates as $q_i = \exp(t_i)$, which are the small parameter for world-sheet instanton expansion. The prepotential can be solved from the double logarithmic solutions.

 The exact three point functions in complex structure coordinates are rational functions of $z_1, z_2, z_3$, with poles at the conifold divisors (\ref{coni4.2}), and are related to the three point functions in flat coordinates $C_{z_i z_j z_k} =(w_0)^2 \frac{\partial t_l}{\partial z_i} \frac{\partial t_m}{\partial z_j} \frac{\partial t_n}{\partial z_k} C_{t_l t_m t_n}$. We solve the the prepotential and determine the exact three point functions in complex structure coordinate 
\begin{eqnarray} \label{threepoint4.5}
&& C_{111} =  \frac{1-432z_1+486z_1z_2 }{54\cdot (432z_1)^3  \Delta_1},  ~~~~
C_{112} =\frac{(1-432z_1)^2 + 648 z_1z_2 (1 - 432 z_1)  - 4z_3(432z_1)^2 }{216\cdot (432)^3\cdot z_1^2 z_2\Delta_1},
\nonumber \\
&& C_{113} =\frac{(1 -432 z_1)^2 + (1 - 432 z_1) (432z_1) z_2 + 248832 z_1^2 z_3 }{144 \cdot (432)^3\cdot z_1^2 z_3\Delta_1},
~~~ C_{122} =\frac{(1-432 z_1)^2+12 z_3(432 z_1)^2  }{ (432)^3\cdot z_2\Delta_1}, \nonumber \\
&&  C_{123} =\frac{(1 - 432 z_1)^3  +  432 z_1 z_2 (1 - 432 z_1)^2  + 4 (1 -432 z_1) (432z_1)^2 z_3 - 6 (432z_1)^3z_2 z_3  }{ (432)^4 \cdot z_1 z_2 z_3\Delta_1},  \nonumber \\
&& C_{133} =\frac{(1 -432 z_1)^3  + 432 z_1 z_2 (1 -432 z1)^2  + 4 (1 -432 z_1) (432z_1)^2 z_3
+ 3 (432 z_1)^3 z_2z_3   }{ (432)^4 \cdot z_1 z_3^2 \Delta_1},  \nonumber \\
&& C_{222} =\frac{1}{ (432)^4  z_2^2 \Delta_1\Delta_2 }  \{
1-4z_3+18 z_2 z_3+432 z_1 [-3+(8-40 z_2-9 z_2^2) z_3+16 z_3^2] \nonumber \\
&& ~~~~  -(432 z_1)^2 [-3+(8-12 z_2-27 z_2^2) z_3+16 (1+9 z_2) z_3^2]    \\
&& ~~~~  -(432 z_1)^3 [1+(-4+8 z_2+27 z_2^2) z_3+4 (-4-72 z_2+135 z_2^2)z_3^2+64 z_3^3]   \}, 
 \nonumber \\
&& C_{223} =\frac{1}{ (432)^4 z_2 \Delta_1\Delta_2}  \{
8-9 z_2  - 432 z_1(32-44 z_2+9 z_2^2) +(432 z_1)^2(8-9 z_2) (6 - 3 z_2 - 8 z_3)    \nonumber \\
&& ~~~~ +(432 z_1)^3 [8 (-3+8 z_3+16 z_3^2) + z_2 (52-144 z_3)+27 z_2^2 (-1+4 z_3)]   \}, 
 \nonumber \\
&& C_{233} =\frac{1}{ (432)^4  z_2 z_3 \Delta_1\Delta_2 }  \{
-2+11 z_2-9 z_2^2 +8 z_3 +432 z_1 (4- z_2) (2-11 z_2+9 z_2^2-8 z_3)  \nonumber \\
&& ~~~~  + (432 z_1)^2 [-10+48 z_3-32 z_3^2 +z_2 (70-40 z_3)+3 z_2^2 (-29+6 z_3) +27 z_2^3 ] \nonumber \\
&& ~~~~  + (432 z_1)^3 [ 4 (1-4 z_3)^2+z_2 (-35+40 z_3-112 z_3^2)+ z_2^2 (58-36 z_3) +27 z_2^3 (-1+z_3) ]   \}, 
 \nonumber \\
&& C_{333} =\frac{1}{ (432)^4   z_3^2 \Delta_1\Delta_2 }  \{
-5+14 z_2-9 z_2^2  +4 z_3 +432 z_1 (4- z_2) (5-14 z_2+9 z_2^2-4 z_3)  \nonumber \\
&& ~~~~  + (432 z_1)^2 [-25 +24 z_3 +48 z_3^2 +z_2 (94+56 z_3)  -3 z_2^2 (32+21 z_3) +27 z_2^3 ] \nonumber \\
&& ~~~~  - (432 z_1)^3 [ 27 z_2^3 (1+2 z_3)-2 z_2^2 (32+63 z_3)+z_2 (47+56 z_3-80 z_3^2)+2 (-5+8 z_3+48 z_3^2) ]   \}. \nonumber 
\end{eqnarray}

As before, we use the tilde symbol to denote the involution transformation of replacing $z_i$ with $x_i$ in (\ref{invo4.3}), e.g. $\tilde{z_i} =x_i$. We check that the three point functions transform as a tensor with a minus sign under the involution transformation. 
 
For the genus one amplitude, the formula is similar to the previous example as in (\ref{genusoneamplitude}). In the large volume limit, we have the leading asymptotic behavior $\log(\det(G^{-1})) = \sum_{i} \log(z_i)+\mathcal{O}(z_i) $, and the genus one amplitude $\mathcal{F}^{(1)}\sim -\frac{1}{24} \sum_{i=1}^3 \log(z_i) \int_M c_2 J_i$, where $J_i$ is the mirror Kahler class corresponding to complex structure modulus $z_i$. The second Chern class number of the classical geometry have been known in the literature, see e.g. \cite{Klemm:2012}, and the coefficients $s_i$  of $\log(z_i)$ are determined accordingly $s_i = \int_M c_2 J_i + 12$. We find the constants
\begin{eqnarray} \label{snumber4.6}
s_1 = 104,~~~ s_2= 36,~~~s_3=48. 
\end{eqnarray}

To compute the involution transformation of the genus one amplitude, we note that the Kahler potential is invariant, and the determinant of the metric transforms as $\det(\tilde{G}) =\det (\frac{\partial z_i}{\partial x_j}) \det(G)$. It is easy to compute  
\begin{eqnarray}
\det (\frac{\partial z_i}{\partial x_j}) = (\frac{1-432 z_1}{432 z_1})^3, 
\end{eqnarray} 
So the involution invariant of the genus one amplitude $\tilde{\mathcal{F}^{(1)}}=  \mathcal{F}^{(1)}$ is equivalent to the constrain 
\begin{eqnarray}
s_1-s_2-2s_3 +28=0,
\end{eqnarray}
which is satisfied by (\ref{snumber4.6}).

At higher genus $g\geq 2$, the analysis is similar to the previous example of elliptic fibration over $\mathbb{P}^2$. We use the involution transformation properties of the higher genus amplitudes, namely  $\tilde{\mathcal{F}}^{(g)} = (-1)^{g-1} \mathcal{F}^{(g)}$, to fix the holomorphic ambiguities. The remaining unfixed holomorphic ambiguities can be parametrized by rational functions $f(z_1,z_2,z_3)$, which satisfy the same transformation property $\tilde{f} = (-1)^{g-1} f$ as the higher genus topological string amplitudes.  We can define the linear space of holomorphic ambiguities at genus $g$ similarly as in \cite{Huang:2015sta}, but with one extra parameter in the base  
\begin{eqnarray} \label{Vspaces4.9}
V_0^{(g,m,n,l)} &:=& \{  f  ~|~  f =  \frac{p(z_1) z_2^n z_3^l }{(\Delta_1\Delta_2)^{2g-2}} ,  \textrm{ where $p(z_1)$ is a polynomial degree $m$ in $z_1$} \}.  \nonumber \\
V_1^{(g,m,n,l)}  &:=& \{  f  ~|~  f \in V_0^{(g,m,n,l)}  \textrm{ and }\tilde{f} \in V_0^{(g,m,n,l)} \}.   \nonumber\\
V_2^{(g,m,n,l)}  &:=& \{  f  ~|~  f \in V_1^{(g,m,n,l)}  \textrm{ and } \tilde{f} = (-1)^{g-1} f  \}.  \nonumber \\
V_3^{(g,m,n,l)}  &:=& \{  f  ~|~  f \in V_1^{(g,m,n,l)}  \textrm{ and } \tilde{f}  = (-1)^{g} f  \}. 
\end{eqnarray}
 
We should also impose the regularity condition at small elliptic fiber limit $z_1\sim \infty$. The asymptotic behavior of the period is universal and we find the power series solution with smallest scaling exponent is $w_0\sim z_1^{-\frac{1}{6}}$, so that together with the denominator at genus $g$, they scale as $(\frac{w_0}{\Delta_1\Delta_2})^{2g-2}\sim z_1^{-\frac{25}{3}(g-1)}$. So imposing the regularity condition near $z_1\sim \infty$ fixes the degree of $z_1$ in the holomorphic ambiguities in (\ref{Vspaces4.9}) to be $m=[\frac{25}{3}(g-1)]$. 

We can work out the involution symmetry transformation 
\begin{eqnarray}
&& (\frac{1}{432}-z_1)^{k_1}z_1^{k_2} \frac{z_2^nz_3^l}{(\Delta_1\Delta_2)^{2g-2}} \nonumber \\ 
&\rightarrow &
  (\frac{1}{432}-z_1)^{k_2+8(g-1) -n-2l }z_1^{k_1-8(g-1) +n+2l } \frac{(-1)^n z_2^nz_3^l}{(\Delta_1\Delta_2)^{2g-2}}.  
\end{eqnarray}
So to find involution symmetric function up to a sign we should consider the case $k_1-k_2 = 8(g-1) -n-2l$. We can also add an extra factor of $\frac{1}{432}-2z_1$, which gives a minus sign under the involution transformation. We further define two linear spaces as 
\begin{eqnarray} \label{Vpm4.11}
V_+^{(g,m,n,l)}  &:=& \{   \textrm{ linear space generated by  } (\frac{1}{432}-z_1)^{k_1}z_1^{k_2} \frac{z_2^nz_3^l}{(\Delta_1\Delta_2)^{2g-2}} \in  V_0^{(g,m,n,l)}  \nonumber \\ && \textrm{ with  }  k_1-k_2 = 8(g-1) -n-2l ~\},
\nonumber \\
V_-^{(g,m,n,l)}  &:=& \{   \textrm{ linear space generated by  } (\frac{1}{432}-z_1)^{k_1}z_1^{k_2} \frac{(\frac{1}{432}-2z_1) z_2^nz_3^l}{(\Delta_1\Delta_2)^{2g-2}} \in  V_0^{(g,m,n,l)}  \nonumber \\ && \textrm{ with  }  k_1-k_2 = 8(g-1) -n-2l ~\}.   
\end{eqnarray}

Similarly as in the previous example, we find the $V_+^{(g,m,n,l)}, V_-^{(g,m,n,l)}$ provide the linearly independent explicit base for the space $V_2^{(g,m,n,l)}, V_3^{(g,m,n,l)}$. We can identity $V_2^{(g,m,n,l)} = V_+^{(g,m,n,l)}, V_3^{(g,m,n,l)} = V_-^{(g,m,n,l)}$ in the case of $n+g$ an odd integer, while $V_2^{(g,m,n,l)} = V_-^{(g,m,n,l)}, V_3^{(g,m,n,l)} = V_+^{(g,m,n,l)}$ in the case of $n+g$ an even integer. 

Similar as in the previous example, the involution symmetry together with the regularity condition $z_1\sim \infty$ is conjectured to be equivalent to the fiber modularity constrains, studied in \cite{Klemm:2012, Alim:2012ss}. The space $V_2^{(g,[\frac{25}{3}(g-1)] ,n,l)}$ automatically encodes the  fiber modularity constrains.

We should also impose the gap conditions for the higher genus topological string amplitudes near the conifold points. The remaining unfixed holomorphic ambiguities should have no pole at the two conifold divisors, so should actually be polynomials of $z_i$.  After taking into account the involution symmetry and all boundary conditions, except for the Gopakumar-Vafa invariants at large volume point, we find the space of unfixed holomorphic ambiguity $X^{(g)}$ at genus $g$ to be 
\begin{eqnarray}
X^{(g)} = \left\{
\begin{array}{cl}
\sum_{n,l=0}^{\infty} \oplus (V_-^{(1,[\frac{g-1}{3}], 2n,l)} \oplus V_+^{(1, [\frac{g-1}{3}] , 2n+1,l)} )   ,    & ~~  \textrm{if  $g$ is even}    ;   \\
\sum_{n,l=0}^{\infty} \oplus (V_+^{(1, [\frac{g-1}{3}] , 2n,l)} \oplus V_-^{(1, [\frac{g-1}{3}] , 2n+1,l)} )   ,    & ~~  \textrm{if  $g$ is odd}   .   
\end{array}
\right. 
\end{eqnarray}   
The space is finite dimensional, and it is quite straightforward to compute the dimension of the space using the explicit basis  (\ref{Vpm4.11}).

\subsection{(Quasi-)modularity properties} 

We can expand the topological string amplitudes in base coordinates as 
\begin{eqnarray} 
\mathcal{F}^{(g)} = \sum_{m,n=0}^{\infty} P^{(g)}_{(m,n)}(q_1) \big( \frac{q_1}{\eta(q_1)^{24}}\big)^{\frac{m}{2}+n} q_2^m q_3^n. 
\end{eqnarray} 
Here the coefficients $P^{(g)}_{(m,n)}$ are quasi-modular forms of weight $6m+12n+2g-2$ and satisfy certain modular anomaly equations \cite{Alim:2012ss, Klemm:2012}. The mathematical proof of many observations about modularity properties in topological string are recently provided in \cite{Haghighat:2015qdq}. 

We can discuss the genus zero case for some details. The leading term contains a cubic polynomial of flat coordinates $t_i=\log(q_i)$ with $i=1,2,3$, so it is actually not independent of the base parameters $q_2, q_3$ and should be written as $P^{(0)}_{(0,0)}(q_1,q_2,q_3)$. To decouple the fiber and base classes, as in the previous example, we should  shift the base coordinate. Here the shifts are
\begin{eqnarray} \label{shift4.15}
\log(q_2) \rightarrow \log(q_2) -\frac{\log(q_1)}{2}, ~~~~ \log(q_3)\rightarrow \log(q_3) -\log(q_1). 
\end{eqnarray}
After the shifts, the derivative is modular  
\begin{eqnarray}
 \frac{\partial ^3 }{\partial^3 t_1} P^{(0)}_{(0,0)}(q_1,q_2,q_3) =2 E_4(q_1),
\end{eqnarray} 
so that the leading term $P^{(0)}_{(0,0)}(q_1,q_2,q_3)$ is effectively a modular form with -2 weight. 

We fix the modular forms $P^{(0)}_{(m,n)}$ for genus zero and some low degrees $m,n$. These are the formulas 
\begin{eqnarray} \label{genuszero4.17}
&& P^{(0)}_{(1,0)} = E_4, ~~~ P^{(0)}_{(0,1)} = -E_4E_6, ~~~  P^{(0)}_{(2,0)} = \frac{E_4 (E_2 E_4+2 E_6)}{24} ,
\nonumber \\ && 
 P^{(0)}_{(1,1)} = \frac{E_4 (8 E_2 E_4 E_6 + 31 E_4^3+105 E_6^2)}{48} ,~~~ 
 P^{(0)}_{(0,2)} = - \frac{E_4 E_6 (17 E_4^3+7 E_6^2)}{96}, 
 \nonumber \\ &&
P^{(0)}_{(3,0)} =  \frac{E_4 (54 E_2^2 E_4^2+216 E_2 E_4 E_6 +109 E_4^3+197 E_6^2)}{15552}, 
 \nonumber \\ &&
P^{(0)}_{(2,1)} =  \frac{E_4 E_6 (2 E_2^2 E_4^2+8 E_2 E_4 E_6-5 E_4^3- 5E_6^2)}{288},
 \nonumber \\ &&
P^{(0)}_{(1,2)} = \frac{E_4 }{55296} \big[ 768 E_2^2 E_4^2 E_6^2 +E_2E_6 (7584  E_4^4  +20832  E_4 E_6^2 )
\nonumber \\ && ~~~+15935 E_4^6+161186 E_4^3 E_6^2+70175 E_6^4 \big], 
 \nonumber \\ &&
P^{(0)}_{(0,3)} =-\frac{E_4 E_6 (9349 E_4^6+16630 E_4^3 E_6^2+1669 E_6^4)}{373248}. 
\end{eqnarray}
Here the formulas of $P^{(0)}_{(m,0)}$ for the case of half K3 model has been known for a long time in \cite{Minahan:1997}. The genus zero modular anomaly equation is 
\begin{eqnarray}
\frac{\partial P^{(0)}_{(m,n)}}{\partial E_2} =  \frac{1}{24} (\sum_{i=0}^m\sum_{j=0}^n)^\prime  
[i(m-i)-j(m-i)-i(n-j)] P^{(0)}_{(i,j)} P^{(0)}_{(m-i,n-j)},
\end{eqnarray}
where the prime denotes the exclusion of the cases $(i,j)=(0,0)$ and $(i,j)=(m,n)$ in the sum. The factor $i(m-i)-j(m-i)-i(n-j)$ in the equation comes from the intersection of corresponding homology classes in the base. From the equation we can see that $\frac{\partial P^{(0)}_{(0,n)}}{\partial E_2} =0 $, so the coefficients $P^{(0)}_{(0,n)}$ of zero degree in $q_2$ are actually modular forms.

We can work out some expansion formulas similar to the previous example of elliptic fibration over $\mathbb{P}^2$ in our previous paper \cite{Huang:2015sta}. First we should define the expansion parameters proportional to the base complex structure moduli 
\begin{eqnarray} \label{Z23-4.14}
Z_2 := \frac{z_2}{(1-432 z_1) E_4(q_1)^{\frac{3}{2}}}, ~~~~Z_3 := \frac{z_3}{(1-432 z_1)^2 E_4(q_1)^3}. 
\end{eqnarray} 
We expand around small $Z_2$ and $Z_3$, and fix the coefficients as modular forms of fiber parameter $q_1$ of appropriate weights as the followings 
\begin{eqnarray}
z_1(1-432z_1) &=& \frac{\eta^{24}}{E_4^3} \big[1 +E_6^2 Z_3 - (77 E_4^3 E_6 + 211 E_6^3) \frac{Z_2Z_3}{144} +2E_6^4 Z_3^2   -E_6 (10703 E_4^6 \nonumber \\  &&  
+237506 E_4^3 E_6^2+332399 E_6^4) \frac{Z_2Z_3^2}{41472}  +5 E_6^6 Z_3^3 
 +\mathcal{O}(Z^4) \big],  \label{expan4.11} \\
 \frac{\Delta_1\Delta_2}{(\frac{1}{432} -z_1)^4} &=&  1 -E_6 Z_2 -4(E_4^3 +E_6^2) Z_3 + (E_6^2-E_4^3)\frac{Z_2^2}{4}+\frac{3}{2} (17 E_4^3 E_6+7 E_6^3)Z_2 Z_3 \nonumber \\ &&   
+2 (E_4^3+E_6^2) (3 E_4^3+E_6^2)  Z_3^2 + (247 E_4^6-6038 E_4^3 E_6^2-1985 E_6^4) \frac{Z_2^2 Z_3}{288}
   \nonumber  \\  && - E_6 (E_4^3 - E_6^2) (2125 E_4^3+179 E_6^2)  \frac{Z_2 Z_3^2}{72} - 4 E_4^3 ( E_4^6 - E_6^4  ) Z_3^3    \nonumber   
 \\ &&  +\mathcal{O}(Z^4),      \label{expan4.12}  \\
 1-864 z_1 &=& \frac{1}{E_4^{\frac{3}{2}}} \big[ E_6  +  E_6(E_6^2 -E_4^3 ) \frac{Z_3}{2} 
 +(E_4^3-E_6^2) (77 E_4^3+211 E_6^2)   \frac{Z_2Z_3}{288} \nonumber \\&&  
  - E_6 (E_4^3 - E_6^2) (E_4^3+7 E_6^2) \frac{Z_3^2}{8} +\frac{7}{82944} 
 (E_4^3-E_6^2) (3113 E_4^6+36686 E_4^3 E_6^2 \nonumber \\ && +43145 E_6^4)   
  Z_2Z_3^2  
-   E_6 (E_4^3-E_6^2) (E_4^6+6 E_4^3 E_6^2+33 E_6^4) \frac{Z_3^3}{16}  \nonumber   
 \\ &&
  +\mathcal{O}(Z^4) \big] ,   \label{expan4.13}  \\
w_0 &=&  E_4^{\frac{1}{4}} \big[1 + \frac{5}{144} (E_4^3-E_6^2) Z_3+ \frac{5}{72} E_6 (E_6^2 - E_4^3) Z_2 Z_3 + 
\frac{5}{82944 } 
(E_4^3-E_6^2) \nonumber \\ &&  \cdot (221 E_4^3+931 E_6^2)  Z_3^2 
-\frac{35}{6912} 
E_6 (E_4^3-E_6^2) (35 E_4^3+61 E_6^2)Z_2 Z_3^2  \nonumber \\ && 
+ \frac{5}{107495424}  
(E_4^3-E_6^2) (160225 E_4^6+825214 E_4^3 E_6^2+2747041 E_6^4) Z_3^3 
 \nonumber \\ && +\mathcal{O}(Z^4) \big].   \label{expan4.14} 
\end{eqnarray}
We conjecture that excluding the pre-factors, the coefficient of $Z_2^nZ_3^l$ is a $SL(2,\mathbb{Z})$ modular form of weight $6n+12l$ for the expansions  of $z_1(1-432z_1)$, $\frac{\Delta_1\Delta_2}{(\frac{1}{432} -z_1)^4}$, $w_0$, and of weight $6n+12l+6$ for the expansion of $1-864 z_1$. 

In the half K3 limit $Z_3\rightarrow 0$, the expansions truncate at finite order, as well known in the literature. In particular, for the cases of $z_1(1-432z_1)$,  $1-864 z_1$, and $w_0$, only the leading terms survive. While for the case of $\frac{\Delta_1\Delta_2}{(\frac{1}{432} -z_1)^4}$, we have additional non-vanishing coefficients for the $Z_2$ and $Z_2^2$ terms as well.

Another interesting limit is $Z_2\rightarrow 0$, where the Picard-Fuchs equations (\ref{PF4.1}) reduce to a two-parameter model, which have been studied before e.g. in \cite{Lian:1995}. In this case there is a rational map between the J-functions of the mirror parameters $q_1, q_3$ and complex structure parameters $z_1, z_3$
\begin{eqnarray} \label{rational4.19}
z_1 &=& 2 \frac{1/J(q_1) +1/ J(q_1q_3) -1728 /(J(q_1)J(q_1q_3))}{1+\sqrt{(1-1728/J(q_1)) (1-1728/J(q_1q_3)) )}}, 
\nonumber \\ 
z_1^2 z_3 &=& \frac{1}{J(q_1)J(q_1q_3)},
\end{eqnarray}
where the J-function $J(q) = \frac{E_4 (q)^3}{\eta(q)^{24}} = \frac{1728 E_4(q)^3} {E_4(q)^3 -E_6(q)^2}$.  The relation was conjectured in \cite{Kachru:1995, Klemm:1995}, and was proven by Lian and Yau in \cite{Lian:1995}. 

We can prove our modularity conjecture in the special limit of $Z_2\rightarrow 0$ using the above relation (\ref{rational4.19}). From the relation we can eliminate $J(q_1q_3)$, change variable $z_3=(1-432 z_1)^2 E_4(q_1)^3 Z_3$ according to (\ref{Z23-4.14}), and solve the rational equation for $z_1$. There are 4 solutions for $z_1$, and we pick the one with the correct asymptotic behavior
\begin{eqnarray} \label{solution4.20}
1-864z_1 = \frac{((E_4^3-E_6^2)\sqrt{1-4E_6^2Z_3} - (E_4^3-E_6^2) + 2E_6^2 E_4^3 Z_3 )^{\frac{1}{2} }}{E_4^{\frac{3}{2}}E_6 \sqrt{2Z_3}   }  ,
\end{eqnarray}
where the argument for Eisenstein series is $q_1$. We make the following expansion for small $Z_3$ 
\begin{eqnarray}  \label{powerseries4.21}
\sqrt{1-4E_6^2 Z_3} = 1-2E_6^2 Z_3 -2E_6^4 Z_3^2 f(E_6^2Z_3), 
\end{eqnarray}
where we denote $f(E_6^2Z_3)= 1+ 2 E_6^2 Z_3+ 5 E_6^4 Z_3^2 +\cdots $, a power series of $E_6^2 Z_3$.  

We can then compute (\ref{solution4.20}) 
\begin{eqnarray} 
1-864z_1 = \frac{E_6}{E_4^{\frac{3}{2}}} \sqrt{1-(E_4^3-E_6^2) Z_3 f(E_6^2Z_3)}. 
\end{eqnarray} 
So the coefficient of $Z_3^l$ in the expansion (\ref{expan4.13}) is a modular form of weight $12l+6$ and contains a factor of $E_6 (E_4^3-E_6^2)$.  

The modularity of the coefficient of $Z_3^l$ in the expansion (\ref{expan4.11}, \ref{expan4.12}) follows straightforwardly by the following computations in the $Z_2\rightarrow 0$ limit 
\begin{eqnarray}
z_1(1-432z_1) &=& \frac{\eta^{24}}{E_4^3} [1+E_6^2 Z_3 f(E_6^2Z_3)] , \nonumber \\
(\frac{\Delta_1\Delta_2}{(\frac{1}{432}-z_1)^4})^{\frac{1}{2}} &=&  1- 4 [(\frac{1}{432}-z_1)^2+ z_1^2] E_4^3 Z_3 + 16 z_1^2 (\frac{1}{432}-z_1)^2 E_4^6 Z_3^2 .
\end{eqnarray}

To demonstrate  the modularity of the coefficient of $Z_3^l$ in the expansion of $w_0$ in (\ref{expan4.14}), we note the following formula which appears in  the proof of (\ref{rational4.19}) by Lian and Yau in \cite{Lian:1995}, 
\begin{eqnarray} \label{w04.24}
w_0 = (E_4(q_1) E_4(q_1q_3) )^{\frac{1}{4}}. 
\end{eqnarray}
We can also expand Eisenstein series in terms of  inverse of J-function
\begin{eqnarray} \label{E44.25}
E_4(q) ^{\frac{1}{4}} &=&  \sum_{k=0}^{\infty} \binom{6k}{3k}  \binom{3k}{2k} \frac{1}{J(q)^k} \big( \frac{2}{1+\sqrt{1-1728/J(q)}}  \big) ^k \nonumber \\
&=&  1 + \frac{60}{J(q)} + \frac{39780}{J(q)^2}  +  
\frac{38454000 }{J(q)^3}  +\mathcal{O}(\frac{1}{J(q)^4} ). 
\end{eqnarray}
From the relations (\ref{rational4.19}) we can also eliminate $z_1$ and solve for $J(q_1q_3)$. Again there are two solutions to the rational equation, and we pick the one with the correct asymptotic behavior 
\begin{eqnarray} \label{J4.26}
\frac{1}{J(q_1q_3)} &=& -(E_4 ^3 -E_6 ^2 ) \frac{\sqrt{1-4E_6^2Z_3} -1+2 E_6^2 Z_3} {3456 E_6^4 Z_3} 
\nonumber \\ &=& \frac{E_4^3 -E_6^2}{1728} Z_3 f(E_6^2Z_3),
\end{eqnarray}
where the omitted argument of the Eisenstein series in the right-hand-side is $q_1$ as before, and $f(E_6^2Z_3)$ is the power series appearing in (\ref{powerseries4.21}). Now from the three equations (\ref{w04.24}), (\ref{E44.25}), (\ref{J4.26}), we deduce the coefficient of $Z_3^l$ in the expansion of $w_0$ in (\ref{expan4.14}) is a modular form of weight $12l$ and furthermore contains a factor of $E_4^3 -E_6^2$. 

We can expand the parameters $Z_2, Z_3$ in terms of the base coordinates $q_2, q_3$ similarly as  in the previous example \cite{Huang:2015sta}. We define the normalized Kahler parameters 
\begin{eqnarray} \label{Q23renor4.32}
Q_2 := ( \frac{q_1}{\eta(q_1)^{24}})^{\frac{1}{2}} q_2,~~~~~~  Q_3 := \frac{q_1}{\eta(q_1)^{24}} q_3.
\end{eqnarray}
We expand $Z_2, Z_3$ and fix the coefficients as quasi-modular forms of fiber parameter $q_1$ 
\begin{eqnarray} \label{Z2ex4.33}
Z_2 &=& Q_2 \big[ 1 +  (E_2 E_4-E_6)\frac{Q_2}{12} +(24 E_2 E_4 E_6 -41 E_4^3-127 E_6^2) \frac{Q_3}{144} 
 +  (6 E_2^2 E_4^2-5 E_4^3   \nonumber \\ &&  -E_6^2)\frac{Q_2^2} {576}  + E_6 (2 E_2^2 E_4^2 - 4 E_2 E_4 E_6 +139 E_4^3+151 E_6^2)\frac{Q_2Q_3}{144}  
+ \big( 2304 E_2^2 E_4^2 E_6^2 \nonumber \\ &&  - E_2E_4E_6(22368  E_6^2 + 2976  E_4^3 )  +7673 E_4^6
  +103406 E_4^3 E_6^2  +77849 E_6^4\big)\frac{Q_3^2}{165888}
\nonumber \\ &&  +\mathcal{O}(Q^3) ~\big],  \\  \label{Z3ex4.34}
Z_3 &=& Q_3 \big[ 1 -  (E_2 E_4-E_6)\frac{Q_2}{12} - ( 31 E_4^3 + 113 E_6^2)  \frac{Q_3}{72} 
 -  (2 E_2^2 E_4^2-8 E_2 E_4 E_6- 5 E_4^3   \nonumber \\ && -5E_6^2) \frac{Q_2^2} {576}  + \big(-24 E_2^2 E_4^2 E_6 +E_2 E_4 (31 E_4^3 +161  E_6^2) +3485 E_4^3 E_6 +3259 E_6^3\big) \frac{Q_2Q_3}{1728}  \nonumber \\ &&
+ (9907 E_4^6+123034 E_4^3 E_6^2+115891 E_6^4)  \frac{Q_3^2}{82944 }
  +\mathcal{O}(Q^3) ~\big]. 
\end{eqnarray}
We conjecture that the coefficients are quasi-modular forms and satisfy the modular anomaly equation
\begin{eqnarray} \label{modularanomaly4.35}
\frac{\partial Z_{2,3}}{\partial E_2} &=& \frac{1}{12} \big[Q_2^2 (\partial_{Q_2} Z_{2,3})  (\partial_{Q_2} P^{(0)}) -Q_2Q_3 (\partial_{Q_2} Z_{2,3})  (\partial_{Q_3} P^{(0)}) \nonumber \\ && - Q_2Q_3 (\partial_{Q_3} Z_{2,3})  (\partial_{Q_2} P^{(0)})    \big]. 
\end{eqnarray}
Here the modular anomaly equations are the same for both $Z_2$ and $Z_3$. We use the notation  $P^{(0)} := \sum_{(m,n)} P^{(0)}_{(m,n)}(q_1) Q_2^mQ_3^n$, which is the generating function for the quasi-modular forms appearing in the expansion of the genus zero prepotential in (\ref{genuszero4.17}), excluding the zero base degree 
$(m,n)\neq (0,0)$. 

We note that in the limit $Q_2\rightarrow 0$, the right hand side of (\ref{modularanomaly4.35}) vanishes so the coefficients of $Q_3^n$ in the expansion of $Z_3$ are actually purely modular forms. We can also easily prove this using the relations (\ref{rational4.19}). Eliminating $z_1$ and solving for $Z_3$, we find 
\begin{eqnarray}
Z_3 = \frac{1}{\eta(q_1)^{24} J(q_1q_3)} \frac{1}{(1+  \frac{E_6(q_1)^2 }{\eta(q_1)^{24} J(q_1q_3) } )^2} .
\end{eqnarray}
Since $\frac{1}{ J(q_1q_3)}$ is a power series of $q_1q_3$, it is clear that we can expand $Z_3$ as power series of $Q_3$ and the coefficients of $Q_3^n$ are modular forms of $q_1$ of weight $12(n-1)$.  

We can also consider the $z_3\rightarrow 0$ limit. In this case the fiber modular parameter $q_1$ only depends on the corresponding complex structure parameter $z_1$. One can solve the Picard-Fuchs equation and obtain the formula, see e.g \cite{Klemm:2012}, 
\begin{eqnarray}
 \frac{\sqrt{q_1}q_2}{\eta(q_1) ^{12}}  = \frac{2z_2}{E_4(q_1) ^{\frac{3}{2}}+E_6(q_1) }   \exp\big[ \sum_{m=1}^{\infty} c_m(z_1) z_2^m \big],
\end{eqnarray}
where $c_m(z_1) =\frac{\mathcal{L}_m w_0(z_1)} {w_0(z_1)}$ with the differential operator $\mathcal{L}_m=\frac{(-1)^m}{m\cdot m!}\prod_{k=1}^m (\theta_{z_1} -k+1) $. It is shown in $\cite{Klemm:2012}$ that $c_m(z_1) = w_0^{-6m} (1-432z_1)^{-m}  P_{6m}(E_2,E_4,E_6)$, where $P_{6m}(E_2,E_4,E_6)$ is a quasi-modular form of weight $6m$, and is linear in $E_2$.

We can rewrite the formulas in the normalized variables $Z_2$ and $Q_2$ in (\ref{Z23-4.14}, \ref{Q23renor4.32}) as
\begin{eqnarray} 
Q_2 = Z_2 \exp\big[ \sum_{m=1}^{\infty} P_{6m}(E_2,E_4,E_6) Z_2^m \big]. 
\end{eqnarray}
It is clear that we can inverse the relation and write $Z_2$ as a power series of $Q_2$ where the coefficient of $Q_2^m$ is a  quasi-modular form of weight $6m$.  Furthermore, the modular anomaly equation (\ref{modularanomaly4.35}) for $Z_2$ in the $z_3\rightarrow 0$ limit  has actually already appeared in the proof of modular anomaly equation for the prepotential for half K3 model in \cite{Klemm:2012}.

\subsection{The genus one refined amplitude $\mathcal{F}^{(1,0)}$}

One may study the half K3 model by taking the $z_3\rightarrow 0$ limit. However, the power series solution $w_0$ in (\ref{period4.5}) of Picard-Fuchs equation near the large volume point is not a constant in the $z_3\rightarrow 0$ limit, which means the half K3 model by itself is not actually a consistent two-parameter local model. One still have to apply the B-model method of holomorphic anomaly equation to the three-parameter moduli space of $z_1, z_2, z_3$ to compute the topological string amplitudes, and then taking the $z_3\rightarrow 0$ limit afterward. 

In \cite{Huang:2013} we studied the (refined) half K3 model using the modular anomaly equation. The modular anomaly equation is recursive in both genus and wrapping number of the base manifold, and the modular ambiguities are modular forms of $SL(2,\mathbb{Z})$ of the elliptic fiber parameter $q_1=\exp(t_1)$. At low orders we can fix the modular ambiguities from the vanishing conditions of the Gopakumar-Vafa invariants. 
  
The B-model method of holomorphic anomaly equation is better than that of \cite{Huang:2013}, since the holomorphic anomaly equation is recursive only in genus, and we can fix the holomorphic ambiguities with the gap conditions at the conifold divisors and the regularity conditions at the small elliptic fiber limit, other than using the Gopakumar-Vafa invariants from the large volume point. Once we fix the topological string amplitude at a given genus, we can expand around large volume point to find the quasi-modular forms of the elliptic fiber for all wrapping numbers of the base. A proposal for the polynomial form of higher genus amplitude for half K3 model is also considered in \cite{Sakai:2011}. However the proposal \cite{Sakai:2011} is not based on the complex structure moduli space in the B-model approach, and as such one can not apply the gap conditions at the conifold divisors.  On the other hand, the recent works on elliptic genus of E-strings \cite{Haghighat:2014, Kim:2014, Cai:2014} find all-genus formulas for half K3 model at a finite base degree, and are complimentary to the B-model method here. The beautiful calculations of \cite{Haghighat:2014} construct a 2d quiver gauge theory and use the techniques of \cite{GG:2013, Benini:2013a, Benini:2013} to compute the path integrals. The techniques have been also applied to more Calabi-Yau models in \cite{HKLV:2014}.

Here we follow the approach in section \ref{refinedsec3.7} to study the genus one refined amplitude $\mathcal{F}^{(1,0)}$. We make the following ansatz for the compact model 
\begin{eqnarray} \label{ansatz4.14}
\mathcal{F}^{(1,0)} = \frac{1}{24}[ \log(\Delta_1\Delta_2) - c_1 \log(z_1) - c_2 \log(z_2)- c_3 \log(z_3)] +c_0 K ,
\end{eqnarray}
similar to the equation (\ref{ansatz3.107}). Again we shall impose the fiber modularity constrain to fix the constants. First we expand the amplitude in terms of flat coordinates of the base manifold 
\begin{eqnarray} \label{expan4.15}
\mathcal{F}^{(1,0)} = \sum_{m,n=0}^{\infty} P^{(1,0)}_{(m,n)}(q_1) \big( \frac{q_1}{\eta(q_1)^{24}}\big)^{\frac{m}{2}+n} q_2^m q_3^n. 
\end{eqnarray} 

The leading term has linear dependence on flat coordinates $t_i=\log(q_i)$ with $i=1,2,3$, so we should  shift the base coordinate according to (\ref{shift4.15}) similarly as in the genus zero case. After the shifts, the leading term should satisfy the modularity constrain 
\begin{eqnarray}
q_1 \frac{\partial }{\partial q_1} P^{(1,0)}_{(0,0)}(q_1,q_2,q_3) \sim E_2(q_1). 
\end{eqnarray} 
We find that this constrain fixes two of the four constants in the ansatz  (\ref{ansatz4.14}), namely  
\begin{eqnarray} 
c_0= -2 + \frac{c_2}{4} + \frac{c_3}{2} , ~~~ c_1= -4 + c_2 + 2 c_3 . 
\end{eqnarray}
 Here the second condition above also ensures the involution symmetry is satisfied by the refined amplitude  $\tilde{\mathcal{F}}^{(1,0)} = \mathcal{F}^{(1,0)}$.

We can use the results from half K3 model to further fix the constants. Here we denote the refined Gopakumar-Vafa numbers $n^{g_L,g_R }_{(d_1,d_2,d_3)}$ where $d_1, d_2, d_3$ are the degrees of Kahler parameters from the complex coordinates $z_1, z_2, z_3$ respectively, so that $d_1$ is the elliptic fiber degree and $d_2, d_3$ are the base degrees. From the calculations of half K3 model, we can obtain the numbers for all $n^{g_L,g_R }_{(d_1,d_2, 0)}$ with $d_2>0$. We can use in particular $n^{0,1}_{(1,1, 0)}=-2$, which gives the constrain $c_2-c_3=12$. On the other hand, we find that the topological amplitudes of half K3 model is independent of the remaining unfix constant.  
In particular, for the expansion (\ref{expan4.15}) of the genus one refined case $\mathcal{F}^{(1,0)}$, we have the following results at some low wrapping numbers
\begin{eqnarray} \label{formulas4.19}
P^{(1,0)}_{(1,0)} &=& -  \frac{E_2 E_4 }{24}, \nonumber \\
P^{(1,0)}_{(2,0)}  &=&  -   \frac{1}{1152} (4 E_2^2 E_4^2+7 E_4^3+8 E_2 E_4 E_6+ 5 E_6^2), 
\nonumber \\ 
P^{(1,0)}_{(3,0)}  &=&  -   \frac{1}{124416}  [54 E_2^3 E_4^3+ 235 E_2 E_4^4 +216 E_2^2 E_4^2 E_6 \nonumber \\ && +776 E_4^3 E_6+287 E_2 E_4 E_6^2+160 E_6^3 ] ,
\end{eqnarray}
where the argument for the Eisenstein series is the elliptic fiber parameter $q_1$.  These agree with the results in \cite{Huang:2013}. 
 
 To completely fix the constants, we need some one more data point from curve counting calculations. In particular, we have $n^{0,1}_{(1, 0, 0)}= 8$ and complete fix all constants in the ansatz (\ref{ansatz4.14}) as
 \begin{eqnarray}
 c_0 = -\frac{23}{2},~~ c_1 =-42, ~~ c_2=-\frac{14}{3}, ~~ c_3= -\frac{50}{3}. 
 \end{eqnarray}
 With the complete formula, we can then compute all numbers $n^{0,1}_{(d_1,d_2,d_3)}$. We list the data up to some finite degrees in table \ref{F1GVtablegenus01} in Appendix \ref{AppendixB1}. Many of the entries can be checked by curve counting calculations.

\subsection{Elliptic fibrations over $\mathbb{F}_0$ and $\mathbb{F}_2$}

We do more calculations with the examples of elliptic fibration over other Hirzebruch surfaces $\mathbb{F}_0$ and $\mathbb{F}_2$, which are $\mathbb{P}^1$ fibration over $\mathbb{P}^1$. These are three-parameter models with complex structure coordinates $z_1, z_2, z_3$ in the mirror. In our notation, among the flat coordinates $t_i\sim \log(z_i)$, the elliptic fiber Kahler modulus is $t_1$ while the base Kahler moduli are $t_2, t_3$, with the corresponding degrees $d_1,d_2,d_3$ in the GV numbers and topological string partition functions. In the limit of large base $\mathbb{P}^1$ modulus $t_2\sim \infty$, the details of the $\mathbb{P}^1$ fibrations decoupled, so these geometries are the same. Indeed in our calculations we have the GV numbers $n_{d_E, 0 ,d_3}^g$ and partition function $Z_{0,d_3}$ are the same for all elliptic fibrations over $\mathbb{F}_n, n=0,1,2$. However, surprisingly, the refined GV numbers at genus one $n_{d_E, 0 ,d_3}^{g_L,g_R}$ with $(g_L,g_R)=(0,1)$ are different for the elliptic fibrations over $\mathbb{F}_0$ and $\mathbb{F}_1$. This suggests that refined numbers in the large $t_2$ limit still retains information of the compact geometry, and is worth of further investigations.

Since the method is similar to the previous case of elliptic fibration over $\mathbb{F}_1$, we will be rather brief in the details. First we consider the elliptic fibration over $\mathbb{F}_0$. The Picard-Fuchs equations for the periods are 
\begin{eqnarray}
\mathcal{L}_1 &=& \Theta_1(\Theta_1 - 2\Theta_3 -2\Theta_2) -12 z_1 (6\Theta_1+5) (6\Theta_1+1) , \nonumber \\  
\mathcal{L}_2 &=& \Theta_3^2  - z_3 (2\Theta_3+2\Theta_2 -\Theta_1) (2\Theta_3+2\Theta_2 -\Theta_1+1)   , \nonumber \\ 
\mathcal{L}_3 &=&\Theta_2^2  - z_2 (2\Theta_3+2\Theta_2 -\Theta_1) (2\Theta_3+2\Theta_2 -\Theta_1+1) ,
\end{eqnarray} 
where $\Theta_i = z_i \partial_{z_i}$ are the logarithmic derivatives. There is an apparent symmetry between the two base moduli $t_2$ and $t_3$. 

The involution transformation is 
\begin{eqnarray}
(z_1, z_2, z_3) \rightarrow (x_1,x_2,x_3) \equiv  (\frac{1}{432} -z_1,  \frac{z_2z_1^2}{(\frac{1}{432}-z_1)^2},  \frac{z_3z_1^2}{(\frac{1}{432}-z_1)^2} ). 
\end{eqnarray}
It exchanges the two conifold divisors up to a factor 
\begin{eqnarray}
\Delta_2(z_1,z_2,z_3)  &=& 1 - 8(z_2+z_3)+16 (z_2-z_3)^2,  \nonumber \\
\Delta_1 (z_1, z_2, z_3) &=& (\frac{1}{432}-z_1)^4 \Delta_2 (x_1,x_2,x_3) . 
\end{eqnarray}

The base complex structure parameters can be expanded as quasi-modular forms similar to the $\mathbb{F}_1$ case  as in (\ref{Z2ex4.33}, \ref{Z3ex4.34}). For the $\mathbb{F}_0$ we define  
\begin{eqnarray}
Z_i := \frac{z_i}{(1-432 z_1)^2 E_4(q_1)^3}, ~~~ Q_i := \frac{q_1}{\eta(q_1)^{24}} q_i, ~~~~ i=2,3
\end{eqnarray}
The expansion formula for $Z_2$ is then
\begin{eqnarray} \label{Zexpansion6.49}
Z_2 &=&  Q_2 + \frac{1}{72} (-31 E_4^3 - 113 E_6^2) Q_2^2  + \frac{1}{6} (-3 E_4^3 + E_2 E_4 E_6 - 10 E_6^2) Q_2 Q_3  
+ \frac{1}{82944}(9907 E_4^6  \nonumber \\ &&  + 123034 E_4^3 E_6^2 + 115891 E_6^4) Q_2^3+ 
 \frac{1}{432} (-15 E_4^6 - 31 E_2 E_4^4 E_6 + 12 E_2^2 E_4^2 E_6^2 - 
    245 E_4^3 E_6^2  \nonumber \\ && - 281 E_2 E_4 E_6^3 + 560 E_6^4) Q_2^2 Q_3 + 
 \frac{1}{576} (83 E_4^6 - 31 E_2 E_4^4 E_6 + 8 E_2^2 E_4^2 E_6^2 + 927 E_4^3 E_6^2  \nonumber \\ && - 
    153 E_2 E_4 E_6^3 + 894 E_6^4) Q_2 Q_3^2 +\cdots , 
\end{eqnarray}
while the formula for $Z_3$ is obtained by switching the indices 2 and 3 in the above equation. The modular anomaly equation is 
\begin{eqnarray} \label{anomaly6.50}
\frac{\partial Z_{2,3}}{\partial E_2} = - \frac{1}{12} Q_2Q_3 \big[ (\partial_{Q_2} Z_{2,3})  (\partial_{Q_3} P^{(0)})  + (\partial_{Q_3} Z_{2,3})  (\partial_{Q_2} P^{(0)})    \big]. 
\end{eqnarray}

We can similarly compute the three-point functions and the genus zero GV invariants. The genus one amplitudes including the refined case, $\mathcal{F}^{(0,1)}$ and  $\mathcal{F}^{(1,0)}$ can be also fixed. The formula for $\mathcal{F}^{(0,1)}$ is the same form as in (\ref{genusoneamplitude}), with the corresponding topological numbers
$h^{1,1}=3, \chi=-480, s_1=104, s_2=s_3=36$. While for the refined amplitude, we use the fiber modularity, the symmetry between the two base moduli and the additional information $n^{(0,1)}_{1,0,0}=8$ from curve counting. The formula is fixed to be 
\begin{eqnarray} 
\mathcal{F}^{(1,0)} = \frac{1}{24}[ \log(\Delta_1\Delta_2) + 42 \log(z_1) + \frac{19}{2} \log(z_2) + 
\frac{19}{2}  \log(z_3)]  -\frac{23}{2} K ,
\end{eqnarray} 
  
From the topological string amplitudes, we can then extract the refined GV numbers up to genus one. We list these numbers in Appendix \ref{AppendixC1}.  Comparing with the tables in Appendix \ref{AppendixB1}, indeed we see that the GV numbers for the rows of base degrees $(0,d_3)$ are the same for elliptic fibrations over $\mathbb{F}_0$ and $\mathbb{F}_1$ for the unrefined cases $(g_L, g_R)=(0,0), (1,0)$, while they are different for the refined case of $(g_L,g_R)=(0,1)$.

It has been known that the local $\mathbb{F}_0$ and $\mathbb{F}_2$ models are related, e.g. in \cite{Huang:2013}. This is true also with the elliptic fibration. We have the relations between Gopakumar-Vafa numbers  
\begin{eqnarray} \label{F0F2}
(n_{d_E,d_2,d_3}^{j_L,j_R})_{ \mathbb{F}_2} = \left\{
\begin{array}{cl}
0,     &   ~d_3<d_2;   \\
(n_{d_E,d_2,d_3-d_2}^{j_L,j_R})_{ \mathbb{F}_0},              &  ~d_3\geq d_2.    
\end{array}
\right.
 \end{eqnarray}
 For completeness we also provide the Picard-Fuchs differential operators 
 \begin{eqnarray}
 \mathcal{L}_1 &=& \Theta_1(\Theta_1 - 2\Theta_3 ) -12 z_1 (6\Theta_1+5) (6\Theta_1+1) , \nonumber \\  
\mathcal{L}_2 &=& \Theta_3(\Theta_3-2\Theta_2)  - z_3 (2\Theta_3 -\Theta_1) (2\Theta_3 -\Theta_1+1)   , \nonumber \\ 
\mathcal{L}_3 &=&\Theta_2^2  - z_2 (2\Theta_2 -\Theta_3) (2\Theta_2 -\Theta_3+1) . 
 \end{eqnarray}
 The involution transformation and the conifold divisors are 
\begin{eqnarray}
(z_1, z_2, z_3) \rightarrow (x_1,x_2,x_3) \equiv  (\frac{1}{432} -z_1, z_2,  \frac{z_3z_1^2}{(\frac{1}{432}-z_1)^2} ). 
\end{eqnarray}
\begin{eqnarray}
\Delta_2(z_1,z_2,z_3)  &=& (1-4z_3)^2-64z_3^2z_2 ,  \nonumber \\
\Delta_1 (z_1, z_2, z_3) &=& (\frac{1}{432}-z_1)^4 \Delta_2 (x_1,x_2,x_3) , \nonumber \\
\Delta_3 &=& 1-4z_2. 
\end{eqnarray}
There is an additional discriminant divisor $\Delta_3$, comparing with the elliptic fibrations over $\mathbb{F}_0$ and  $\mathbb{F}_1$. The genus one formulas can be fixed as
\begin{eqnarray}
\mathcal{F}^{(0, 1)} &=& \frac{1}{2}(3+h^{1,1} -\frac{\chi}{12}) K + \frac{1}{2} \log\det G^{-1} -\frac{1}{12} \log (\Delta_1\Delta_2)   -\frac{1}{4} \log (\Delta_3) \nonumber \\&&  - \frac{1}{24} [104\log(z_1)+36\log(z_2)+60\log(z_3)]  ,  \nonumber \\
\mathcal{F}^{(1,0)} &=& \frac{1}{24}[ \log(\Delta_1\Delta_2) + 42 \log(z_1) + \frac{19}{2} \log(z_2) + 
19  \log(z_3)]  -\frac{23}{2} K,
\end{eqnarray}
where the hodge number $h^{1,1}=3$ and Euler character $\chi=-480$ are the same as the elliptic fibration over $\mathbb{F}_0$. 

The formulas for the partition function can be determined similarly and are also related as
\begin{eqnarray} \label{F0F2partition}
(Z_{d_2,d_3})_{ \mathbb{F}_2} = \left\{
\begin{array}{cl}
0,     &   ~d_3<d_2;   \\
(Z_{d_2,d_3-d_2})_{ \mathbb{F}_0},              &  ~d_3\geq d_2.    
\end{array}
\right.
 \end{eqnarray}

 \subsection{Derivation of modular anomaly equation: general case} 
We have seen that the modular anomaly equation for the topological string partition function can be appealingly derived from Witten's wave function interpretation. The alternative and more complicated approach, by carefully analyzing the Picard-Fuchs equation and BCOV holomorphic anomaly equation, is provided for the simplest case of elliptic fibration over $\mathbb{P}^2$ in section \ref{subsecproofgenus01}. This approach utilizes some results from Appendix D in our previous  paper \cite{Huang:2015sta}. Here we generalize the approach in section \ref{subsecproofgenus01} to the case of general base manifold, but restricted to the conventional unrefined case. We shall provide generalizations and some derivations of key formulas, without checking all the details that work out in the $\mathbb{P}^2$ case. We will see how the index $\frac{1}{2}[\beta]\cdot ([\beta] -c_1(B))$ in the anomaly equation comes about.

We denote the complex structure parameters, the corresponding mirror maps and exponentials of the elliptic fiber and the base $B$ as $z_e, z_i$, $t_e,t_i$ and $q_e,q_i$, where $i=1,\cdots, h_{1,1}(B)$. As in the Appendix D in \cite{Huang:2015sta}, we also need to define an auxiliary modular parameter $\tilde{q}$ which is purely a function of complex structure parameter of the elliptic fiber 
\begin{eqnarray}
J(\tilde{q}) = \frac{1}{z_e(1-432z_e)} . 
\end{eqnarray} 
In the small fiber (or large base) limit $z_i\rightarrow 0$, this is the same as the physical elliptic fiber parameter $q_e$. Similar to the $\mathbb{P}^2$ case in \ref{defineE23.145}, we will be working entirely in the holomorphic limit, and define the notations of $E_2$ derivatives 
\begin{eqnarray} \label{E2eq5.2}
\mathcal{L}_{E_2(q)} f := \partial_{E_2(q)} f(q, z_i), ~~~~~\partial_{E_2(q)} f := \partial_{E_2(q)} f(q, q_i), 
\end{eqnarray}
where $q$ can be either $\tilde{q}$ or $q_e$. We should caution the reader that $\mathcal{L}_{E_2}$ notation is not introduced in the previous paper \cite{Huang:2015sta}. In particular, in the Appendix D.2 in \cite{Huang:2015sta}, we still used the notation $\partial_{E_2(q)}$, while implicitly kept $z_i$ fixed in taking $E_2$ derivative. From now on, to avoid confusion, we strictly distinguish the two different notations (\ref{E2eq5.2}).  

In general, we can write the ansatz for the solutions of the Picard-Fuchs system which are power series of the base parameters as 
\begin{eqnarray} \label{ansatz5.3}
w = \sum_{n_1,n_2, \cdots ,n_{h_{1,1}(B)} =0}^{\infty} c_{n_1,n_2, \cdots, n_{h_{1,1}(B)} } (z_1)\prod_{i=1}^{h_{1,1}(B)} z_i^{n_i} . 
\end{eqnarray}
First we consider the leading term $c_0(z_1)\equiv c_{0, \cdots, 0 } (z_1)$. There are $h_{1,1}(B)+1$ differential equations. In the small fiber limit $z_i\rightarrow 0$,  only one equation (\ref{Le}) is non-trivial, and the Picard-Fuchs operator is universal $\mathcal{L}_e \rightarrow \theta_e^2 -12z_e (6\theta_e+1) (6\theta_e+5)$, where $\theta_e =z_e\partial_{z_e}$.  The purely power series solution has $c_0(z_1) = E_4(\tilde{q})^{\frac{1}{4}}$, while the single logarithmic solution of the fiber parameter has $c_0(z_1) = \tilde{t} E_4(\tilde{q})^{\frac{1}{4}}$. One can indeed easily check that they are annihilated by the fiber Picard-Fuchs operator $\mathcal{L}_e$ in this limit. 

The higher order terms in (\ref{ansatz5.3}) can be determined recursively by the other $h_{1,1}(B)$ differential equations. This is worked out in details for the $\mathbb{P}^2$ case in the Appendix D.2 in \cite{Huang:2015sta}. Mirror symmetry guarantees that the recursions are consistent with each other and with the fiber differential equation. To illustrate the procedure, we further work out the case of elliptic fibration over $\mathbb{F}_1$ model,  where the Picard-Fuchs system is in (\ref{PF4.1}). To confer with the previous notation, we still use $z_1$ as the fiber parameter and $z_2, z_3$ as the base parameters in this discussion of the $\mathbb{F}_1$ model. We see the first equation in (\ref{PF4.1}) is what we call the fiber equation. The other two equations give the following recursion relations 
\begin{eqnarray} \label{recursion5.4}
c_{m+1,n}(z_1) &=& \frac{n-m}{(m+1)^2} (\theta_1-m-2n)  c_{m,n}(z_1), \\ \nonumber 
c_{m,n+1}(z_1) &=& \frac{1}{(n+1-m)(n+1)}  (\theta_1-m-2n) (\theta_1-m-2n-1) c_{m,n}(z_1). 
\end{eqnarray} 
The recursion relations determine all higher order terms consistently from the initial coefficient $c_{0,0}(z_1)$. In fact, due to the factor $n-m$ in the first recursion, we immediately infer that $c_{m,n}(z_1) =0$ when $m>n$. For the non-vanishing case $m\leq n$, we can write down the general formula 
\begin{eqnarray}
c_{m,n}(z_1) =\frac{1}{m!^2 n! (n-m)!} \prod_{k=0}^{m+2n-1} (\theta_{z_1}-k)c_{0,0}(z_1) , ~~~ m\leq n,
\end{eqnarray}
which apparently solves the two recursion relations (\ref{recursion5.4}). 

We can compute the mirror map for the fiber parameter $t_1$ as the ratio of the two solutions with initial conditions $c_0(z_1) = \tilde{t} E_4(\tilde{q})^{\frac{1}{4}}, E_4(\tilde{q})^{\frac{1}{4}}$. A significant simplification occurs in the limit $z_3\rightarrow 0$, which is the half K3 case and was considered in \cite{Klemm:2012}. In this limit, the higher order terms all vanish since $c_{m,0}(z_1) =0$ for $m>0$. So the mirror map $t_1$ is the same as the auxiliary parameter $\tilde{t}$. However, for the generically compact Calabi-Yau models, this is no longer true. 

Now we return to the discussion of general case. We do not provide a general proof here, but many properties of the $\mathbb{P}^2$ should apply. In particular, up to an overall factor, the coefficient $c_{n_1,n_2, \cdots, n_{h_{1,1}(B)} } (z_1)$ can be written as a quasi-modular form of $\tilde{q}$, and it is linear in the $E_2(\tilde{q})$. The coefficients from the two initial conditions  $c_0(z_1) = \tilde{t} E_4(\tilde{q})^{\frac{1}{4}}, E_4(\tilde{q})^{\frac{1}{4}}$ are related, from which we get a nice formula 
\begin{eqnarray} \label{E2derivative9.6}
\mathcal{L}_{E_2(\tilde{q})} (t_e-\tilde{t})^{-1} =\frac{1}{12}. 
\end{eqnarray}
In the case of $\mathbb{P}^2$ model, we then derived the relation between $E_2(\tilde{q})$ and  $E_2(q_e)$ derivatives, and the actions on complex structure parameters and mirror maps. We believe the following formulas are universal to all models
\begin{eqnarray}  \label{univer9.7}
&& \mathcal{L}_{E_2(\tilde{q})} P_k(q_e(\tilde{q},z_i),z_i) = \mathcal{L}_{E_2(q_e)} P_k(q_e,z_i)  +\frac{k}{12} (t_e-\tilde{t})P_k(q_e,z_i),
\nonumber \\
&& \mathcal{L}_{E_2(q_e)} z_i = \mathcal{L}_{E_2(q_e)} z_e = \mathcal{L}_{E_2(q_e)} t_e =\mathcal{L}_{E_2(q_e)} w_0 =0, 
\end{eqnarray}
where $P_k$ is a series of $z_i$ whose series coefficients are rational functions of weight $k$ of Eisenstein series $E_2, E_4, E_6$, with possible fractional powers. For convenience of notation, from now on, we denote $\mathcal{L}_{E_2} \equiv  \mathcal{L}_{E_2(q_e)}$ and $\partial_{E_2} \equiv  \partial_{E_2(q_e)}$ unless the modular parameter is otherwise explicitly specified. 

The generalization of the last equation in (\ref{results3.145}) involves the classical intersection numbers in the base. It is well known that the classical part of the prepotential can be written by the triple intersection numbers of the Calabi-Yau manifold.  Here we have 
\begin{eqnarray}
\mathcal{F}^{(0)}_{classical} = \frac{c_e}{6} t_e^3 +\sum_{k} \frac{1}{2} a_k t_e^2 t_k + \sum_{i, j} \frac{1}{2} c_{ij} t_e t_i t_j, 
\end{eqnarray} 
where  $c_e, a_k, c_{ij}$ are the intersection numbers introduced in section \ref{subsecTopoproper}.  The upper index matrix $c^{ij}$ is the inverse matrix of $c_{ij}$, while $a^k= c^{ki}a_i$. A curve class $[\beta]$ in the base  can be written as $[\beta] = \sum_{k} d_k[\tilde{\mathcal{C}}^k]$, which is the curve classes that generate the Mori cone with the Kahler parameter $t_k$ and the corresponding non-negative integer degree $d_k$. The intersections with itself and with the canonical class $c_1(B)$ of the base are 
\begin{eqnarray}
[\beta]\cdot [\beta] = \sum_{i,j} d_id_j c^{ij},~~~~ [\beta]\cdot c_1(B) =\sum_k d_k a^k
\end{eqnarray}

We claim the generalization of  the last equation in (\ref{results3.145}) is 
\begin{eqnarray} \label{gene9.10}
\mathcal{L}_{E_2} t_i  = \frac{1}{12}\sum_j c^{ij} \partial_{t_j}P^{(0)}.
\end{eqnarray}
The arguments follow the $\mathbb{P}^2$ case in Appendix D.2 in our previous paper \cite{Huang:2015sta}. We first consider the solution to the Picard-Fuchs equation with the leading term $X_0\log(z_i)$, where $X_0$ is the pure power series solution. We can write the ansatz 
\begin{eqnarray}
X_i =E_4(\tilde{q})^{\frac{1}{4}} \log(z_i) +\xi(z_e) + \mathcal{O}(z_1,z_2, \cdots z_{h_{11}(B)}) . 
\end{eqnarray} 
Then the fiber differential equation (\ref{Le}) becomes  
\begin{eqnarray}
[\theta_e^2 -12z_e (6\theta_e+1) (6\theta_e+5)] \xi(z_e) =a^i \theta_e E_4(\tilde{q})^{\frac{1}{4}} . 
\end{eqnarray} 
From the $\mathbb{P}^2$ case we know there is an exact solution to this differential equation 
\begin{eqnarray} \label{solu9.13}
\xi(z_e) = -\frac{a^i}{2} E_4(\tilde{q})^{\frac{1}{4}} \log[\frac{\tilde{q}(1-432z_e)}{z_e}]. 
\end{eqnarray}
We note that we can add any linear combination of $E_4(\tilde{q})^{\frac{1}{4}} \log(\tilde{q})$ to $\xi(z_e)$, which is still a solution. The solution (\ref{solu9.13}) is chosen to cancel the logarithmic $\log(z_e)$ term, due to $z_e\sim q_e \sim \tilde{q}$ in the leading order. Since $q_i =\exp(\frac{X_i}{X_0})$, we see at the leading order, the combination of $q_i q_e^{\frac{a^i}{2}}$ appears if we expand around large volume limit. Inversely, the parameters $z_i$'s can be expanded as power series of $q_i q_e^{\frac{a^i}{2}}$ with coefficients as rational functions and fractional powers of quasi-modular forms. For example, for the $\mathbb{F}_1$ model,  we have the expansions (\ref{Z2ex4.33}), (\ref{Z3ex4.34}) with the reparametrization (\ref{Q23renor4.32}).

Again the recursion relations from the other Picard-Fuchs equations dictate that the higher coefficients in the series expansion of the base complex parameters are quasi-modular forms which are linear in $E_2(\tilde{q})$. So we can compute the derivatives $\mathcal{L}_{E_2(\tilde{q})} t_i$ and therefore $\mathcal{L}_{E_2(q_e)} t_i$, which could be possibly the instanton part of another  double logarithmic solution. Assuming this is indeed the case, the question is then what the classical part of the solution is. Generally, the higher coefficients with the leading term $X_0\log(z_k)$ for different $z_k$'s are not related. So the necessary condition here is that the classical part for matching $\mathcal{L}_{E_2} t_i$ involves only $t_i$ parameter but not the other base mirror maps. We see the following correct combination of partial derivative of the classical prepotential 
\begin{eqnarray}
\sum_j c^{ij} \partial_{t_j} \mathcal{F}^{(0)}_{classical} = \sum_j \frac{1}{2} c^{ij} a_j t_e^2  + \sum_{j,k} c^{ij} c_{jk} t_e t_k = \sum_j \frac{1}{2} c^{ij} a_j t_e^2  +t_et_i. 
\end{eqnarray}
So without checking the details, we infer that the only possibility is (\ref{gene9.10}), where the coefficient $\frac{1}{12}$ is taken from the calculations in the $\mathbb{P}^2$ case. 

We can then derive the relation between $\mathcal{L}_{E_2}$ and $\partial_{E_2}$ as in the $\mathbb{P}^2$ case 
\begin{eqnarray} \label{rela9.12}
\mathcal{L}_{E_2} f &= &\partial_{E_2} f +\sum_{i} (\partial_{t_i} f ) \mathcal{L}_{E_2} t_i  \nonumber \\
&=& \partial_{E_2} f +\frac{1}{12}\sum_{i,j} c^{ij} (\partial_{t_i} f)( \partial_{t_j}P^{(0)}).
\end{eqnarray}
where $\partial_{t_i} f  \equiv \partial_{t_i} f(t_e,t_1,t_2, \cdots, t_{h^{11}(B)}) $ is  the partial derivative as a function of flat coordinates.

We can indeed explicitly check the validity of (\ref{gene9.10}) when the base is a Hirzebruch surface $\mathbb{F}_n, n=0,1,2$. For the case of $\mathbb{F}_1$ model, the relevant expansion formulas are fixed by perturbative calculations in (\ref{Z2ex4.33}), (\ref{Z3ex4.34}), while similarly for the $\mathbb{F}_0$ model in  (\ref{Zexpansion6.49}). The modular anomaly equations (\ref{modularanomaly4.35}, \ref{anomaly6.50}) are equivalent to (\ref{gene9.10}) with the available intersection numbers in section \ref{subsecTopoproper}, when we convert to the $\mathcal{L}_{E_2}$ convention using (\ref{rela9.12}). 

In the case of $\mathbb{P}^2$ model, the partial derivative with respect to $E_2(\tilde{q})$ of the equation (\ref{gene9.10}) can be explicitly computed. Using the relation in the first line in (\ref{univer9.7}) and the fact that the both sides of the equation (\ref{gene9.10}) have modular weight -2, we claim that the following result is universal for all models
\begin{eqnarray}
\mathcal{L}_{E_2} (\partial_{t_i} P^{(0)}) =0 , ~~~~ i=1,2, \cdots ,  t_{h^{11}(B)}. 
\end{eqnarray}
We note that $\mathcal{L}_{E_2}$ does not commute with $\partial_{t_i}$ since they operate on different sets of variables, while $\partial_{E_2}$ does commute with $\partial_{t_i}$. Using the relation (\ref{rela9.12}) and integrate over $t_i$, we arrive at the genus zero modular anomaly equation 
\begin{eqnarray}
\partial_{E_2} P^{(0)} = -\frac{1}{24} \sum_{j,k} c^{jk} (\partial_{t_j} P^{(0)}) (\partial_{t_k} P^{(0)}). 
\end{eqnarray}

Now we consider higher genus. Using the relation (\ref{rela9.12}), the genus zero part in the anomaly equation can be absorbed by converting the left hand side to the $\mathcal{L}_{E_2}$ notation. For genus $g\geq 1$, we shall derive 
\allowdisplaybreaks[0]
\begin{eqnarray} \label{highergenus9.15}
\mathcal{L}_{E_2} P^{(g)} &=& -\frac{1}{24} \sum_{i,j} \sum_{n=1}^{g-1} c^{ij} (\partial_{t_i} P^{(n)} ) (\partial_{t_j} P^{(g-n)} ) + \frac{1 }{24} \sum_i a^i \partial_{t_i} P^{(g-1)} \nonumber \\ && -  \frac{1}{24} \sum_{i,j} c^{ij} \partial_{t_i} \partial_{t_j}  P^{(g-1)} . 
\end{eqnarray}  
\allowdisplaybreaks
First we provide the generalization of the useful formulas (\ref{E23.149}, \ref{E23.150}, \ref{E2second3.154}). We have 
\begin{eqnarray}   \label{usefor9.16}
\mathcal{L}_{E_2} \partial_{t_\alpha} z_\beta  &=& - \frac{1}{12}\sum_{i,j} c^{ij} (\partial_{t_i} z_\beta )  (\partial_{t_j} \partial_{t_\alpha}  P^{(0)}) , \\
\mathcal{L}_{E_2} \partial_{t_\alpha} \log w_0  
&=& \frac{1 }{12}\delta_{\alpha}^e  - \frac{1}{12}\sum_{ij} c^{ij} (\partial_{t_i}  \log w_0 )  (\partial_{t_j} \partial_{t_\alpha}  P^{(0)}) , \\
\mathcal{L}_{E_2} \partial_{t_\alpha} \partial_{t_\beta} z_\gamma  &=& \frac{1}{6} \delta^e_\alpha \delta^e_\beta (\partial_{t_e} - \sum_i \frac{a^i}{2} \partial_{t_i}) z_\gamma -\frac{1}{12}\sum_{i,j} c^{ij} [ (\partial_{t_i} z_\gamma )  (\partial_{t_j} \partial_{t_\alpha}  \partial_{t_\beta}  P^{(0)})  
\nonumber \\ &&
+ (\partial_{t_i} \partial_{t_\alpha} z_\gamma )  (\partial_{t_j} \partial_{t_\beta}  P^{(0)})  
 + (\partial_{t_i} \partial_{t_\beta} z_\gamma )  (\partial_{t_j} \partial_{t_\alpha}  P^{(0)})] 
\end{eqnarray} 
where the greek letters $\alpha, \beta, \gamma \cdots$ could label the fiber index $e$ or the base indices $i,j = 1,2,\cdots h_{1,1}(B)$.  The term $ \sum_i a^i\partial_{t_i}$ in the last equation comes from the fact that the complex parameters $z_i$'s can be expanded as power series of $q_i q_e^{\frac{a^i}{2}}$ with zero modular weight coefficients. For example, for the $\mathbb{F}_1$ model,  we have the expansions (\ref{Z2ex4.33}), (\ref{Z3ex4.34}) with the reparametrization (\ref{Q23renor4.32}). 

For the genus one amplitude, the only anomaly comes from the determinant of the coordinate transformation matrix between complex structure parameters and their mirror maps. We can generalize the computations for $\mathbb{P}^2$ case 
\begin{eqnarray}
\mathcal{L}_{E_2} (\mathcal{F}^{(1)}) &=& \frac{1}{2} \mathcal{L}_{E_2} \log(\det(\partial_{t_\alpha} z_\beta))
=\frac{1}{2} \sum_{\alpha, \beta} (\partial_{z_\beta} t_\alpha) \mathcal{L}_{E_2} (\partial_{t_\alpha} z_\beta) \nonumber 
\\ &=& -\frac{1}{24} \sum_{i,j} c^{ij} \partial_{t_i}\partial_{t_j} P^{(0)} ,
\end{eqnarray}
where we have used the formula (\ref{usefor9.16}). Comparing to the amplitude $P^{(1)}$ of only positive base degree contributions, the topological amplitude $\mathcal{F}^{(1)}$ has the extra pieces of classical contributions which is linear in the Kahler parameter, and the instanton contributions from only fiber degree.  According to the well-known geometric interpretation of genus one amplitude as the Ray-Singer torsion, the linear coefficients of the Kahler parameters are determined by the integrals of the second Chern class of the Calabi-Yau space combined with the corresponding curve classes, in (\ref{class2}). For the base classes we have  
\begin{eqnarray} \label{genusone9.20}
\mathcal{F}^{(1)} = -\frac{1}{2}\sum_i a_i (t_i + \frac{a^i}{2} t_e) + c \log\eta(q_e) + P^{(1)}  . 
\end{eqnarray} 
Here the coefficient of $\log\eta(q_e)$, which also contributes a linear term in $t_e$, is determined by the first equation in (\ref{class2}).  In any case, these contributions from the fiber parameter have no modular anomaly. From the useful formula (\ref{gene9.10}), we arrive at the anomaly equation for the positive base degree  genus one amplitude 
\begin{eqnarray}
\mathcal{L}_{E_2} P^{(1)} = \frac{1}{24} \sum_{i} a^i \partial_{t_i} P^{(0)} -\frac{1}{24} \sum_{i,j} c^{i j} \partial_{t_i} \partial_{t_j} P^{(0)} . 
\end{eqnarray} 

For genus $g\geq 2$, the derivations of (\ref{highergenus9.15}) are similar to the $\mathbb{P}^2$ case in subsection \ref{subsecproofgenus01}. We skip the details but only provide the generalization of the key formulas (\ref{E2Sij}, \ref{E2Si}, \ref{E2S}), which we believe should be 
\begin{eqnarray}
\mathcal{L}_{E_2} S^{\alpha \beta } &=&  -\frac{1}{12w_0^2} \sum_{i,j}c^{ij}  (\partial_{t_i} z_\alpha)  (\partial_{t_j} z_\beta),  \\
\mathcal{L}_{E_2} S^{\alpha} &=&  -\frac{1}{12w_0^2} \sum_{i,j}c^{ij} (\partial_{t_i} z_\alpha)  (\partial_{t_j} \log w_0),  
\\
\mathcal{L}_{E_2} S &=&  -\frac{1}{24w_0^2} \sum_{i,j}c^{ij}  (\partial_{t_i} \log w_0)(\partial_{t_j} \log w_0).  
\end{eqnarray} 
Similarly as in the genus one case, the term $a^i \partial_{t_i} P^{(0)}$ in (\ref{highergenus9.15}), due to the intersection of curve class of Kahler parameter $t_i$ with the canonical class of the base, comes from the classical contributions in the genus one topological string amplitude, i.e. the linear terms of $t_i$ in (\ref{genusone9.20}).

\section{Geometric Calculations}

In this section, we compute the refined GV numbers of elliptic fibrations in low base degree.

\subsection{Base Degree 0}
\label{basedeg0}

The algebro-geometric calculation of the refined Gopakumar-Vafa numbers for base degree~0 was already done in Section\ref{sec:geometry}.   In this section, we repeat part of that calculation using the language of \cite{GV2,KKV}, in preparation for calculations for nonzero base degree in Section~\ref{sec:baseg0}.

We first recall from Section~\ref{sec:geometry} that for any multiple of the fiber class $f$ we have identified $\widehat{\mathcal{M}}_{kf}\to \mathcal{M}_{kf}$ with the elliptic fibration $\pi:X\to B$ itself.  We also recall that the refined GV numbers could be extracted from the decomposition (\ref{eq:pushdown}), (\ref{eq:decomp})
\begin{equation}\label{eq:decompsec6}
R\pi_*\mathbb{C}[3]=P[1]\oplus IC(R^1\pi_*\mathbb{C}|_{B_0})\oplus P[-1],
\end{equation}
where $P$ is the perverse sheaf $\mathbb{C}[2]$ on $B$ and $ IC(R^1\pi_*\mathbb{C}|_{B_0})$ is the IC sheaf associated to the rank 2 local system of $H^1$ of the elliptic fiber over the complement $B^0$ of the discriminant curve.  We now compute the refined GV numbers using the methods of \cite{GV2,KKV} and relate to (\ref{eq:decompsec6}).

Letting $R_B,\ R_X$ be the Lefschetz $SL(2)$ representations of $B$ and $X$ respectively,  so we see that the $SL(2)\times SL(2)$
representation is of the form $[1/2,R_B]\oplus [0,R]$ for some representation $R$ and restricts to $R_X$ on the diagonal $SL(2)$.

Before continuing, let's pause to make contact with (\ref{eq:decompsec6}), from which we have seen that ${}^pR^{-1}\pi_*\mathbb{C}[3]\simeq {}^pR^{1}\pi_*\mathbb{C}[3]\simeq \mathbb{C}[2]$.  This completely determines the $SL(2)_L$ action by the hard Lefschetz actions which on ${}^pR^{-1}$ is the isomorphism
\begin{equation}\label{eq:hardl}
{}^pR^{-1}\pi_*\mathbb{C}[3]\simeq {}^pR^{1}\pi_*\mathbb{C}[3]\simeq \mathbb{C}[2],
\end{equation}
and is the trivial action on ${}^pR^{0}\pi_*\mathbb{C}[3]\simeq  IC(R^1\pi_*\mathbb{C}|_{B_0})$ (since $ {}^pR^2\pi_*\mathbb{C}[3]=0$).

Thus the terms in (\ref{eq:hardl}) correspond to $j_L=1/2$, with $j_R$ determined by the perverse sheaf $\mathbb{C}[2]$ on $B$, whose Poincar\'e polynomial matches the $SL(2)_R$ Lefschetz action on $B$.  This matches the first term $[1/2,R_B]$ above.

The since the hard Lefschetz action is trivial on ${}^pR^{0}\pi_*\mathbb{C}[3]\simeq  IC(R^1\pi_*\mathbb{C}|_{B_0})$, we have $j_L=0$.  Letting $R$ be the $SL(2)_R$-representation matching the Poincar\'e polynomial of $ IC(R^1\pi_*\mathbb{C}|_{B_0})$, we see that this term corresponds to the representation $[0,R]$.  Thus the geometric method matches the methods of \cite{GV2,KKV} perfectly up to this point.

Continuing, we know that the diagonal $SU(2)$ is $[1/2][R_B]+[R]$, which we equate to $R_X$.  
Therefore the representation is
\be
\left[\frac12,R_B\right]\oplus \left[0,R_X-\left(\left[\frac12\right]R_B\right)\right].
\label{gvdb0}
\ee
Note that $R_X-\left(\left[\frac12\right] R_B\right)$ is an actual $SL(2)$ representation rather than just a virtual 
representation.

Before completing the computation in our examples, we pause to make contact with the geometric methods of Section~\ref{sec:geometry}.   Letting $P_B(y)$ be the Poincar\'e polynomial of $\mathbb{C}_B[2]$ (which is $y^{-2}$ times the ordinary Poincar\'e polynomial of $B$), and letting $R(y)$ be the Poincar\'e polynomial of the IC sheaf, we extract from (\ref{eq:hardl}) the generating Laurent polynomial
\begin{equation}\label{eq:diag}
\left(u+u^{-1}\right)P_B(y)+R(y).
\end{equation}
From this, the ordinary (unrefined) Poincar\'e polynomial of $\mathbb{C}_X[3]$ is obtained by setting $u=y$.  Identifying Laurent polynomials with characters of $SL(2)$ representations, (\ref{eq:diag}) after $u=y$ becomes $[1/2][R_B]+[R]=[R_X]$, which completes the mathematical justification of the methods of \cite{GV2,KKV} in this situation.

\smallskip
For $B=\mathbb{P}^2$, we compute
\be
R_B=\left[1\right],\ R_M=\left[\frac32\right]+\left[\frac12\right]+546\left[0\right],
\ee
so by (\ref{gvdb0}) the refined GV representation is
\be
\left[\frac12,1\right]\oplus546\left[0,0\right].
\label{refinedBPSP}
\ee

\smallskip
For $B=\mathbb{F}_0$ or $B=\mathbb{F}_1$, we compute
\be
R_B=\left[1\right]\oplus\left[0\right],\ R_M=\left[\frac32\right]+2\left[\frac12\right]+488\left[0\right].
\label{Fdata}
\ee
So by (\ref{gvdb0}) the refined GV representation is
\be
\left[\frac12,1\right]\oplus\left[\frac12,0\right]\oplus488\left[0,0\right].
\label{refinedBPSF}
\ee

In particular, the refined GV numbers in class $(0,d)$ are given by (\ref{refinedBPSP}) for $B=\mathbb{P}^2$ and by
(\ref
{refinedBPSF}) for $B=\mathbb{F}_0$ or $B=\mathbb{F}_1$, for any $d>0$.  This
is in complete agreement with the results obtained using
refined holomorphic anomaly methods in Sections~\ref{refinedsec3.7} and \ref{refinedfano}.  The results in both the spin and integer bases
are displayed in Appendix~\ref{AppendixA1} for $B=\mathbb{P}^2$,
in Appendix~\ref{AppendixB0} for $B=\mathbb{F}_1$, and in
Appendix~\ref{AppendixC0} for $B=\mathbb{F}_0$.

\subsection{Base Curves of Genus~0}\label{sec:baseg0}

In this section, we give a geometric computation of the refined Gopakumar-Vafa numbers for 
certain base curves of genus~0 and $d_E=1$.  It is more convenient to present our calculations using the language and techniques of \cite{GV2,KKV} rather than using the formulation in Section~\ref{sec:geometry}.  We invite the interested reader to work out the translation.

Let $\beta'\in H_2(B,\mathbb{Z})$ be a genus~0 curve class.  We will
compute the refined numbers in the class $\beta=(\beta',1)$ if either
$\beta'$ is irreducible and $c_1(B)\cdot\beta'\ge2$ or $\beta'=\beta'_1+\beta'_2$
with $\beta'_i$ an irreducible genus~0 curve class with 
$c_1(B)\cdot\beta'_i\ge2$ for $i=1,2$.  These
assumptions are satisfied for $B=\mathbb{P}^2$ and $d_B=1$ or $d_B=2$, $B=
\mathbb{F}_0$ or $B=\mathbb{F}_1$ and $\beta'$ the class of a fiber of the Hirzebruch
surface, or $B=\mathbb{F}_0\simeq\mathbb{P}^1\times\mathbb{P}^1$ and $\beta'$
the class of the diagonal.

Under these assumptions, we claim that the curves $C$ of class $\beta$ are 
precisely the union of curves $C'$ of class $\beta'$ together 
with an elliptic
fiber meeting it.  Furthermore, all such curves are connected.

The argument is a straightforward extension of an argument in \cite{Huang:2015sta}.
For such curves, we have
\be
(\beta\cdot E)_M=-(\beta'\cdot c_1(B))_B+1<0.
\label{negint}
\ee
So if $C$ is a curve of class $\beta$, then a component $\tilde{C}$ of $C$ is 
necessarily contained in $E$.  

If $\beta'$ is an irreducible class, it follows 
that $\tilde{C}$ has class $\beta'$ and $C$ consists of the union of $\tilde{C}$ together 
with an elliptic fiber meeting $\tilde{C}$, and both parts of the claim
follow immediately for such classes.  

In the case $\beta'=\beta'_1+\beta'_2$,
we have that $C$ contains an irreducible curve $\tilde{C}$ which is 
in the class $\beta'$, $\beta'_1$, or $\beta'_2$.
If $\tilde{C}$ is in the class $\beta'$, we are already done by repeating the
previous argument.  Otherwise,
without loss of generality we can assume that $\tilde{C}$ is in the class $\beta'_1$, and we change notation, denoting $\tilde{C}$ by $C'_1$.
Write $C=C'_1+\hat{C}$ where $\hat{C}$ is in the curve class $(\beta'_2,1)$.  
Then
$c_1(B)\cdot \hat{C}=c_1(B)\cdot\beta'_2+1<0$, and we see by invoking the
irreducible case again that $\hat{C}$ is the union of a curve $C'_2$
of class $\beta'_2$ 
and an elliptic fiber meeting it.  The first part of the
claim follows with $C'=C'_1\cup C'_2$.
 For the connectedness, we need only observe that
$C'_1$ and $C'_2$ intersect.  Otherwise the union $C'_1\cup C'_2$ is a curve 
of arithmetic genus $-1$ in the genus~0 class $\beta'$, a contradiction.

We now compute the map $\widehat{\mathcal{M}}_\beta\to\mathcal{M}_\beta$ which takes 
a stable sheaf $F$ of class $\beta$ and $\chi(F)=1$ to its support curve
(formally an element of the Chow variety of cycle classes).  

We begin
with the Chow variety.  Denoting an elliptic fiber by $f$, our curves
are of the form $C=C'+f$.  We have a fibration
\begin{equation} 
\mathcal{M}_\beta\to B, \qquad C'+f\to q\in C'\cap f.
\label{chowxtob}
\end{equation}
Since the fiber of (\ref{chowxtob}) over $q\in B$ is the projective space of all curves of class $\beta'$ which contain $q$, and we know the Lefschetz
representation of $B$ in our examples, we can compute the Lefschetz
representation of the Chow variety.

We next
describe $\widehat{\mathcal{M}}_\beta$.  First, pick a point $q\in B$.
This determines a fiber $f_q$.  Next, as in the case $d_B=0$, for any $p\in f_q$, we get a sheaf $F'$ supported on the fiber $f_q$ with $\chi(F')=1$.

Next, take any curve $C'$ of class $\beta'$ 
containing $q$, noting as above that these are parametrized by a projective
space.  Letting $C=C'\cup f_p$, we have a sheaf $F$ on $C$ obtained by gluing $F'$ on $f_q$ to the sheaf $\mathcal{O}_{C'}$ on $C'$ at the point $q$.   Taking Euler characteristics in the normalization exact sequence
\be
0\to F\to F'\oplus \mathcal{O}_{C'}\to \mathcal{O}_q\to 0
\ee
and using $\chi(\mathcal{O}_{C'})=1$ from the genus zero assumption, we get $\chi(F)=1$ and so $F\in \widehat{\mathcal{M}}_\beta$.
 It is straightforward to check that every sheaf in
$\widehat{\mathcal{M}}_\beta$ is of this form.  We conclude that the Lefschetz representation
of $\widehat{\mathcal{M}}_\beta$ is the tensor product of the Lefschetz of $X$ with the
Lefschetz of the projective space of $C'$ (for any fixed $q$).

\smallskip
Let's work this out in our examples.

\medskip\noindent
$B=\mathbb{P}^2,\ d_B=1$.  The lines in $\mathbb{P}^2$ containing a fixed
$p\in \mathbb{P}^2$ are parametrized by $\mathbb{P}^1$, with Lefschetz
$[1/2]$, and the Lefschetz of $B$ is $[1]$.  So the Lefschetz
of $\mathcal{M}_\beta$ in this case is
\be
\left[1\right]\left[\frac12\right]=\left[\frac32\right]\oplus
\left[\frac12\right].
\label{chowd1}
\ee
Similarly, the Lefschetz for $\widehat{\mathcal{M}}_\beta$ is
\be
\left[\frac12\right]\left(\left[\frac32\right]\oplus\left[\frac12\right]\oplus
546\left[0\right]\right)
=
\left[2\right]\oplus
2\left[1\right]\oplus
\left[0\right]\oplus
546\left[\frac12\right]
\label{stabled1}
\ee
This forces the refined Gopakumar-Vafa numbers to be given by
\be
\left[\frac12,\frac32\right]\oplus
\left[\frac12,\frac12\right]\oplus
546\left[0,\frac12\right],
\label{refined11}
\ee
the unique representation whose right-spin partner of $j_L=1/2$ is 
(\ref{chowd1}) and whose restriction to the diagonal $SL(2)$ is
(\ref{stabled1}).  In other words
\be
N_\beta^{j_L,j_R}=\left\{
\begin{array}{cl}
1&(j_L,j_R)\in\{(1/2,3/2),(1/2,1/2)\\
546&(j_L,j_R)=(0,1/2)\\
0&{\rm otherwise}
\end{array}
\right.
\ee
This is in complete agreement with the results obtained
using the refined holomorphic anomaly in
Section~\ref{refinedsec3.7}.

\smallskip
The calculations in the other cases are entirely similar:

\medskip  
For $\mathbb{P}^2$ with
$d_B=2$, the curves $C'$ containing
$q$ are 
parametrized by a $\mathbb{P}^4$, with Lefschetz $[2]$.  The Hirzebruch 
surfaces have Lefschetz $[1]\oplus[0]$ while $X$ has Lefschetz
$[3/2]\oplus[1/2]\oplus 488[0]$ (\ref{Fdata}).  We repeat the calculation
of the refined Gopakumar-Vafa numbers
and have displayed the result in both the spin and integer bases
in Appendix~\ref{AppendixA1}.  We find
complete agreement with the results obtained using
refined holomorphic anomaly methods in Section~\ref{refinedsec3.7}, which
are recorded in the tables in Appendix~\ref{AppendixA2}.

\medskip
For $B=\mathbb{F}_n,\ n=0,1$, the Lefschetz of $B$ is $[1]+[0]$.  Starting
with $\beta'$ the fiber class of $\mathbb{F}_n$, the curve $C'$ through
a fixed $q\in B$ is unique.  So the Lefschetz of $\mathcal{M}_\beta$
is  $[1]+[0]$ in this case, while $\widehat{\mathcal{M}}_\beta$ is isomorphic to $X$,
with Lefschetz $[3/2]+2[1/2]+488[0]$.  
Finally, in the case $B=\mathbb{F}_0\simeq\mathbb{P}^1\times\mathbb{P}^1$
and $\beta'$ the diagonal class, the curves $C'$ containing $q$ are parametrized
by $\mathbb{P}^3$, with Lefschetz representation $[3/2]$.  

We repeat the calculation of the refined Gopakumar-Vafa numbers
and have displayed the results in both the spin and integer bases
in Appendix~\ref{AppendixB0} for $B=\mathbb{F}_1$ and in 
Appendix~\ref{AppendixC0} for $B=\mathbb{F}_0$.  We find
complete agreement with the results obtained using
refined holomorphic anomaly methods in Section~\ref{refinedfano}, which
are recorded in the tables in Appendix~\ref{AppendixB1} for $B=\mathbb{F}_1$
and Appendix~\ref{AppendixC1} for $B=\mathbb{F}_0$.  

\medskip
In other curve classes for $B=\mathbb{P}^2,\mathbb{F}_0$, or $\mathbb{F}_1$,
the moduli spaces of stable sheaves are not expected to be so simple, and
in particular can be singular and/or can depend on the complex structure 
moduli of $M$.  So there is no
obvious reason to expect from geometry alone that refined Gopakumar-Vafa numbers
should exist in all cases, as there could be orientation dependence, as explained in Section~\ref{sec:geometry}.

\section{Conclusion}
The refined Gopakumar-Vafa numbers do not yet have a precise geometric definition, while the physical definition 
coming from the 5d Lorentz representation on the space of states of the M2 brane wrapping holomorphic curves 
reinterprets concepts from heterotic strings on the Type II side but seems to give a precise and well defined 
route  to understand  the GVI's and their refinement. For a detailed more recent account from the physical point 
of view see~\cite{Dedushenko:2014nya}. Most calculations of the GVI's that check their properties on compact Calabi-Yau spaces  
are however done using indirect means. i.e. using the holomorphic anomaly and automorphic properties of the 
corresponding topological string amplitudes~\cite{MR2596635,Marino:1998,Grimm:2007tm,Hosono:2007vf,Haghighat:2008ut,Haghighat:2009nr,Klemm:2012,Alim:2012ss,Huang:2015sta}.    
    
The current geometric obstacles are twofold: first, the non uniqueness of orientations and the absence of a plausible ansatz when the mixed Hodge module 
underlying the relevant perverse sheaf is not a pure Hodge module.  It is possible that the first obstacle can be overcome, since the 
canonical orientation of Joyce and Upmeier may prove to be consistent with physics after further study.  But the second obstacle on 
purity will require entirely new ideas.

However, we want to raise another possibility: perhaps the quantization of the M2- brane or the D2-D0 brane moduli spaces that should correspond 
to $\widehat{\mathcal{M}}_{\beta,s}$ requires additional choices corresponding to an orientation.  Following ideas of \cite{GV2}, suppose we 
try to work out the twisted supersymmetric quantum mechanics SQM on the singular space $\widehat{\mathcal{M}}_{\beta,s}$ that 
corresponds  to the zero modes of the deformations of the branes. This  certainly requires to extend the analysis  
from the case of line bundles on smooth curves considered in~\cite{GV2} to more general sheaves $F$ as we have 
emphasized in this paper.  First order deformations that  coordinatize $\widehat{\mathcal{M}}_{\beta,s}$ should 
corresponds to $Ext^1(F,F)$, i.e. open string modes that deform the brane location and the $U(1)$ connections on the brane.  
In the smooth case this reduces  to the tangent space of $\widehat{\mathcal{M}}_{\beta}$. In the proper formulation of the 
SQM the basis elements of the tangent space are represented  by fermions and more general by fermions coming from $Ext^*(F,F)$.  
Then the path integral will involve a Pfaffian, which can be described by replacing $F$ by the universal sheaf 
on $\widehat{\mathcal{M}}_\beta\times X$.  On $\widehat{\mathcal{M}}_\beta$ the square of the corresponding Pfaffian 
bundle is the determinant line bundle $\det(Ext^*(F,F))$ which expands to
\begin{equation}\label{eq:det}
\det(Ext^0(F,F))\otimes \det(Ext^1(F,F))^*\otimes \det(Ext^2(F,F))\otimes \det(Ext^3(F,F))^*,
\end{equation}
which is just the canonical bundle $K_{\widehat{\mathcal{M}}_{\beta,s}}$ associated to the d-critical locus described in Section~\ref{subsec:dcrit}.  For example, if $\widehat{\mathcal{M}}_{\beta,s}$ is smooth, then $\det(Ext^1(F,F))$ is the tangent bundle of $\widehat{\mathcal{M}_\beta}$ and $\det(Ext^2(F,F))$ is the cotangent bundle of $\widehat{\mathcal{M}_\beta}$,  from which it follows that the respective factors in (\ref{eq:det}) are $\mathcal{O}_{\widehat{\mathcal{M}}_\beta},K_{\widehat{\mathcal{M}}_\beta}, K_{\widehat{\mathcal{M}}_\beta},\mathcal{O}_{\widehat{\mathcal{M}}_\beta}$, where $K_{\widehat{\mathcal{M}}_\beta}$ is the ordinary canonical bundle.  Accordingly, $\det(Ext^*(F,F))$ is the square of the ordinary canonical bundle, which is just $K_{\widehat{\mathcal{M}}_{\beta,0}}$.  In general, the Pfaffian must be a square root of $K_{\widehat{\mathcal{M}}_{\beta,s}}$, which is precisely an orientation.  This choice could conceivably depend on a choice of  a regularization of the singular 
configurations in the path integral.

In this work  have further proposed a refined holomorphic equation for elliptic fibrations, which agrees with the refined 
Gopakumar-Vafa numbers of particular classes, in particular base one.  Since agreement is not universal, the refined 
holomorphic anomaly equations need a correction in general.  This is the topic of further investigation.

\vspace{0.2in} {\leftline {\bf Acknowledgments}}

We thank Chris Dodd, Georg Oberdieck, Rahul Pandharipande  and Wati Taylor for discussions. MH thanks Bao-ning Du, Yuji Sugimoto, Xin Wang for collaborations on related papers.  The work of SK was supported in part by NSF grants 1502170 and 1802242. The work of MH was supported in parts by the national Natural Science Foundation of China (Grants No.11675167 and No.11947301).

\appendix

\section{Refined GV numbers for elliptic fibration over $\mathbb{P}^2$}
\label{AppendixA}

\subsection{The refined GV numbers for some low degrees}  \label{AppendixA1}
The complete refined GV numbers in the spin basis can be computed by algebro-geometric methods for some low degrees with $d_E>0$. We list the data for $\sum_{j_L,j_R} N^{j_L,j_R }_\beta [j_L, j_R] $ in the following. 
\begin{eqnarray}
\beta=(d_B,d_E)=(0,d_E): && ~~  [\frac{1}{2},1]+546 [0,0] , \nonumber \\
\beta=(d_B,d_E)=(1,1): && ~~  [\frac{1}{2}, \frac{3}{2}]+[\frac{1}{2}, \frac{1}{2}]+ 546[0,\frac{1}{2}]  , \nonumber \\
\beta=(d_B,d_E)=(2,1): && ~~  [\frac{1}{2} , 3]+ [\frac{1}{2}, 2] + [\frac{1}{2}, 1] + 546 [0, 2] . 
\label{eq:lowdeg}
\end{eqnarray}
We transform the refined GV numbers from the spin basis to the integer basis according to the formula (\ref{changebasis}), in order to compare with B-model calculations. The results are listed in the following tables. 

\begin{table} [H]
\begin{center}
\begin{tabular} {|c|c|c|c|} \hline $g_L \backslash g_R$  & 0 & 1 & 2 \\  \hline 0 & 540 & 8 & -2 \\  \hline 1 & 3 & -4 & 1 \\  \hline \end{tabular}
\caption{The refined GV numbers $n^{g_L,g_R }_\beta$ for the homology class $\beta=(d_B,d_E)=(0,d_E)$ in the integer basis. }
\label{GVtabledegree01}
\end{center} 
\end{table}

\begin{table} [H]
\begin{center}
 \begin{tabular} {|c|c|c|c|c|} \hline $g_L \backslash g_R$  & 0 & 1 & 2 & 3 \\  \hline 0 & -1080 & 524 & 12 & -2 \\  \hline 1 & -6 & 11 & -6 & 1 \\  \hline \end{tabular}
\caption{The refined GV numbers $n^{g_L,g_R }_\beta$ for the homology class $\beta=(d_B,d_E)=(1,1)$ in the integer basis. }
\label{GVtabledegree11}
\end{center} 
\end{table}

\begin{table} [H]
\begin{center}
 \begin{tabular} {|c|c|c|c|c|c|c|c|} \hline $g_L \backslash g_R$  & 0 & 1 & 2 & 3 & 4 & 5 & 6 \\  \hline 0 & 2700 & -10760 & 11170 & -4112 & 434 & 24 & -2 \\  \hline 1 & 15 & -80 & 148 & -128 & 56 & -12 & 1 \\  \hline \end{tabular}
\caption{The refined GV numbers $n^{g_L,g_R }_\beta$ for the homology class $\beta=(d_B,d_E)=(2,1)$ in the integer basis. }
\label{GVtabledegree21}
\end{center} 
\end{table}

\subsection{The refined GV numbers for some low genera} \label{AppendixA2}

We can also extract the refined Gopakumar-Vafa numbers from the exact B-model formula.  In the following we list the tables of GVN's for some low genera. The case of $g_R=0$ corresponds to no refinement and has been studied in our previous paper \cite{Huang:2015sta}. For completeness we also include them here. Due to space constraints, in each table we provide the GVN's up to some finite degrees, although in this approach we can in principle  compute for all homology classes $\beta=(d_B,d_E)$.

\begin{table} [H]
\begin{center}
  \begin{tabular} {|c|c|c|c|c|c|c|} \hline $d_B \backslash d_E$  & 0 & 1 & 2 & 3 & 4 & 5 \\  \hline 0 &  & 540 & 540 & 540 & 540 & 540 \\  \hline 1 & 3 & -1080 & 143370 & 204071184 & 21772947555 & 1076518252152 \\  \hline 2 & -6 & 2700 & -574560 & 74810520 & -49933059660 & 7772494870800 \\  \hline 3 & 27 & -17280 & 5051970 & -913383000 & 224108858700 & -42712135606368 \\  \hline 4 & -192 & 154440 & -57879900 & 13593850920 & -2953943334360 & 603778002921828 \\  \hline \end{tabular}
  \caption{The refined GV numbers $n^{g_L,g_R }_{(d_B,d_E)}$ for the genus $(g_L,g_R)=(0,0)$. }
\label{GVtablegenus00}
\end{center} 
\end{table}

\begin{table} [H]
\begin{center}
  \begin{tabular} {|c|c|c|c|c|c|c|} \hline $d_B \backslash d_E$  & 0 & 1 & 2 & 3 & 4 & 5 \\  \hline 0 &  & 3 & 3 & 3 & 3 & 3 \\  \hline 1 & 0 & -6 & 2142 & -280284 & -408993990 & -44771454090 \\  \hline 2 & 0 & 15 & -8574 & 2126358 & 521856996 & 1122213103092 \\  \hline 3 & -10 & 4764 & -1079298 & 152278986 & -16704086880 & -3328467399468 \\  \hline 4 & 231 & -154662 & 48907815 & -9759419622 & 1591062421074 & -186415241060547 \\  \hline \end{tabular}
  \caption{The refined GV numbers $n^{g_L,g_R }_{(d_B,d_E)}$ for the genus $(g_L,g_R)=(1,0)$. }
\label{GVtablegenus10}
\end{center} 
\end{table}

\begin{table} [H]
\begin{center}
  \begin{tabular} {|c|c|c|c|c|c|c|} \hline $d_B \backslash d_E$  & 0 & 1 & 2 & 3 & 4 & 5 \\  \hline 0 &  & 8 & 8 & 8 & 8 & 8 \\  \hline 1 & -4 & 524 & 4812 & 352294248 & 59107602110 & 3947178383364 \\  \hline 2 & 35 & -10760 & 1416596 & -95518872 & -93784873094 & 17417472357576 \\  \hline 3 & -386 & 193604 & -43121628 & 5704148756 & -636789454340 & 30691634708004 \\  \hline 4 & 5161 & -3477472 & 1076763820 & -205276478472 & 32984800267144 & -4518742293670008 \\  \hline \end{tabular}
   \caption{The refined GV numbers $n^{g_L,g_R }_{(d_B,d_E)}$ for the genus $(g_L,g_R)=(0,1)$. }
\label{GVtablegenus01}
\end{center} 
\end{table}

\begin{table} [H]
\begin{center}
  \begin{tabular} {|c|c|c|c|c|c|c|} \hline $d_B \backslash d_E$  & 0 & 1 & 2 & 3 & 4 & 5 \\  \hline 0 &  & 0 & 0 & 0 & 0 & 0 \\  \hline 1 & 0 & 0 & 9 & -3192 & 412965 & 614459160 \\  \hline 2 & 0 & 0 & -36 & 20826 & -5904756 & -47646003780 \\  \hline 3 & 0 & 27 & -16884 & 4768830 & -818096436 & 288137120463 \\  \hline 4 & -102 & 57456 & -15452514 & 2632083714 & -320511624876 & 18550698291252 \\  \hline \end{tabular}
   \caption{The refined GV numbers $n^{g_L,g_R }_{(d_B,d_E)}$ for the genus $(g_L,g_R)=(2,0)$. }
\label{GVtablegenus20}
\end{center} 
\end{table}

\begin{table} [H]
\begin{center}
  \begin{tabular} {|c|c|c|c|c|c|c|} \hline $d_B \backslash d_E$  & 0 & 1 & 2 & 3 & 4 & 5 \\  \hline 0 &  & -4 & -4 & -4 & -4 & -4 \\  \hline 1 & 0 & 11 & -2077 & 120998 & -499290011 & -96715491383 \\  \hline 2 & 0 & -80 & 29748 & -4403178 & 906250240 & 4096876264726 \\  \hline 3 & 165 & -63523 & 11346927 & -1225197859 & 92649730400 & -9503106690067 \\  \hline 4 & -7448 & 4228100 & -1121213060 & 185100882210 & -23541424385384 & 2199111416534168 \\  \hline \end{tabular}
  \caption{The refined GV numbers $n^{g_L,g_R }_{(d_B,d_E)}$ for the genus $(g_L,g_R)=(1,1)$. }
\label{GVtablegenus11}
\end{center} 
\end{table}

\begin{table} [H]
\begin{center}
 \begin{tabular} {|c|c|c|c|c|c|c|} \hline $d_B \backslash d_E$  & 0 & 1 & 2 & 3 & 4 & 5 \\  \hline 0 &  & 0 & 0 & 0 & 0 & 0 \\  \hline 1 & 0 & 0 & -24 & 5702 & -506080 & 234890706 \\  \hline 2 & 0 & 0 & 186 & -71844 & 12595603 & -97981973539 \\  \hline 3 & 0 & -396 & 193676 & -41678250 & 5241914434 & -518731287570 \\  \hline 4 & 3646 & -1763928 & 403642031 & -57868004916 & 5836810849841 & -359244664016416 \\  \hline \end{tabular}
\caption{The refined GV numbers $n^{g_L,g_R }_{(d_B,d_E)}$ for the genus $(g_L,g_R)=(2,1)$. }
\label{GVtablegenus21}
\end{center} 
\end{table}

\section{Refined GV numbers for elliptic fibration over $\mathbb{F}_1$}

\subsection{The GVN for some low degrees}  \label{AppendixB0}
The complete refined GV numbers in the spin basis can be computed by algebro-geometric methods for some low degrees with $d_E>0$. We list the data for $\sum_{j_L,j_R} N^{j_L,j_R }_\beta [j_L, j_R] $ in the following. 
\begin{eqnarray}
\beta=(d_E,d_2,d_3)=(d_E,0,0): && ~~  [\frac{1}{2},1]+[\frac{1}{2},0]+488 [0,0] , \nonumber \\
\beta=(d_E,d_2,d_3)=(1,0,1): && ~~  [\frac12,1]\oplus
[\frac12,0]\oplus
488[0,0]   \nonumber \\
\label{eq:lowdegF1}
\end{eqnarray}
We transform the refined GV numbers from the spin basis to the integer basis according to the formula (\ref{changebasis}), in order to compare with B-model calculations. The results are listed in the following tables. 

\begin{table} [H]
\begin{center}
\begin{tabular} {|c|c|c|c|} \hline $g_L \backslash g_R$  & 0 & 1 & 2 \\  \hline 0 & 480 & 8 & -2 \\  \hline 1 & 4 & -4 & 1 \\  \hline \end{tabular}
\caption{The refined GV numbers $n^{g_L,g_R }_\beta$ for the homology class $\beta=(d_E,d_2,d_3)=(d_E,0,0)$ in the integer basis. }
\label{GVtabledegreed00}
\end{center} 
\end{table}

\begin{table} [H]
\begin{center}
\begin{tabular} {|c|c|c|c|} \hline $g_L \backslash g_R$  & 0 & 1 & 2 \\  \hline 0 & 480 & 8 & -2 \\  \hline 1 & 4 & -4 & 1 \\  \hline \end{tabular}
\caption{The refined GV numbers $n^{g_L,g_R }_\beta$ for the homology class $\beta=(d_E,d_2,d_3)=(1,0,1)$ in the integer basis. }
\label{GVtabledegree101}
\end{center} 
\end{table}

\subsection{The refined GV numbers for some low genera}  \label{AppendixB1}
In the following we list the tables of GVN's for genus zero and one for the elliptic fibration over $\mathbb{F}_1$. The new results here are the refined case of $(g_L,g_R)=(0,1)$. 

\begin{table} [H]
\begin{center}
    \begin{tabular} {|c|c|c|c|c|c|c|} \hline $(d_2,d_3) \backslash d_E$  & 0 & 1 & 2 & 3 & 4 & 5 \\  \hline (0,0) &  & 480 & 480 & 480 & 480 & 480 \\  \hline (0,1) & -2 & 480 & 282888 & 17058560 & 477516780 & 8606976768 \\  \hline (1,0) & 1 & 252 & 5130 & 54760 & 419895 & 2587788 \\  \hline (0,2) & 0 & 0 & 480 & 17058560 & 8606976768 & 1242058447872 \\  \hline (1,1) & 3 & -960 & 118170 & 186280704 & 20229416355 & 1010446016832 \\  \hline (2,0) & 0 & 0 & -9252 & -673760 & -20534040 & -389320128 \\  \hline (0,3) & 0 & 0 & 0 & 480 & 477516780 & 1242058447872 \\  \hline (1,2) & 5 & -1920 & 339120 & -68861720 & 28474940475 & 58862632312080 \\  \hline (2,1) & 0 & 0 & -10260 & 1569600 & 1634529240 & 220572616320 \\  \hline (3,0) & 0 & 0 & 0 & 848628 & 115243155 & 6499779552 \\  \hline (0,4) & 0 & 0 & 0 & 0 & 480 & 8606976768 \\  \hline (1,3) & 7 & -2880 & 565200 & -137832960 & 24736992255 & -3999822251328 \\  \hline (2,2) & -6 & 2400 & -473640 & 58929120 & -45370251660 & 6216836921280 \\  \hline (3,1) & 0 & 0 & 0 & 1347520 & -112869600 & -227921708160 \\  \hline (4,0) & 0 & 0 & 0 & 0 & -114265008 & -23064530112 \\  \hline \end{tabular}
       \caption{The refined GV numbers $n^{g_L,g_R }_{(d_E,d_2,d_3)}$ for the genus $(g_L,g_R)=(0,0)$. }
\label{F1GVtablegenus00}
\end{center} 
\end{table}

\begin{table} [H]
\begin{center}
   \begin{tabular} {|c|c|c|c|c|c|c|} \hline $(d_2,d_3) \backslash d_E$  & 0 & 1 & 2 & 3 & 4 & 5 \\  \hline (0,0) &  & 4 & 4 & 4 & 4 & 4 \\  \hline (0,1) & 0 & 4 & -948 & -568640 & -35818260 & -1059654720 \\  \hline (1,0) & 0 & -2 & -510 & -11780 & -142330 & -1212930 \\  \hline (0,2) & 0 & 0 & 4 & -568640 & -1059654720 & -286327464192 \\  \hline (1,1) & 0 & -8 & 1866 & 162816 & -280202390 & -34775596920 \\  \hline (2,0) & 0 & 0 & 760 & 205320 & 11361360 & 317469648 \\  \hline (0,3) & 0 & 0 & 0 & 4 & -35818260 & -286327464192 \\  \hline (1,2) & 0 & -16 & 7128 & -210916 & 255670518 & 329307423480 \\  \hline (2,1) & 0 & 0 & 1020 & 82920 & -202216320 & -59382182160 \\  \hline (3,0) & 0 & 0 & 0 & -246790 & -76854240 & -6912918432 \\  \hline (0,4) & 0 & 0 & 0 & 0 & 4 & -1059654720 \\  \hline (1,3) & 0 & -24 & 11880 & -398272 & 1607476394 & 96076910040 \\  \hline (2,2) & 0 & 20 & -7472 & 1637424 & 334441748 & 1044271059552 \\  \hline (3,1) & 0 & 0 & 0 & -410640 & -49590440 & 78732120480 \\  \hline (4,0) & 0 & 0 & 0 & 0 & 76413833 & 27863327760 \\  \hline \end{tabular}
   \caption{The refined GV numbers $n^{g_L,g_R }_{(d_E,d_2,d_3)}$ for the genus $(g_L,g_R)=(1,0)$. }
\label{F1GVtablegenus10}
\end{center} 
\end{table}

\begin{table} [H]
\begin{center}
   \begin{tabular} {|c|c|c|c|c|c|c|} \hline $(d_2,d_3) \backslash d_E$  & 0 & 1 & 2 & 3 & 4 & 5 \\  \hline (0,0) &  & 8 & 8 & 8 & 8 & 8 \\  \hline (0,1) & 1 & 8 & 128176 & 15504320 & 647562510 & 15480542656 \\  \hline (1,0) & 0 & -2 & -510 & -11780 & -142330 & -1212930 \\  \hline (0,2) & 0 & 0 & 8 & 15504320 & 15480542656 & 3322992777984 \\  \hline (1,1) & -4 & 464 & 3762 & 352886448 & 59189746310 & 3953388356784 \\  \hline (2,0) & 0 & 0 & 5890 & 651720 & 26840460 & 641953488 \\  \hline (0,3) & 0 & 0 & 0 & 8 & 647562510 & 3322992777984 \\  \hline (1,2) & -20 & 4768 & -439344 & 1220028 & 64842259866 & 309306456337320 \\  \hline (2,1) & 0 & 0 & 6150 & 549840 & 477454500 & 81099456480 \\  \hline (3,0) & 0 & 0 & 0 & -1594310 & -287389440 & -20118277152 \\  \hline (0,4) & 0 & 0 & 0 & 0 & 8 & 15480542656 \\  \hline (1,3) & -56 & 16752 & -2239440 & 278129536 & -16024698142 & -2000902621296 \\  \hline (2,2) & 35 & -9560 & 1168556 & -75551112 & -96066035774 & 13950362465136 \\  \hline (3,1) & 0 & 0 & 0 & -1977200 & -110465080 & 139194539680 \\  \hline (4,0) & 0 & 0 & 0 & 0 & 435299183 & 108933415440 \\  \hline \end{tabular}
   \caption{The refined GV numbers $n^{g_L,g_R }_{(d_E,d_2,d_3)}$ for the genus $(g_L,g_R)=(0,1)$. }
\label{F1GVtablegenus01}
\end{center} 
\end{table}

\section{Refined GV numbers for elliptic fibration over $\mathbb{F}_0$}

\subsection{The GVN's for some low degrees}  \label{AppendixC0}
The complete refined GV numbers in the spin basis can be computed by algebro-geometric methods for some low degrees with $d_E>0$. We list the data for $\sum_{j_L,j_R} N^{j_L,j_R }_\beta [j_L, j_R] $ in the following. 
\begin{eqnarray}
\beta=(d_E,d_2,d_3)=(d_E,0,0): && ~~  [\frac{1}{2},1]+[\frac{1}{2},0]+488 [0,0] , \nonumber \\
\beta=(d_E,d_2,d_3)=(1,0,1): && ~~  [\frac12,1]\oplus
[\frac12,0]\oplus
488[0,0]  , \nonumber \\
\beta=(d_E,d_2,d_3)=(1,1,1): && ~~  [\frac12,2]\oplus
2[\frac12,1]\oplus [\frac12,0]\oplus
488[0,1]  \nonumber \\
\label{eq:lowdegF0}
\end{eqnarray}
We transform the GVN's from the spin basis to the integer basis according to the formula (\ref{changebasis}), in order to compare with B-model calculations. The results are listed in the following tables. 

\begin{table} [H]
\begin{center}
\begin{tabular} {|c|c|c|c|} \hline $g_L \backslash g_R$  & 0 & 1 & 2 \\  \hline 0 & 480 & 8 & -2 \\  \hline 1 & 4 & -4 & 1 \\  \hline \end{tabular}
\caption{The refined GV numbers $n^{g_L,g_R }_\beta$ for the homology class $\beta=(d_E,d_2,d_3)=(d_E,0,0)$ in the integer basis. }
\label{GVtabledegreeF0d00}
\end{center} 
\end{table}

\begin{table} [H]
\begin{center}
\begin{tabular} {|c|c|c|c|} \hline $g_L \backslash g_R$  & 0 & 1 & 2 \\  \hline 0 & 480 & 8 & -2 \\  \hline 1 & 4 & -4 & 1 \\  \hline \end{tabular}
\caption{The refined GV numbers $n^{g_L,g_R }_\beta$ for the homology class $\beta=(d_E,d_2,d_3)=(1,0,1)$ in the integer basis. }
\label{GVtabledegreeF0101}
\end{center} 
\end{table}

\begin{table} [H]
\begin{center}
 \begin{tabular} {|c|c|c|c|c|c|} \hline $g_L \backslash g_R$  & 0 & 1 & 2 & 3 & 4  \\  \hline 0 & 1440 & -1896 & 442 & 16 & -2  \\  \hline 1 & 12 & -28 & 23 & -8  & 1 \\  \hline \end{tabular}
\caption{The refined GV numbers $n^{g_L,g_R }_\beta$ for the homology class $\beta=(d_E,d_2,d_3)=(1,1,1)$ in the integer basis. }
\label{GVtabledegreeF0111}
\end{center} 
\end{table}

\subsection{The refined GV numbers for some low genera}  \label{AppendixC1}
In the following we list the tables of GVN's for genus zero and one for the elliptic fibration over $\mathbb{F}_0$. The new results here are the refined case of $(g_L,g_R)=(0,1)$. We only list the rows of $d_2\leq d_3$ since the other cases are related by symmetry of the two base moduli.

\begin{table} [H]
\begin{center}
  \begin{tabular} {|c|c|c|c|c|c|c|} \hline $(d_2,d_3) \backslash d_E$  & 0 & 1 & 2 & 3 & 4 & 5 \\  \hline (0,0) &  & 480 & 480 & 480 & 480 & 480 \\  \hline (0,1) & -2 & 480 & 282888 & 17058560 & 477516780 & 8606976768 \\  \hline (0,2) & 0 & 0 & 480 & 17058560 & 8606976768 & 1242058447872 \\  \hline (1,1) & -4 & 1440 & -226080 & 51516800 & 107913873744 & 17263743561792 \\  \hline (0,3) & 0 & 0 & 0 & 480 & 477516780 & 1242058447872 \\  \hline (1,2) & -6 & 2400 & -452160 & 103374720 & -16013531460 & 19768877695872 \\  \hline (0,4) & 0 & 0 & 0 & 0 & 480 & 8606976768 \\  \hline (1,3) & -8 & 3360 & -678240 & 172291200 & -32982096480 & 7524009379008 \\  \hline (2,2) & -32 & 16800 & -4093920 & 789875200 & -115113738240 & 63247732459200 \\  \hline \end{tabular}
          \caption{The refined GV numbers $n^{g_L,g_R }_{(d_E,d_2,d_3)}$ for the genus $(g_L,g_R)=(0,0)$. }
\label{F0GVtablegenus00}
\end{center} 
\end{table}

\begin{table} [H]
\begin{center}
 \begin{tabular} {|c|c|c|c|c|c|c|} \hline $(d_2,d_3) \backslash d_E$  & 0 & 1 & 2 & 3 & 4 & 5 \\  \hline (0,0) &  & 4 & 4 & 4 & 4 & 4 \\  \hline (0,1) & 0 & 4 & -948 & -568640 & -35818260 & -1059654720 \\  \hline (0,2) & 0 & 0 & 4 & -568640 & -1059654720 & -286327464192 \\  \hline (1,1) & 0 & 12 & -4752 & -469072 & 96219816 & -120434126760 \\  \hline (0,3) & 0 & 0 & 0 & 4 & -35818260 & -286327464192 \\  \hline (1,2) & 0 & 20 & -9504 & 298704 & -1107564076 & -185860380240 \\  \hline (0,4) & 0 & 0 & 0 & 0 & 4 & -1059654720 \\  \hline (1,3) & 0 & 28 & -14256 & 497840 & -2143491632 & -197338415160 \\  \hline (2,2) & 9 & -3700 & 711228 & -88757120 & 3031113384 & 556573294728 \\  \hline \end{tabular}
     \caption{The  refined GV numbers $n^{g_L,g_R }_{(d_E,d_2,d_3)}$ for the genus $(g_L,g_R)=(1,0)$. }
\label{F0GVtablegenus10}
\end{center} 
\end{table}

\begin{table} [H]
\begin{center}
 \begin{tabular} {|c|c|c|c|c|c|c|} \hline $(d_2,d_3) \backslash d_E$  & 0 & 1 & 2 & 3 & 4 & 5 \\  \hline (0,0) &  & 8 & 8 & 8 & 8 & 8 \\  \hline (0,1) & 1 & 8 & 172836 & 19077120 & 766804710 & 17922908736 \\  \hline (0,2) & 0 & 0 & 8 & 19077120 & 17922908736 & 3746592520704 \\  \hline (1,1) & 10 & -1896 & 104496 & 39263136 & 365903374416 & 83835642518640 \\  \hline (0,3) & 0 & 0 & 0 & 8 & 766804710 & 3746592520704 \\  \hline (1,2) & 35 & -9560 & 1113312 & -77274912 & -4768648682 & 88685860400160 \\  \hline (0,4) & 0 & 0 & 0 & 0 & 8 & 17922908736 \\  \hline (1,3) & 84 & -26824 & 3930768 & -588234720 & 54893286176 & 4548040016976 \\  \hline (2,2) & 359 & -143720 & 25655016 & -3175372800 & 250122441528 & 172858960782288 \\  \hline \end{tabular}
    \caption{The refined GV numbers $n^{g_L,g_R }_{(d_E,d_2,d_3)}$ for the genus $(g_L,g_R)=(0,1)$. }
\label{F0GVtablegenus01}
\end{center} 
\end{table}

\addcontentsline{toc}{section}{References}

\bibliographystyle{utphys}
\bibliography{RefinedCalabiYau}

\end{document}